\documentclass[10pt,journal,compsoc]{IEEEtran}

\usepackage[utf8]{inputenc}
\usepackage[T1]{fontenc}
\usepackage{graphicx}
\graphicspath{{figures/}}
\usepackage{amsmath}
\usepackage{amssymb}
\usepackage{booktabs}
\usepackage{multirow}
\usepackage{makecell}
\usepackage{array}
\usepackage{url}
\usepackage{xcolor}
\usepackage{stfloats}
\usepackage{placeins}
\usepackage{hyperref}
\definecolor{SectionLinkColor}{HTML}{1565C0}
\definecolor{FigureLinkColor}{HTML}{00838F}
\definecolor{TableLinkColor}{HTML}{6A1B9A}
\definecolor{EquationLinkColor}{HTML}{C62828}
\definecolor{CitationLinkColor}{HTML}{2E7D32}
\definecolor{DefinitionColor}{HTML}{8B1A1A}
\definecolor{CoreStatementColor}{HTML}{0B5394}
\definecolor{ConceptColor}{HTML}{6A1B9A}

\hypersetup{
  colorlinks=true,
  linkcolor=SectionLinkColor,
  citecolor=CitationLinkColor,
  urlcolor=FigureLinkColor,
  breaklinks=true,
  pdfauthor={Jianhong Ma, Wenhao Li, Xixing Li, Chenyu Yan, and Haowen Chen},
  pdftitle={Towards Early Warning of Real-World Vulnerabilities}
}

\newcommand{\hypersection}[1]{\hyperref[#1]{\textcolor{SectionLinkColor}{\S~\ref*{#1}}}}
\newcommand{\hyperfigure}[1]{\hyperref[#1]{\textcolor{FigureLinkColor}{Figure~\ref*{#1}}}}
\newcommand{\hypertable}[1]{\hyperref[#1]{\textcolor{TableLinkColor}{Table~\ref*{#1}}}}
\newcommand{\hyperequation}[1]{\hyperref[#1]{\textcolor{EquationLinkColor}{Equation~\ref*{#1}}}}
\newcommand{\hyperappendix}[1]{\textcolor{SectionLinkColor}{the supplementary material}}
\newcommand{\termdef}[1]{\textcolor{DefinitionColor}{\textbf{#1}}}
\newcommand{\coreclaim}[1]{\textcolor{CoreStatementColor}{\textbf{#1}}}

\newcommand{\namedconcept}[1]{\textcolor{ConceptColor}{\emph{#1}}}
\author{Jianhong Ma, Wenhao Li, Xixing Li, Chenyu Yan, and Haowen Chen%
\IEEEcompsocitemizethanks{%
\IEEEcompsocthanksitem J. Ma and W. Li are with the School of Cyber Science and Engineering, Zhengzhou University, Zhengzhou, China. E-mail: majianhong@zzu.edu.cn and liuniu467@gmail.com.
\IEEEcompsocthanksitem X. Li, C. Yan, and H. Chen are with Information Engineering University, Zhengzhou, China. E-mail: scholar\_xx@whu.edu.cn, yan\_cyu@163.com, and hwc\_zzu@126.com.
\IEEEcompsocthanksitem X. Li is also with the Key Laboratory of Cyberspace Security, Ministry of Education of China.
\IEEEcompsocthanksitem Corresponding author: Xixing Li (scholar\_xx@whu.edu.cn).}}
\title{Towards Early Warning of Real-World Vulnerabilities: An Evidence-Aware Decoupled Method for CVSS Vector Prediction Based on Unstructured Disclosures}

\begin{document}
\maketitle

\begin{center}
\footnotesize
This work has been submitted to the IEEE for possible publication.
Copyright may be transferred without notice, after which this version may no longer be accessible.
\end{center}

\begin{abstract}
Early assessment of Common Vulnerability Scoring System (CVSS) vectors from Open Source Intelligence (OSINT) can reduce the delay between initial public disclosure and official National Vulnerability Database (NVD) scoring.
% 基于开源情报（OSINT）对通用漏洞评分系统（CVSS）向量进行早期评估，可以缩短从首次公开披露到国家漏洞数据库（NVD）正式评分之间的延迟。
Existing predictors, trained mainly on standardized late-stage descriptions, degrade on noisy and incomplete pre-NVD reports.
% 然而，现有预测器主要在标准化的后期描述上训练，应用于含噪且不完整的 NVD 前期报告时性能会退化。
We introduce PARSE, a vulnerability-disclosure lifecycle model, and VPre, an early assessment framework built on two-level Evidence-Aware Decoupling.
% 我们提出了 PARSE 漏洞披露生命周期模型，以及建立在两层证据感知解耦之上的早期评估框架 VPre。
Decoupling Level I, \emph{Security-Evidence Screening}, separates security-relevant evidence from noisy early reports at the report and sentence levels.
% 解耦第一层为安全证据筛选，在报告级与句子级两个层面从含噪早期报告中分离安全相关证据。
Decoupling Level II, \emph{Evidence-Matched Routing}, decouples the prediction task of the eight CVSS metrics into a scattered-evidence class and an explicit-evidence class based on validation-set measurements, matching each with a targeted predictor: large-language-model reasoning for the former, lightweight supervised classification for the latter.
% 解耦第二层为证据匹配路由，依据验证集实测，将八个 CVSS 指标的预测任务解耦为分散证据类与显式证据类，并为每类匹配针对性的预测器：前者由大语言模型推理处理，后者由轻量监督模型分类处理。
On the 851-report held-out test split of the 5,581-report VPre-Bench, VPre achieves a mean absolute error of 0.7239---44.7\% below the reproduced BERT-Small-NVD early-transfer baseline; paired timestamps across the full benchmark show an average 38-day lead before official scoring (observed range 21--60 days).
% 在包含 5,581 份报告的 VPre-Bench 中，VPre 在 851 份报告的留出测试集上取得 0.7239 的平均绝对误差，比复现的 BERT-Small-NVD 早期迁移基线低 44.7\%；完整基准的配对时间戳显示，其平均可在正式评分前 38 天给出估计（观测范围 21--60 天）。
Under the 38:1 PARSE-Triage noise setting, STDU's report-level screening reaches 0.98 precision, 1.00 recall, and 0.99 F1.
% 在 PARSE-Triage 的 38:1 噪声设置下，STDU 的报告级筛选达到 0.98 的精确率、1.00 的召回率和 0.99 的 F1 值。
Controlled ablations support the contributions of both decoupling levels.
% 受控消融实验支持了两层解耦各自的贡献。
\end{abstract}


\section{Introduction}
% 引言。
\label{sec:introduction}
%\noindent\textbf{Motivation: Early CVSS Forecasting Before Standardized Assessment.}
%研究动机：在标准化评估之前进行早期 CVSS 预报。
The volume of publicly disclosed vulnerabilities has increased rapidly, while the manual analysis and scoring processes of the National Vulnerability Database (NVD) have struggled to keep pace~\cite{ref15,ref16}.
%公开披露的漏洞数量快速增长，而国家漏洞数据库（NVD）的人工分析与评分流程难以及时跟上这一增长速度。
Published CVE Records increased by 38.4\% in 2024, while only 6.6\% of the CVEs added during the February--May NVD processing slowdown had been analyzed~\cite{refCVEMetrics,refVulnCheckBacklog2024}.
%2024 年发布的 CVE Records 数量增长了 38.4\%，而在 2 月至 5 月的 NVD 处理放缓期间，新增 CVE 中仅有 6.6\% 完成了分析。
More importantly, a substantial delay often separates the initial public discussion of a vulnerability from the publication of its official NVD assessment.
%更重要的是，从漏洞最初被公开讨论到 NVD 发布正式评估结果之间，往往存在较长的时间延迟。
During this interval, issue trackers, mailing lists, bug reports, and vulnerability disclosure platforms may already contain evidence about attack conditions and security impact, even though no standardized description or official CVSS score is available.
%在这一时间窗口内，问题跟踪器、邮件列表、缺陷报告和漏洞披露平台中可能已经包含攻击条件与安全影响方面的证据，但此时标准化描述和正式 CVSS 评分尚不可用。
This creates an important early-warning opportunity: \coreclaim{\emph{Can CVSS vectors and base scores be reliably estimated from early, unstructured OSINT before CVE assignment and NVD standardization?}}
%这带来了一个重要的早期预警机会：在 CVE 编号分配和 NVD 标准化之前，能否从早期非结构化 OSINT 中可靠地估计 CVSS 向量与基础分数？
Our objective is to perform this prediction in the Pre-$T_0$ stage and to quantify how much earlier the resulting estimates can be produced than official NVD scoring.
%我们的目标是在 Pre-$T_0$ 阶段完成上述预测，并量化其相对于 NVD 正式评分能够提前多长时间给出估计。
\par

%\noindent\textbf{Research Gap and Lifecycle Boundary: From Late-Stage NVD Prediction to Pre-$T_0$ OSINT.}
%研究缺口与生命周期边界：从后期 NVD 预测到 Pre-$T_0$ OSINT。
Most existing automated CVSS prediction methods are trained and evaluated on standardized NVD descriptions, in which vulnerabilities have already been confirmed and their security properties are expressed in concise, expert-authored language~\cite{ref20,ref21,ref24,ref31,ref52,ref66}.
%现有自动化 CVSS 预测方法主要在标准化 NVD 描述上训练和评估；在这些描述中，漏洞已经得到确认，其安全属性也已由专家以简洁语言明确表达。
Early OSINT belongs to a fundamentally different information regime: before $T_0$, no CVE identifier or standardized description is available, and the evidence remains incomplete, uncertain, and highly unstructured~\cite{ref07,ref09}.
%早期 OSINT 属于一种根本不同的信息环境：在 $T_0$ 之前，CVE 编号和标准化描述均不可用，相关证据仍然不完整、不确定且高度非结构化。
This difference directly affects the reliability of cross-phase prediction.
%这种差异会直接影响跨阶段预测的可靠性。
In our reproduced cross-phase evaluation, BERT-Small-NVD achieves an MAE of 0.8918 on standardized NVD descriptions but degrades to 1.3095 when directly transferred to early OSINT, corresponding to a 47\% increase in prediction error.
%在我们复现的跨阶段评估中，BERT-Small-NVD 在标准化 NVD 描述上的 MAE 为 0.8918，但直接迁移到早期 OSINT 后退化至 1.3095，预测误差增加了 47\%。
Early CVSS forecasting is therefore not simply a matter of moving existing methods to an earlier stage; it constitutes a distinct prediction task under a different information regime.
%因此，早期 CVSS 预报并不是简单地将现有方法前移，而是在不同信息环境下进行的一项独立预测任务。
To formalize this temporal and informational mismatch, we introduce the \namedconcept{PARSE vulnerability disclosure lifecycle}: \textbf{P} (Probing), which occurs before $T_0$ and captures the discovery of a bug or anomaly in highly unstructured text; \textbf{A} (Acknowledgment), which extends through the Pre-$T_0$ reporting and confirmation process to $T_0$; \textbf{R} (Reservation), which spans $T_0$--$T_1$ after a CVE identifier is assigned; \textbf{S} (Standardization), which progresses through $T_1$--$T_2$--$T_3$, with disclosure, description standardization, and final scoring as successive milestones; and \textbf{E} (Evaluation), which begins after $T_3$, when the finalized NVD description and official CVSS score are available for downstream evaluation.
% 为形式化描述这种时间与信息条件的不匹配，我们提出 PARSE 漏洞披露生命周期：P 阶段发生在 T0 之前，对应缺陷或异常的发现；A 阶段覆盖 Pre-T0 的报告、复现和确认过程并终止于 T0；R 阶段覆盖 T0 至 T1；S 阶段依次经过 T1、T2 和 T3 三个里程碑；E 阶段始于 T3 之后，此时最终 NVD 描述与官方 CVSS 评分已可用于后续评估。
PARSE provides a unified basis for specifying when prediction occurs and what information is available at that point.
%PARSE 为明确预测发生的阶段以及该阶段可获得的信息提供了统一基础。
VPre targets the P/A phases before $T_0$, whereas most existing CVSS prediction methods operate in the later S/E regime after vulnerability confirmation and description standardization.
%VPre 面向 $T_0$ 之前的 P/A 阶段，而大多数现有 CVSS 预测方法运行在漏洞确认和描述标准化之后的 S/E 阶段。

%\noindent\textbf{Three Core Observations.}
%三个核心观察。
After locating our work within the PARSE lifecycle, we identify three interconnected challenges in early forecasting.
%明确本工作在 PARSE 生命周期中的定位后，我们发现早期预报存在三个相互关联的挑战。
\termdef{First, cross-phase semantic drift:} standardized NVD descriptions are concise and conclusive, whereas Pre-$T_0$ OSINT is incomplete, uncertain, and highly unstructured, making feature mappings learned from late-stage text unreliable on early reports~\cite{ref17}.
%第一，跨阶段语义漂移：标准化 NVD 描述简洁且结论明确，而 Pre-$T_0$ OSINT 不完整、不确定且高度非结构化，使得从后期文本中学习到的特征映射难以可靠应用于早期报告。
\termdef{Second, security-evidence dilution:} early OSINT is characterized by a mixture of genuine vulnerability reports and non-security bugs; moreover, even valid vulnerability reports contain substantial content unrelated to CVSS assessment~\cite{ref05,ref50,ref51,ref65}.
%第二，安全证据稀释：早期 OSINT 的特点是真实漏洞报告与非安全缺陷并存；此外，即使是有效漏洞报告，也包含大量与 CVSS 评估无关的内容。
\termdef{Third, heterogeneous requirements across CVSS metrics:} the eight CVSS v3.1 Base metrics---AV, AC, PR, UI, S, C, I, and A---differ substantially in their evidence characteristics and reasoning requirements, so that a uniform predictor is unlikely to serve all dimensions equally well~\cite{ref13,ref37}.
%第三，CVSS 度量之间存在异质性需求：CVSS v3.1 的八个基础度量（AV、AC、PR、UI、S、C、I、A）在证据特征与推理需求上差异显著，单一统一预测器难以同时在所有维度上取得最优。
Treating all eight metrics as a single prediction problem therefore forces a uniform model to handle qualitatively different forms of evidence, making reliable prediction across all dimensions difficult.
%因此，将八个度量视为同一种预测问题，会迫使统一模型处理性质不同的证据，从而难以在所有维度上实现可靠预测。
Together, these observations show that early CVSS forecasting differs from late-stage prediction at the disclosure, input-evidence, and prediction-target levels.
%综合来看，这些观察表明，早期 CVSS 预报与后期预测在披露阶段、输入证据和预测目标三个层次均存在差异。
\par

%\noindent\textbf{From Core Observations to VPre: Evidence-Aware Decoupling.}
%从核心观察到 VPre：证据感知解耦。
The core idea of \namedconcept{VPre} is to first isolate security-relevant evidence from noisy early reports and then decouple the prediction of the eight CVSS Base metrics according to their evidence characteristics and reasoning requirements.
%VPre 的核心思想是，首先从含噪的早期报告中筛选安全相关证据，随后依据八个 CVSS 基础度量各自的证据特征和推理需求，对其预测任务进行解耦。
We refer to this two-level design as \termdef{Evidence-Aware Decoupling}. At the first level, Security-Evidence Screening decouples evidence preparation from CVSS prediction through report-level triage and sentence-level filtering; at the second, validation-derived evidence density decouples a uniform eight-metric predictor into two specialized prediction routes.
%我们将这一两层设计称为证据感知解耦：第一层为安全证据筛选，通过报告级分诊和句子级过滤将证据准备与 CVSS 预测解耦；第二层依据验证集证据密度，将统一的八指标预测器解耦为两条专门化预测路径。
\textbf{First,} VPre is designed for early P/A-stage OSINT rather than directly transferring a model trained on late-stage NVD descriptions.
%第一，VPre 专门面向早期 P/A 阶段的 OSINT，而不是直接迁移在后期 NVD 描述上训练的模型。
\textbf{Second,} the first level, Security-Evidence Screening, performs report-level security triage and sentence-level VT/CP/O screening, removing irrelevant content before prediction.
%第二，第一层安全证据筛选执行报告级安全分诊和句子级 VT/CP/O 筛选，在预测前移除无关内容。
\textbf{Third,} VPre uses the dimension-level evidence-density function $D(m)$ on validation data to establish a fixed SER/EEC routing table for the eight CVSS Base metrics.
%第三，VPre 在验证数据上使用维度级证据密度函数 $D(m)$，为八个 CVSS 基础度量建立固定的 SER/EEC 路由表。
Under this routing, AV/AC/PR/UI are assigned to the Scattered-Evidence Reasoner (SER), which uses LLMs for dimensions whose evidence is scattered and implicit in early reports, whereas S/C/I/A are assigned to the Explicit-Evidence Classifier (EEC), which uses a fine-tuned lightweight model for dimensions with stable lexical patterns.
%在该路由下，早期报告中证据分散且隐含的 AV/AC/PR/UI 被分配给分散证据推理器（SER），由 LLM 处理；具有稳定词法模式的 S/C/I/A 被分配给显式证据分类器（EEC），由微调轻量级模型处理。
The routing table is derived once from the validation set and reused throughout inference rather than being recomputed for each report.
%该路由表仅从验证集推导一次，并在整个推理过程中重复使用，而不是针对每份报告重新计算。
VPre therefore executes a two-stage pipeline comprising three modules: the Semantic Triage and Denoising Unit (STDU) first performs triage and denoising, after which SER and EEC conduct parallel decoupled prediction and produce the complete CVSS v3.1 vector and base score.
%因此，VPre 执行一个由三个模块组成的两阶段流水线：语义分诊与去噪单元（STDU）首先完成分诊与去噪，随后 SER 和 EEC 并行进行解耦预测，生成完整的 CVSS v3.1 向量和基础分数。
\par
%\noindent\textbf{Comprehensive Evaluation.}
%全面评估。
To comprehensively evaluate VPre, our evaluation addresses three questions central to practical early warning: \emph{prediction accuracy on early OSINT}, \emph{screening reliability in high-noise OSINT streams}, and \emph{the effectiveness of SER/EEC specialization within Evidence-Aware Decoupling on the held-out test set}.
%为全面评估 VPre，实验围绕三个与实际早期预警密切相关的问题展开：早期 OSINT 预测准确性、高噪声 OSINT 流中的筛选可靠性，以及证据感知解耦中 SER/EEC 专门化在留出测试集上的有效性。
First, on VPre-Bench, which contains early OSINT matched to 5,581 real-world vulnerabilities, we compare VPre with directly transferred SOTA methods under the same early-OSINT setting to assess its prediction accuracy.
%首先，在 VPre-Bench 上，我们将 VPre 与直接迁移的 SOTA 方法置于相同的早期 OSINT 设置下进行比较，以检验其预测准确率；该数据集包含与 5,581 个真实世界漏洞相匹配的早期 OSINT。
Second, to approximate real-world deployment, we place VPre in the high-noise OSINT stream represented by PARSE-Triage, where non-security reports outnumber genuine vulnerabilities by approximately 38:1, and test whether it can reliably identify valid vulnerabilities before prediction.
%其次，为了逼近真实世界的部署环境，我们将 VPre 置于 PARSE-Triage 所表示的高噪声 OSINT 信息流中，其中非安全报告与真实漏洞的数量之比约为 38:1，并检验其能否在预测前可靠识别有效漏洞。
Third, under the same STDU-filtered input, we compare the fixed SER/EEC split with SER-only and EEC-only predictors to evaluate the metric-wise decoupling design.
%最后，在相同 STDU 过滤输入下比较固定 SER/EEC 划分与 SER-only、EEC-only，以评估指标级解耦设计。
The results show that VPre achieves an MAE of \textbf{0.7239} on the 851-report held-out test set, reducing MAE by \textbf{44.7\%} relative to the reproduced BERT-Small-NVD early-transfer baseline, while producing estimates an average of \textbf{38 days} before official scoring.
%实验结果表明，VPre 在包含 851 份报告的留出测试集上取得 0.7239 的 MAE，相比复现的 BERT-Small-NVD 早期迁移基线将 MAE 降低 44.7\%，并且平均可在正式评分前 38 天给出估计。
Under the high-noise setting, STDU's report-level screening achieves a Precision of 0.98, a Recall of 1.00, and an F1 score of 0.99; the complete VPre framework also consistently outperforms the unified SER-only and EEC-only alternatives.
%在高噪声设置下，STDU 的报告级筛选的 Precision、Recall 和 F1 分别达到 0.98、1.00 和 0.99；完整的 VPre 框架也持续优于统一的 SER-only 和 EEC-only 替代方案。
\par
%\noindent\textbf{Contributions.}
%主要贡献。
The main contributions of this paper are as follows.
%本文的主要贡献如下。
\begin{enumerate}
\item \termdef{PARSE Lifecycle and Early-Forecasting Formulation:} We formalize the semantic and informational evolution of vulnerability disclosures and define Pre-$T_0$ CVSS prediction as a distinct early-warning task with explicit temporal and input boundaries.
%PARSE 生命周期与早期预报问题定义：我们形式化刻画漏洞披露过程中语义与信息的演化，并将 Pre-$T_0$ CVSS 预测定义为一项具有明确时间边界和输入边界的独立早期预警任务。
\item \termdef{Complementary Benchmark Resources:} We construct and publicly release PARSE-Triage for high-noise vulnerability triage and VPre-Bench for temporally controlled early-report/CVSS evaluation.
%互补基准资源：我们构建并公开 PARSE-Triage，用于高噪声漏洞分诊，以及 VPre-Bench，用于经过时间控制的早期报告与 CVSS 评估。
\item \termdef{Evidence-Aware Decoupling:} We propose VPre with two explicit levels: Decoupling Level I, Security-Evidence Screening, separates evidence preparation from prediction through report- and sentence-level screening; Decoupling Level II, Evidence-Matched Routing, uses validation-derived evidence density to divide a uniform eight-metric predictor into specialized SER/EEC routes.
%证据感知解耦：VPre 首先通过报告级与句子级筛选将证据准备与预测解耦，随后利用验证集证据密度，将统一的八指标预测器解耦为专门化的 SER/EEC 路由。
\item \termdef{Early-Regime Empirical Validation:} Through matched early-OSINT comparisons, high-noise triage experiments, and module ablations, we show that VPre achieves an MAE of 0.7239 on the 851-report held-out test set, reduces MAE by 44.7\% relative to the reproduced BERT-Small-NVD early-transfer baseline, and produces CVSS estimates an average of 38 days before official scoring based on paired VPre-Bench timestamps, with an observed range of 21--60 days.
%早期信息环境下的实证验证：通过匹配的早期 OSINT 对比、高噪声分诊实验和模块消融，我们表明 VPre 在包含 851 份报告的留出测试集上取得 0.7239 的 MAE，相比复现的 BERT-Small-NVD 早期迁移基线将 MAE 降低 44.7\%，并且根据 VPre-Bench 的配对时间戳，平均可在正式评分前 38 天给出 CVSS 估计，观测范围为 21--60 天。
\end{enumerate}

\section{Related Work}
% 相关工作。
\label{sec:related-work}
% 相关工作

\subsection{Deep Learning for Automated CVSS Prediction}
% 基于深度学习的自动化 CVSS 预测

Early work by Spanos and Angelis~\cite{ref52} used multi-target regression with TF-IDF features to estimate vulnerability severity, providing an early data-driven approach to CVSS prediction.
% Spanos 和 Angelis 的早期工作使用基于 TF-IDF 特征的多目标回归来估计漏洞严重性，为数据驱动的 CVSS 预测提供了早期思路。
Related studies also examined machine-learning-based severity assessment and context-aware CVSS prioritization~\cite{ref46,ref53,ref54}.
% 相关研究还考察了基于机器学习的严重性评估和结合上下文的 CVSS 优先级排序。

K\"uhn et al.~\cite{ref20} fine-tuned BERT-Small, DistilBERT, and BERT-Medium on NVD descriptions to predict the eight CVSS v3.1 metrics as supervised classification tasks.
% Kühn 等人在 NVD 描述上微调了 BERT-Small、DistilBERT 和 BERT-Medium，并将八个 CVSS v3.1 指标作为监督分类任务进行预测。
BERT-Small achieved an MAE(Mean Absolute Error) of 0.8918 and offered the best balance between prediction accuracy and model size in their study.
% 在该研究中，BERT-Small 的 MAE 为 0.8918，并在预测精度和模型规模之间取得了较好的平衡。
These results show that fine-tuned language models can learn useful links between vulnerability descriptions and CVSS metrics.
% 这些结果表明，经过微调的语言模型能够学习漏洞描述与 CVSS 指标之间的有效联系。

AutoCVSS~\cite{ref66} later provided a broad comparison of supervised CVSS prediction models.
% AutoCVSS 随后对监督式 CVSS 预测模型进行了较全面的比较。
Its results include an MAE of 0.5490 for CVEDrill, 0.5540 for ModernBERT, and 0.7620 for DistilBERT-E.
% 其结果包括 CVEDrill 的 MAE 为 0.5490、ModernBERT 为 0.5540，以及 DistilBERT-E 为 0.7620。
However, these models were trained and evaluated mainly on standardized descriptions of confirmed vulnerabilities from NVD.
% 然而，这些模型主要在 NVD 中已经确认并经过标准化的漏洞描述上进行训练和评估。
Their performance on incomplete and less structured early information therefore remains less clear.
% 因此，它们在信息不完整、结构较弱的早期情报上的表现仍不明确。

\subsection{LLMs for Automated CVSS Prediction}
% 基于大语言模型的自动化 CVSS 预测

Recent studies have examined whether LLMs can predict CVSS metrics directly from vulnerability descriptions.
% 近期研究开始考察大语言模型能否直接根据漏洞描述预测 CVSS 指标。
Liu et al.~\cite{ref04} evaluated ChatGPT across several vulnerability-management tasks, including severity assessment, and showed that step-by-step prompting can support security reasoning.
% Liu 等人评估了 ChatGPT 在多项漏洞管理任务中的能力，其中包括严重性评估，并表明分步提示能够支持安全推理。
Isogai et al.~\cite{ref68} compared GPT-4o-mini with fine-tuned BERT models for CVSS base-vector prediction and found that BERT performed better on most metrics.
% Isogai 等人比较了 GPT-4o-mini 与微调 BERT 模型在 CVSS 基础向量预测中的表现，并发现 BERT 在大多数指标上表现更好。

AutoCVSS~\cite{ref66} compared supervised models with several generative models under the same evaluation setting.
% AutoCVSS 在同一评估设置下比较了监督模型和多种生成模型。
GPT-4o with 24 examples achieved an MAE of 0.6080, while the supervised models produced lower errors when enough labeled data were available.
% 使用 24 个示例的 GPT-4o 取得了 0.6080 的 MAE，而在标注数据充足时，监督模型取得了更低的误差。
The study also showed that LLMs remain useful in zero-shot and few-shot settings where labeled examples are limited.
% 该研究也表明，当标注样本有限时，大语言模型在零样本和少样本场景中仍有使用价值。

KBS 2025~\cite{ref31} presented a hybrid method in which GPT-3.5-Turbo first rewrites a vulnerability description into a more standard form and BERT-Small then predicts the CVSS vector.
% KBS 2025 提出了一种混合方法：先由 GPT-3.5-Turbo 将漏洞描述改写为较标准的形式，再由 BERT-Small 预测 CVSS 向量。
The method reported an MAE of 0.166 and an MSE(Mean Squared Error) of 0.299 on NVD data, showing that LLM-based text processing can be combined with a smaller prediction model.
% 该方法在 NVD 数据上报告了 0.166 的 MAE 和 0.299 的 MSE，表明基于大语言模型的文本处理可以与较小的预测模型结合。

Marchiori et al.~\cite{ref37} compared seven LLMs with three embedding models combined with XGBoost for individual CVSS metrics.
% Marchiori 等人针对各个 CVSS 指标，比较了七种大语言模型与三种结合 XGBoost 的嵌入模型。
They found that LLMs performed better on AV and AC, while the embedding-based models performed better on C, I, and A.
% 他们发现，大语言模型在 AV 和 AC 上表现更好，而基于嵌入的模型在 C、I 和 A 上表现更好。
These findings suggest that LLMs and supervised models have different strengths and that one model may not be the best choice for every CVSS metric.
% 这些发现说明，大语言模型和监督模型各有优势，同一种模型未必适合所有 CVSS 指标。

\subsection{Early Vulnerability Intelligence and Report Triage}
% 早期漏洞情报与报告筛选

Early vulnerability intelligence refers to public information that appears before formal vulnerability descriptions and scores are available.
% 早期漏洞情报是指在正式漏洞描述和评分出现之前已经公开的信息。
Earlier studies showed that public signals can support exploit prediction and vulnerability prioritization~\cite{ref09,ref10}.
% 早期研究表明，公开信号可以支持漏洞利用预测和漏洞优先级排序。
Other studies found that vulnerability discovery and disclosure can involve several project sources, such as issue reports and code changes~\cite{ref14}.
% 其他研究发现，漏洞发现和披露可能涉及问题报告、代码修改等多种项目来源。

One related research line focuses on separating security-related reports from ordinary software bugs.
% 一类相关研究关注如何从普通软件缺陷中识别与安全相关的报告。
Peters et al.~\cite{peters2019text} proposed FARSEC, which filters and ranks bug reports while handling class imbalance and security words that may also appear in non-security reports.
% Peters 等人提出了 FARSEC，在处理类别不平衡的同时，对缺陷报告进行筛选和排序，并考虑普通报告中也可能出现安全词汇的问题。
Ghosh et al.~\cite{ghosh2025securityrisk} analyzed 10.9 million GitHub issues from npm projects and trained text classifiers to identify security-related issues without clear security labels.
% Ghosh 等人分析了 npm 项目的 1090 万条 GitHub 问题，并训练文本分类器来识别没有明确安全标签的安全相关问题。
These studies show that report triage is important because early public reports can contain both security issues and ordinary bugs.
% 这些研究说明报告筛选十分重要，因为早期公开报告中可能同时包含安全问题和普通缺陷。

Liu et al.~\cite{liu2025disclosure} studied 21,501 vulnerabilities linked to open-source projects and found that 41.66\% had related GitHub issues before formal disclosure, with a median lead time of 39 days.
% Liu 等人研究了与开源项目相关的 21,501 个漏洞，发现其中 41.66\% 在正式披露前已有相关 GitHub 问题，时间中位数为提前 39 天。
This result confirms that public project discussions can provide useful early information.
% 这一结果说明，公开的项目讨论可以提供有用的早期信息。
However, early reports do not always have reliable labels.
% 然而，早期报告并不总是具有可靠的标签。
Wu et al.~\cite{wu2022dataquality} examined five public security-bug-report datasets and found 749 security reports that had been incorrectly labeled as non-security reports.
% Wu 等人检查了五个公开的安全缺陷报告数据集，发现其中有 749 条安全报告被错误标记为非安全报告。
They also showed that correcting these labels changed the measured performance of prediction models.
% 他们还表明，修正这些标签会改变预测模型的测量结果。

VLAI~\cite{ref67} represents another form of early assessment by predicting severity categories before an official CVSS score is published.
% VLAI 代表了另一种早期评估形式，它在官方 CVSS 分数发布前预测严重性类别。
It uses descriptions that already refer to vulnerabilities and predicts a severity category rather than the full CVSS vector.
% 它使用已经明确指向漏洞的描述，并预测严重性类别，而不是完整的 CVSS 向量。
Overall, current studies address early report triage and early severity assessment as mostly separate tasks.
% 总体来看，现有研究通常将早期报告筛选和早期严重性评估作为两个相对独立的任务。

\subsection{Research Gap and Motivation}
% 研究缺口与研究动机

The studies above reveal three main gaps.
% 上述研究反映出三个主要研究缺口。
First, most deep-learning CVSS models are trained and tested on standardized descriptions of confirmed vulnerabilities, so their reliability on early intelligence remains uncertain.
% 第一，大多数深度学习 CVSS 模型都在已确认漏洞的标准化描述上训练和测试，因此它们在早期情报上的可靠性仍不确定。
Second, LLMs can support text understanding and reasoning, but they do not consistently outperform supervised models across all CVSS metrics.
% 第二，大语言模型能够支持文本理解和推理，但并没有在所有 CVSS 指标上稳定超过监督模型。
Third, early-intelligence studies have mainly focused on finding security-related reports, while CVSS prediction studies usually assume that the input is already a confirmed vulnerability.
% 第三，早期情报研究主要关注识别安全相关报告，而 CVSS 预测研究通常假设输入已经是确认的漏洞。

These gaps show the need for reliable CVSS prediction from early intelligence that may be incomplete, unstructured, or mixed with ordinary software bugs.
% 这些缺口表明，需要研究如何从可能不完整、缺少结构或混有普通软件缺陷的早期情报中可靠地预测 CVSS。
They also suggest that report triage, text preparation, and the different characteristics of CVSS metrics should be considered together.
% 它们也说明，需要同时考虑报告筛选、文本准备以及不同 CVSS 指标的特点
\FloatBarrier

\section{PARSE Lifecycle and Methodology Foundation}
% PARSE 生命周期与方法论基础。
\label{sec:parse-methodology}
This section develops the methodological foundation of VPre. It proceeds in three layers.
%本节分三层展开 VPre 的方法论基础。
\hypersection{sec:parse-lifecycle-model} fixes the temporal and informational boundary of our task by introducing the PARSE vulnerability disclosure lifecycle and anchoring its five phases to four verifiable time nodes. 
% 3.1 通过 PARSE 漏洞披露生命周期，将任务的时间与信息边界固定下来，并将五个阶段锚定到四个可验证的时间节点。
\hypersection{sec:vpre-positioning} then locates VPre within that boundary, fixing the Pre-$T_0$ operating window and the Phase~P / Phase~A correspondence on which the rest of the paper relies. 
% 3.2 在该边界内为 VPre 选定 Pre-$T_0$ 运行窗口，并固定后续章节所依赖的 Phase P / Phase A 对应关系。
Finally, \hypersection{sec:methodology-foundation} turns that positioning into two concrete design choices:  Security-Evidence Screening removes non-security reports and sentences, and dimension-level evidence density $D(m)$ determines a frozen SER/EEC routing table reused at inference time. 
% 最后 3.3 把上述定位落实为两项具体设计选择：安全证据筛选移除非安全报告与非安全句子，维度级证据密度 $D(m)$ 决定推理时复用的冻结 SER/EEC 路由表。
The notation, threshold symbols, and frozen routing rule defined here are reused directly by the implementation in \hypersection{sec:implementation} and by the ablation analysis in \hypersection{sec:rq3}.
%此处定义的符号、阈值以及冻结的路由表，将在 \hypersection{sec:implementation} 的实现与 \hypersection{sec:rq3} 的消融分析中直接复用。

\subsection{The PARSE Lifecycle Model}
% PARSE 生命周期模型。
\label{sec:parse-lifecycle-model}

Most existing automated CVSS prediction studies rely on descriptions of already confirmed vulnerabilities, typically standardized NVD records, thereby modeling each vulnerability at a single, relatively static information state~\cite{ref20,ref21,ref24,ref31,ref52,ref66}.
%现有自动化 CVSS 预测研究大多依赖已确认漏洞的描述，通常是标准化 NVD 记录，因而是在单一、相对静态的信息状态下对每个漏洞进行建模。
As established in the Introduction, early CVSS forecasting is not merely late-stage prediction moved earlier, but a distinct task under an evolving information regime: the available evidence develops from tentative, unstructured reports into confirmed and standardized assessments~\cite{ref14,ref15,ref16,ref17}.
%正如引言所明确的，早期 CVSS 预报并非只是将后期预测前移，而是在持续演化的信息环境下进行的一项独立任务：可用证据会从初步、非结构化的报告逐步发展为已确认并标准化的评估。
This distinction makes an explicit model of the vulnerability disclosure lifecycle necessary to define when forecasting occurs and what information is available at that point.
%这一区别使得显式建模漏洞披露生命周期成为必要，以明确预报发生的时间位置及该时点可获得的信息。
Accordingly, to formalise the temporal and informational boundary of our task, we introduce the \namedconcept{PARSE vulnerability disclosure lifecycle model} with \hypertable{tab:parse-phases} and~\hypertable{tab:temporal-nodes}.
% 因此，为了形式化任务的时间与信息边界，我们提出 \namedconcept{PARSE 漏洞披露生命周期模型}，其要点见 \hypertable{tab:parse-phases}，并由 \hypertable{tab:temporal-nodes} 锚定到可观测的披露里程碑。

\begin{table*}[!tbp]
\centering
\small
\caption{PARSE lifecycle phases and security events.}%PARSE 生命周期周期和安全事件
\label{tab:parse-phases}
\resizebox{\textwidth}{!}{%
\begin{tabular}{lll}
\toprule
PARSE Phase & Temporal Scope & Core Security Event \\
\midrule
\textbf{P} - Probing & Before $T_0$ & Researchers/attackers discover a bug or anomaly \\
\textbf{A} - Acknowledgment & Pre-$T_0$ to $T_0$ & Vendor receives the report, reproduces the issue, and confirms its security relevance \\
\textbf{R} - Reservation & $T_0$--$T_1$ & A CVE ID is assigned at $T_0$; the embargo remains in effect until $T_1$ \\
\textbf{S} - Standardization & $T_1$--$T_2$--$T_3$ & OSINT emerges at $T_1$; the NVD description is standardized at $T_2$ and finalized with official scoring at $T_3$ \\
\textbf{E} - Evaluation & After $T_3$ & Finalized NVD descriptions and CVSS scores become available for downstream evaluation \\
\bottomrule
\end{tabular}%
}
\end{table*}

\begin{table*}[!tbp]
\centering
\small
\caption{Temporal landmarks and available data states.}
\label{tab:temporal-nodes}
\resizebox{\textwidth}{!}{%
\begin{tabular}{lll}
\toprule
Time Node & Event & Data State \\
\midrule
$T_0$ (Embargo) & CNA assigns CVE ID & CVE ID assigned; NVD lacks description or score \\
$T_1$ (Embargo ends) & Vulnerability details disclosed; NVD publishes initial description & CVE ID and initial description exist; \textbf{no CVSS score} \\
$T_2$ (Days/months post-$T_1$) & NVD refines description to standardized version & Description complete; \textbf{score may still be missing} \\
$T_3$ (Days/weeks post-$T_2$) & NVD experts publish official CVSS scores & Both description and scoring are fully finalized \\
\bottomrule
\end{tabular}%
}
\end{table*}

\hypertable{tab:parse-phases} names the five lifecycle phases and their security events, while \hypertable{tab:temporal-nodes} anchors those phases to observable disclosure milestones and the data available at each milestone.
% Table~\ref{tab:parse-phases} 给出五个生命周期阶段及其安全事件，Table~\ref{tab:temporal-nodes} 则将这些阶段锚定到可观测的披露里程碑，并说明每个时刻可获得的数据。

The PARSE model is structured around five phases anchored by four critical time nodes.
% PARSE 模型由五个阶段构成，并由四个关键时间节点进行锚定。
The five phases---\textbf{P} (Probing), \textbf{A} (Acknowledgment), \textbf{R} (Reservation), \textbf{S} (Standardization), and \textbf{E} (Evaluation)---capture the qualitative stages of vulnerability disclosure, each characterized by distinct textual properties and information availability.
% 五个阶段——P（Probing）、A（Acknowledgment）、R（Reservation）、S（Standardization）和 E（Evaluation）——刻画了漏洞披露的定性阶段，各阶段具有不同的文本属性和信息可用性。
The four time nodes---$T_0$, $T_1$, $T_2$, and $T_3$---mark objective, verifiable milestones.
% 四个时间节点——$T_0$、$T_1$、$T_2$ 和 $T_3$——标记了客观且可验证的里程碑。
Phase P occurs before $T_0$ and captures initial probing, while Phase A extends through the Pre-$T_0$ acknowledgment process and ends at $T_0$.
% P 阶段发生在 T0 之前并对应初始探查过程，而 A 阶段覆盖 Pre-T0 的报告、复现和确认过程并终止于 T0。
During this combined P/A window, only raw, unstructured reports exist, with no standardized NVD description and no CVSS score.
% 在这一 P/A 时间窗内，仅存在原始非结构化报告，尚无标准化 NVD 描述或 CVSS 评分。
Phase R spans $T_0$--$T_1$: a CVE ID is assigned at $T_0$, and the vulnerability remains under embargo until $T_1$.
% R 阶段覆盖 T0 至 T1：CVE 编号在 T0 分配，漏洞禁运期持续至 T1。
Phase S progresses through $T_1$--$T_2$--$T_3$: vulnerability details emerge at $T_1$, the NVD description reaches its standardized form at $T_2$, and the official description and CVSS score are finalized at $T_3$.
% S 阶段依次经过 T1、T2 和 T3：漏洞信息在 T1 公开，NVD 描述在 T2 完成标准化，并在 T3 完成官方描述与 CVSS 评分。
Phase E begins after $T_3$, when the finalized description and score become available for downstream evaluation.
% E 阶段始于 T3 之后，此时最终描述和评分可用于后续评估。

\begin{figure*}[!tbp]
\centering
\includegraphics[width=0.83\textwidth]{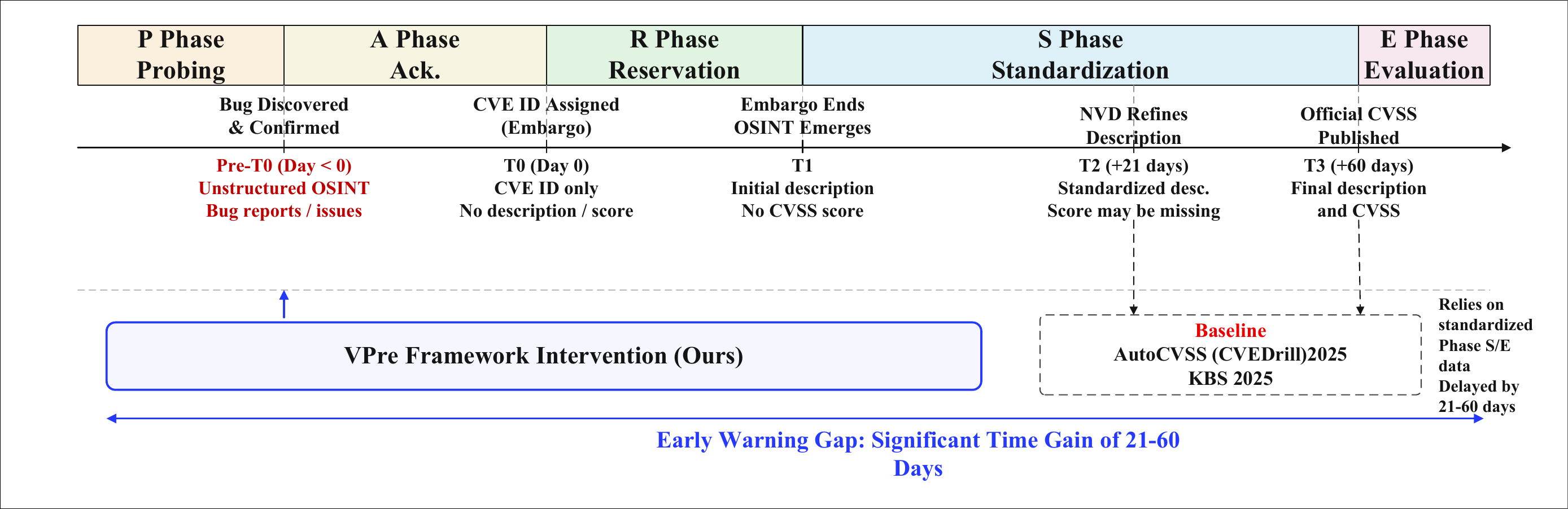}
\caption{PARSE lifecycle and the early intervention point of VPre.}
\label{fig:parse-lifecycle}
\end{figure*}

\hyperfigure{fig:parse-lifecycle} summarizes the lifecycle from left to right.
% Figure~\ref{fig:parse-lifecycle} 从左到右概括了完整生命周期。
The upper colored band names the five PARSE phases; the milestone arrows below it locate bug confirmation, CVE reservation at $T_0$, initial NVD publication at $T_1$, description refinement at $T_2$, and official scoring at $T_3$.
% 上方彩色带标出五个 PARSE 阶段；下方里程碑箭头依次定位漏洞确认、$T_0$ 时刻的 CVE 保留、$T_1$ 时刻的 NVD 初次发布、$T_2$ 时刻的描述完善以及 $T_3$ 时刻的正式评分。
P and A jointly cover the pre-$T_0$ process where VPre operates; R spans $T_0$--$T_1$; S covers the successive $T_1$--$T_2$--$T_3$ milestones; and E begins after $T_3$.
% P 和 A 共同覆盖 VPre 所处理的 T0 之前过程；R 覆盖 T0 至 T1；S 覆盖 T1、T2 和 T3 三个连续里程碑；E 始于 T3 之后。
In the lower comparison lane, the solid VPre interval begins before $T_0$, whereas the dashed AutoCVSS/CVEDrill and KBS baseline box is aligned with the late-S/E regime because those methods depend on standardized descriptions and finalized scores.
% 在下方对比区域中，VPre 在 T0 之前介入，而 AutoCVSS/CVEDrill 与 KBS 基线位于后期 S/E 阶段，因为这些方法依赖标准化描述和最终评分。
The bottom double-headed arrow denotes the timestamp-derived early-warning interval between these operating points; paired VPre-Bench samples yield a mean of 38 days and an observed range of 21--60 days.
% 底部双向箭头表示这些运行时点之间由时间戳计算的提前预警区间；VPre-Bench 的配对样本给出 38 天的平均值和 21--60 天的观测范围。
It is a disclosure-time difference, not model execution time.
% 该数值表示披露时间差，而不是模型执行时间。
This layered structure---five phases driven by four nodes---reveals the central asymmetry of the paper: existing baselines operate on late-stage, information-rich data, whereas VPre intervenes before a CVE ID exists.
% 这种由四个节点驱动五个阶段的分层结构揭示了本文的核心不对称性：现有基线使用后期且信息丰富的数据，而 VPre 在 CVE ID 出现之前介入。

The main point of \hyperfigure{fig:parse-lifecycle} is the temporal mismatch between training data and intended deployment.
% Figure~\ref{fig:parse-lifecycle} 的核心在于展示训练数据与预期部署时点之间的时间错位。
VPre is designed for the left side of the timeline, whereas NVD-trained baselines are naturally aligned with the right side after standardization and official scoring.
% VPre 面向时间线左侧，而基于 NVD 训练的基线自然对应右侧的标准化与正式评分之后。

\subsection{VPre's Operating Window within PARSE}
%Vpre 在PARSE模型中的位置
\label{sec:vpre-positioning}
\textbf{Why position matters.}
The PARSE model in \hypersection{sec:parse-lifecycle-model} does not by itself determine where CVSS prediction should occur.
% \hypersection{sec:parse-lifecycle-model} 中的 PARSE 模型本身并不决定 CVSS 预测应发生在哪里。
Any CVSS prediction task, however, must answer two questions that the lifecycle makes explicit: at which disclosure phase does prediction occur, and what information is then available.
% 然而任何 CVSS 预测任务都必须回答生命周期所显式提出的两个问题：在哪一披露阶段做预测，以及该阶段可获得哪些信息。
This subsection fixes VPre's position along both axes.
% 本小节沿这两个坐标固定 VPre 的位置。
It also explains why that position is the one that existing methods leave empty.
% 并解释为何现有方法恰好留下了这一窗口。

\textbf{Phase and time-node.}
VPre targets the Phase~P and Phase~A window before $T_0$, where no CVE identifier, no NVD description, and no CVSS score yet exist.
% VPre 面向 $T_0$ 之前的 P/A 阶段窗口，此时尚不存在 CVE 编号、NVD 描述与 CVSS 评分。
Phase~R and later phases are out of scope: those are where the K\"uhn et al., AutoCVSS, and KBS 2025 baselines naturally operate.
% Phase R 及之后阶段不在范围内：那是 K\"uhn 等、AutoCVSS 与 KBS 2025 等基线自然运行的阶段。

\textbf{Information regime.}
Within this window, the only inputs are raw, unstructured disclosures that may describe security vulnerabilities, functional bugs, or non-security content, with no standardised labels.
% 在该窗口内，唯一可用的输入是可能描述安全漏洞、功能缺陷或非安全内容的原始非结构化披露，且没有任何标准化标签。
The prediction target is the full CVSS v3.1 vector and base score.
% 预测目标是完整的 CVSS v3.1 向量与基础分。
This combination of inputs and target is what we mean by \emph{Pre-$T_0$ CVSS forecasting}.
% 这一输入与目标的组合，正是我们所说的 \emph{Pre-$T_0$ CVSS 预报}。

\textbf{Why this position.}
Two reasons motivate VPre's operating window.
% 选择这一窗口有两点原因。
First, Phase~P/A provides the earliest warning window: a CVE ID is assigned at $T_0$, vulnerability information is progressively standardized through $T_1$--$T_2$--$T_3$, and official scoring is finalized at $T_3$.
% 第一，Phase P/A 提供了最早的预警窗口：CVE 编号在 T0 分配，漏洞信息随后依次经过 T1、T2 和 T3 的标准化过程，并在 T3 完成官方评分。
Second, the unique information regime of this window---tentative reports, absent ground-truth labels, and the mixture of security and non-security content---is precisely what existing NVD-trained methods cannot handle.
% 第二，该窗口独特的信息环境——含推测的报告、缺失的真值标签、以及安全与非安全内容的混杂——恰好是现有 NVD 训练方法无能为力的地方。
VPre fills that gap; it does not compete with later-phase methods on their own input regime.
% VPre 填补这一空缺，但并不在后期方法自己的输入阶段上与它们竞争。

\subsection{Methodology Foundation: Evidence-Aware Decoupling}
%方法论基础：证据感知解耦。
\label{sec:methodology-foundation}

Directly transferring models trained on curated NVD descriptions to early OSINT reports fails because the two textual regimes differ fundamentally in information structure (\hypersection{sec:semantic-drift}).
% 将基于人工整理后的 NVD 描述训练的模型直接迁移到早期 OSINT 报告会失效，因为两种文本形态在信息结构上存在本质差异（Section~\ref{sec:semantic-drift}）。
The core idea of VPre therefore proceeds in two steps: it first screens security-relevant evidence from noisy early reports, and then decouples the prediction task according to the evidence characteristics and reasoning demands of the eight CVSS base metrics.
% 因此，VPre 的核心思想分为两步：首先从含噪的早期报告中筛选出安全相关证据，随后依据八个 CVSS 基础度量各自的证据特征与推理需求，对其预测任务进行解耦。
We term this principle \textbf{Evidence-Aware Decoupling}: prediction is decoupled along the structure of the evidence instead of being handled by one uniform predictor.
% 我们将这一原则称为证据感知解耦：沿证据的结构进行解耦，而非用统一预测器处理。
The principle operates at two levels.
% 该原则在两个层次运作。
Decoupling Level I, \emph{Security-Evidence Screening}, separates security-relevant evidence from the noisy early OSINT stream at two granularities and contracts the input into a concentrated evidence set $R'$ (\hypersection{sec:sentence-certainty}).
% 第一层为安全证据筛选：在报告级与句子级两个粒度上，从含噪的早期 OSINT 流中分离安全相关证据，并将输入收缩为集中的证据集 $R'$（Section~\ref{sec:sentence-certainty}）。
Decoupling Level II, \emph{Evidence-Matched Routing}, routes the prediction task of each metric to the specialized predictor matched to its evidence characteristics and reasoning demands (\hypersection{sec:evidence-density-assignment}).
% 第二层为证据匹配路由：依据各度量各自的证据特征与推理需求，将其预测任务路由至与之匹配的专门化预测器（Section~\ref{sec:evidence-density-assignment}）。

\subsubsection{Why direct transfer fails: Semantic Drift}
% 直接迁移为何失效：语义漂移。
\label{sec:semantic-drift}
%Why direct transfer fails: Semantic Drift
%动机 本节先讲为什么 现有基于NVD训练的模型不可直接迁移
The vulnerability disclosure process is not merely a chronological sequence---it is a trajectory of linguistic transformation.
% 漏洞披露过程并非单纯的时间序列，而是一条语言形态持续转变的轨迹。
Late S/E-stage NVD descriptions and early P/A-stage OSINT belong to fundamentally different linguistic registers.
% PARSE 后期 S/E 阶段的 NVD 描述与早期 P/A 阶段的 OSINT 属于本质不同的文本形态。
A late-stage NVD entry is a carefully curated, deterministic security statement: every sentence serves a specific diagnostic purpose, written in standardized terminology by experts who already know the vulnerability's full scope (e.g., \emph{``Improper input validation in the HTTP parser allows remote attackers to execute arbitrary code via a crafted packet''}).
% 一条后期 NVD 条目是精心策划的、确定性的安全陈述：每句话服务于特定的诊断目的，由已经了解漏洞全貌的专家使用标准化术语撰写（如“HTTP 解析器中的不当输入验证允许远程攻击者通过构造的数据包执行任意代码”）。
In contrast, an early-stage raw OSINT report is an evolving, multi-voice conversation: a developer reports unexpected behavior (\emph{"the server crashes when I send a malformed request"}), another contributor speculates about exploitability (\emph{"this might be reachable from the network"}), yet another suggests a workaround (\emph{"should we add a bounds check here?"}), all interspersed with stack traces, version logs, and tangential discussion.
% 相比之下，早期原始 OSINT 报告是一场不断演进的多方对话：开发者报告异常行为（“发送畸形请求时服务器崩溃”），另一位贡献者推测其可利用性（“这可能能够从网络访问”），还有人提出规避方案（“是否应在此处增加边界检查？”）；其间还夹杂堆栈跟踪、版本日志和无关讨论。
This is not a difference of vocabulary---it is a difference in \emph{information structure}: late-stage text is curated and conclusive, while early-stage text is exploratory, contradictory, and predominantly non-security in content.
% 这并非词汇差异，而是信息结构差异：后期文本经过人工整理与审定、结论明确，早期文本则具有探索性、可能相互矛盾，并且其主要内容并非安全信息。
Existing models trained solely on late-stage NVD descriptions learn a specific mapping from concise, expert-authored security prose to CVSS vector labels~\cite{ref20}.
% 仅使用后期 NVD 描述训练的现有模型会学习一种特定映射，即从简洁的专家安全表述映射到 CVSS 向量标签。
When the existing model is directly applied to early OSINT---without any adaptation to this radically different textual environment---the learned mapping breaks down.
% 当同一模型未经任何适配便直接应用于文本环境截然不同的早期 OSINT 时，已学习的映射会失效。
This is the phenomenon we term \textbf{semantic drift}.
% 我们将这一现象称为语义漂移。
The empirical evidence is unambiguous: the reproduced \textbf{BERT-Small-NVD} checkpoint achieves an MAE of 0.8918 on its native late-stage NVD test set.
% 实证证据十分明确：复现的 \textbf{BERT-Small-NVD} checkpoint 在原生后期 NVD 测试集上的 MAE 为 0.8918。
When the identical checkpoint is evaluated on early-stage VPre-Bench reports without retraining or architectural modification, its MAE rises to 1.3095, corresponding to a 47\% increase in prediction error.
% 当完全相同的 checkpoint 在不重新训练、也不修改架构的情况下用于早期 VPre-Bench 报告时，其 MAE 上升至 1.3095，预测误差增加 47\%。
Importantly, this degradation is not uniform across CVSS dimensions.
% 重要的是，这种性能退化在不同 CVSS 维度上并不均匀。
Dimensions whose prediction relies on explicit lexical cues present in both registers (e.g., Availability, where "DoS" and "crash" appear in both early and late stages) suffer less.
% 依赖两种文本形态中均存在的显式词法线索进行预测的维度受影响较小，例如可用性维度中的“DoS”和“crash”同时出现在早期与后期文本中。
Dimensions requiring inference about attack preconditions, privilege states, or user involvement---information that is explicit in late-stage curated prose but scattered or absent in early-stage exploratory discussion---deteriorate far more severely.
% 需要推断攻击前提、信任边界或影响范围的维度退化更为严重；这些信息在后期人工整理文本中被明确表达，但在早期探索性讨论中往往分散或缺失。

\begin{figure*}[!tbp]
\centering
\includegraphics[width=0.86\textwidth]{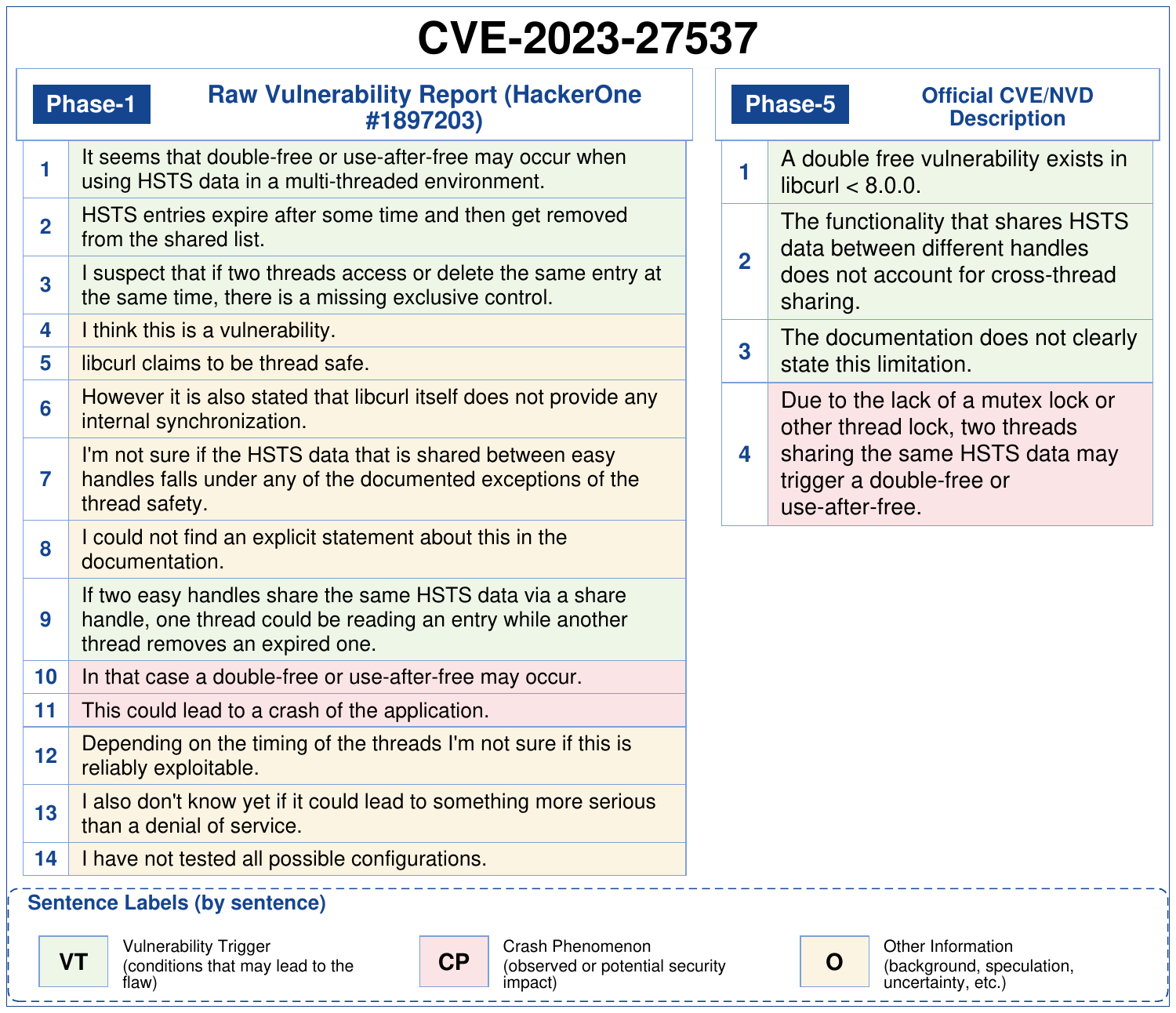}
\caption{Illustrative comparison of early-report and standardized CVE/NVD text for CVE-2023-27537, with sentence-role annotations.}
\label{fig:semantic-drift}
\end{figure*}

\hyperfigure{fig:semantic-drift} compares an early-report excerpt with a standardized CVE/NVD description of the same case, CVE-2023-27537.
% Figure~\ref{fig:semantic-drift} 对比同一案例 CVE-2023-27537 的早期报告摘录与标准化 CVE/NVD 描述。
The left panel discusses shared HSTS data in libcurl, concurrent access or removal by multiple threads, and possible double-free or use-after-free behavior, together with uncertainty about thread-safety guarantees, exploitability, and untested configurations.
% 左侧讨论 libcurl 中共享的 HSTS 数据、多线程并发访问或删除，以及可能的双重释放或释放后使用问题，同时保留了对线程安全保证、可利用性和未测试配置的不确定表述。
The right panel condenses the affected version range, the cross-thread sharing limitation, and the lack of mutual exclusion into a more concise description of the memory-safety problem.
% 右侧将受影响版本范围、跨线程共享限制以及缺少互斥保护等内容凝练为对内存安全问题的简洁描述。
The comparison therefore illustrates a change in information organization and evidential certainty across disclosure stages, rather than a difference between two vulnerabilities or merely between document lengths.
% 因此，该对比展示的是披露阶段之间信息组织方式与证据确定性的变化，而不是两个不同漏洞或单纯的文档长度差异。
The colors distinguish vulnerability-trigger evidence (VT), consequence evidence (CP), and other information (O).
% 颜色区分漏洞触发证据（VT）、后果证据（CP）与其他信息（O）。
The figure uses the example-specific wording ``Crash Phenomenon'' for CP; in the method, CP denotes the broader Consequence Phenomenon category defined in Table~\ref{tab:stdu-labels}.
% 图中的 CP 使用该示例的表述“Crash Phenomenon”；在本文方法中，CP 表示 Table~\ref{tab:stdu-labels} 定义的更广义的 Consequence Phenomenon（后果现象）类别。
These illustrative O labels are not deletion decisions: the STDU rule in \hyperequation{eq:filtered-evidence} removes only high-confidence O predictions and retains low-confidence O sentences.
% 这些示意性的 O 标签不代表删除决策：Equation~\ref{eq:filtered-evidence} 中的 STDU 规则只删除高置信度的 O 预测句，并保留低置信度 O 句子。
This difference in information organization helps explain why a feature mapping learned from concise NVD prose can become unreliable when applied directly to early reports.
% 这种信息组织差异有助于解释：从简洁 NVD 文本学习到的特征映射，为何在直接应用于早期报告时可能失效。

The key visual message of \hyperfigure{fig:semantic-drift} is that early OSINT contains useful security evidence, but it is embedded in a noisier and less conclusive register than NVD text.
% Figure~\ref{fig:semantic-drift} 的关键视觉信息是：早期 OSINT 确实包含有用的安全证据，但这些证据嵌入在比 NVD 文本噪声更多、结论性更弱的语言形式中。
Semantic drift thus exposes not one but two structural gaps between OSINT and NVD.
% 因此，语义漂移在 OSINT 与 NVD 之间暴露的不是单一缺口，而是两个结构性缺口。
First, noise exists at two granularities: at the report level, non-security reports (ordinary functional, usability, build, or documentation defects) dominate the raw stream---in the PARSE-Triage setting, non-security reports outnumber genuine vulnerabilities by approximately 38:1---and at the sentence level, even accepted reports contain stack traces, version references, and tangential discussion that carry no CVSS-relevant signal; feeding raw reports to any predictor therefore injects noise before prediction begins.
% 第一，噪声存在于两个粒度：报告级上，非安全报告（普通功能、可用性、构建或文档类缺陷）在原始信息流中占据主体——在 PARSE-Triage 设置下，非安全报告与真实漏洞的数量之比约为 38:1；句子级上，即使被接收的报告也包含堆栈跟踪、版本引用和无关讨论等不携带 CVSS 相关信号的内容。因此，将原始报告直接送入任何预测器都会在预测开始前注入噪声。
Second, even among security-relevant sentences, the eight CVSS metrics differ in how explicitly their evidence is expressed, so a single predictor cannot serve all of them uniformly.
% 第二，即便在安全相关句子中，八个 CVSS 指标的证据显式程度也存在差异，单一预测器无法均匀地处理所有指标。
These two gaps motivate the two levels of Evidence-Aware Decoupling introduced above: Security-Evidence Screening separates security-relevant evidence from this two-granularity noise to recover a concentrated evidence set (\hypersection{sec:sentence-certainty}), and Evidence-Matched Routing partitions the eight metrics by evidence density to match each cluster to an appropriate predictor (\hypersection{sec:evidence-density-assignment}).
% 这两个缺口正是上文引入的两层证据感知解耦的动机：安全证据筛选将安全相关证据从上述两级噪声中分离出来，得到集中的证据集（Section~\ref{sec:sentence-certainty}）；证据匹配路由按证据密度划分八个指标，使每个簇匹配到合适的预测器（Section~\ref{sec:evidence-density-assignment}）。

\subsubsection{Decoupling Level I: Security-Evidence Screening}
% 解耦第一层：安全证据筛选。
\label{sec:sentence-certainty}
%第一层解耦：安全证据筛选（报告级 + 句子级两个粒度），承接 3.3.1 升格后的两级噪声动机
Security-Evidence Screening responds to the two-granularity noise identified in \hypersection{sec:semantic-drift} with two matching decisions.
% 安全证据筛选针对 Section~\ref{sec:semantic-drift} 指出的两级噪声，作出两个与之对应的决策。
At the report level, non-security reports---ordinary functional, usability, build, or documentation defects that can never yield a CVSS vector---are removed from the stream before any prediction is attempted, so that downstream modules consume only cases that can plausibly carry security evidence.
% 报告级上，非安全报告——即根本无法产生 CVSS 向量的普通功能、可用性、构建或文档类缺陷——在任何预测发生之前就被移出信息流，使下游模块只消费可能携带安全证据的案例。
At the sentence level, only sentences predicted as O with high confidence are removed from each accepted report; VT/CP predictions and low-confidence O sentences are retained, so that prediction reads a conservatively screened evidence set rather than the raw noisy narrative.
% 句子级上，仅删除被高置信度预测为 O 的句子；VT/CP 预测句和低置信度 O 句子均予以保留，使预测读取经保守筛选的证据集而非原始含噪叙述。
We term this two-granularity separation \termdef{Security-Evidence Screening}, the first level of Evidence-Aware Decoupling.
% 我们将这种两个粒度上的分离称为安全证据筛选，即证据感知解耦的第一层。
The remainder of this subsection specifies the sentence-level decision; the report-level decision is methodological in nature and its concrete realization is described with the implementation in \hypersection{sec:stdu}.
% 本小节余下部分具体说明句子级决策；报告级决策属于方法论层面，其具体实现随 Section~\ref{sec:stdu} 一并描述。

Not every sentence in an early OSINT report carries a security signal.
% 早期 OSINT 报告中的句子并非都与安全相关。
We first define three semantic categories at the sentence level.
% 我们首先定义了句子的三个语义类别：
\textbf{VT} (Vulnerability Trigger) describes how a vulnerability is activated---e.g., \emph{``crafted request''} and \emph{``malformed packet''}.
% VT（Vulnerability Trigger，漏洞触发条件）描述漏洞如何被激活，例如“crafted request”和“malformed packet”。
\textbf{CP} (Consequence Phenomenon) describes the observable security consequence---e.g., \emph{``crash''}, \emph{``out-of-bounds read''}, and \emph{``RCE''}.
% CP（Consequence Phenomenon，后果现象）描述可观测的安全后果，例如“crash”“out-of-bounds read”和“RCE”。
Everything else---stack traces, version references, configuration notes, and general discussion---falls into \textbf{O} (Other) and carries no CVSS-relevant signal.
% 其余内容——堆栈跟踪、版本引用、配置说明和一般讨论——归为 O（Other，其他），不携带 CVSS 预测所需的信号。
These three labels form the basis of the sentence-level screening decision.
%这三类标签构成句子级筛选决策的基础。

The core idea of the sentence-level decision is straightforward: downstream CVSS prediction models do not need to read the entire raw report---they only need the subset of sentences that directly pertain to security.
% 句子级决策的核心思想很直接：下游 CVSS 预测模型无须读取整份原始报告，只需读取其中与安全直接相关的句子子集。
Contracting the input from a noisy, speculative narrative to a security-focused evidence set increases the concentration of task-relevant evidence before prediction.
%将输入从含噪的推测性叙述收缩为聚焦安全信息的证据集，可以在预测前提高任务相关证据的集中度。
To implement this at scale, we fine-tune a lightweight DistilBERT-based sentence classifier---\hyperref[para:vuln-extract-classifier]{VulnExtractClassifier}---for three-way classification (VT/CP/O)~\cite{ref56}.
% 为实现规模化处理，我们基于 DistilBERT 微调轻量句子分类器 VulnExtractClassifier（点击可跳转至 implementation 章节的详细实现），用于 VT/CP/O 三分类。
For each sentence $s_i$, the classifier outputs a probability distribution over the three categories.
% 对于每个句子 $si$，分类器输出其在三个类别上的概率分布。
The screening decision deliberately avoids a symmetric ``keep only confident VT/CP'' rule: deleting a sentence that actually carries attack preconditions or consequences is far more harmful than keeping an additional non-security sentence, because lost evidence cannot be recovered downstream.
% 筛选决策有意避免了"仅保留高置信 VT/CP"的对称规则：误删一个实际携带攻击前提或后果的句子，其代价远高于多保留一个非安全句子，因为丢失的证据无法在下游恢复。
The filtering action therefore removes a sentence only when the classifier is confidently certain that it is irrelevant, as specified in \hyperequation{eq:filtered-evidence}:
% 因此，只有当分类器确有把握判定句子无关时，过滤动作才将其删除，如 Equation~\ref{eq:filtered-evidence} 所示：
\begin{equation}
\label{eq:filtered-evidence}
\begin{aligned}
R' &= \left\{s_i\in R\mid \hat y_i\neq O
      \;\lor\;P(O\mid s_i)\leq\tau_s\right\},\\
\hat y_i &= \arg\max_{k\in\{O,VT,CP\}} P(k\mid s_i).
\end{aligned}
\end{equation}
That is, a sentence is deleted only when its predicted label is O \emph{and} the O-class probability exceeds the threshold $\tau_s$; predicted VT/CP sentences and low-confidence O sentences are all retained.
% 也就是说，只有当预测标签为 O 且 O 类概率超过阈值 $\tau_s$ 时，句子才被删除；预测为 VT/CP 的句子和低置信度的 O 句子全部保留。
The threshold controls how strict the deletion condition is: a higher $\tau_s$ demands stronger evidence of irrelevance before deleting, retaining more sentences and higher evidence recall at the cost of admitting more noise; a lower $\tau_s$ deletes more aggressively, yielding purer but potentially evidence-sparse input.
% 该阈值控制删除条件的严格程度：较高的 $\tau_s$ 要求在删除前有更强的"无关"证据，保留更多句子、获得更高的证据召回，但代价是引入更多噪声；较低的 $\tau_s$ 删除更激进，得到更纯净但可能损失证据的输入。
The threshold $\tau_s$ is tuned exclusively on the validation split of the manually annotated sentence data; the same split also governs model selection (the training split fits the classifier, the validation split selects the checkpoint and $\tau_s$, and the held-out test split is evaluated only once after all choices are frozen), so no threshold decision ever touches the test data.
% 阈值 $\tau_s$ 仅在人工标注句子数据的验证划分上调节；同一划分体系也约束模型选择（训练划分拟合分类器，验证划分选择检查点与 $\tau_s$，留出测试划分在所有选择冻结后仅评估一次），因此任何阈值决策都不会接触测试数据。
In summary, Security-Evidence Screening operates at two granularities---discarding non-security reports and removing only confidently irrelevant sentences---and contracts the raw stream into the concentrated evidence set $R'$; it constitutes the first level of Evidence-Aware Decoupling and prepares the input for the second, metric-wise predictor decoupling level.
% 总结来说，安全证据筛选在报告级与句子级两个粒度上运作——剔除非安全报告、仅删除确凿无关的句子——将原始信息流收缩为集中的证据集 $R'$；它构成证据感知解耦的第一层，并为第二层指标级预测器解耦准备输入。

% Keep this transition at its natural spacing even when IEEEtran balances
% the surrounding two-column page under a different TeX engine.
\raggedbottom
\subsubsection{Decoupling Level II: Evidence-Matched Routing}
% 解耦第二层：证据匹配路由。
\label{sec:evidence-density-assignment}

The Security-Evidence Screening in \hypersection{sec:sentence-certainty}---discarding non-security reports, classifying each remaining sentence as VT, CP, or O, and deleting only confidently irrelevant ones---produces a filtered sentence set $R'$ with substantially reduced noise.
% Section~\ref{sec:sentence-certainty} 的安全证据筛选剔除非安全报告，并将剩余句子分类为 VT、CP 或 O，仅删除确凿无关的句子，从而得到噪声明显降低的过滤句子集 $R'$。
However, removing irrelevant content is only the first step.
% 但去除无关内容只是第一步。
Even within $R'$, a fundamental asymmetry remains.
% 即使在 $R'$ 内部，仍然存在一种根本性的不对称。
CVSS v3.1 defines eight base metrics---Attack Vector (AV), Attack Complexity (AC), Privileges Required (PR), User Interaction (UI), Scope (S), Confidentiality (C), Integrity (I), and Availability (A)---and these eight dimensions draw on different types and densities of textual evidence.
% CVSS v3.1 定义了八个基础指标——攻击向量（AV）、攻击复杂度（AC）、权限要求（PR）、用户交互（UI）、影响范围（S）、机密性（C）、完整性（I）和可用性（A）——这些维度依赖不同类型和密度的文本证据。
Impact dimensions such as C, I, and A are often signaled by explicit lexical anchors such as "information disclosure," "tampering," or "denial of service."
% C、I、A 等影响类维度通常由“信息泄露”“篡改”或“拒绝服务”等显式词法锚点指示。
In contrast, AV, AC, PR, and UI often require linking multiple clauses about the attack precondition, triggering context, privilege state, and user participation.
% 相比之下，AV、AC、PR 和 UI 往往需要关联多个分句，以推断攻击前提、触发环境、权限状态和用户参与情况。
Blindly feeding the same text to the same model for every dimension would therefore be doubly wasteful: it would spend expensive LLM inference on dimensions a lightweight classifier can handle deterministically, while simultaneously denying genuinely complex dimensions the focused reasoning they require.
% 对每个维度都将相同文本盲目输入同一模型会造成双重浪费：既在轻量分类器能够确定处理的维度上消耗昂贵的 LLM 推理，又使真正复杂的维度得不到所需的集中推理。

\begin{figure}[!tbp]
\centering
\includegraphics[width=0.98\columnwidth]{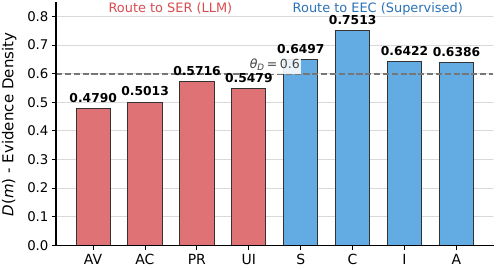}
\caption{Stable two-cluster evidence-density pattern across the eight CVSS base metrics.}
\label{fig:evidence-density}
\end{figure}

The framework therefore moves from sentence-level evidence selection to metric-level predictor decoupling, and requires a quantitative way to characterize this asymmetry.
% 至此，框架从句子级证据选择转向指标级预测器解耦，需要一种定量方式来刻画上述不对称。
We use metric-level evidence density to describe how consistently a CVSS metric is expressed through explicit, recurring, and locally identifiable cues in early OSINT, rather than the number or percentage of sentences that mention it.
% 本文所说的指标级证据密度，是指一个 CVSS 指标在早期报告中是否经常通过明确、重复且可局部识别的线索表达，而不是提及该指标的句子数量或比例。
Metrics with high evidence density tend to exhibit stable evidence patterns, whereas low-density metrics rely more on evidence scattered across sentences and on contextual relations.
% 高证据密度指标通常具有较稳定的证据表达模式；低证据密度指标的证据则更分散，并且更依赖跨句上下文关系。
We quantify this dataset-level property using $D(m)$, a report-balanced statistic derived from the normalized angular distances between retained sentences and the fixed CVSS definition anchor of each metric:
% 本文使用 $D(m)$ 对这一数据集级属性进行量化。该分数根据保留句子与各指标固定 CVSS 定义锚点之间的归一化角距离计算，并采用报告平衡的汇总方式：

\begin{equation}
\label{eq:evidence-density}
\begin{aligned}
\delta(s_i,m)
&=\frac{2}{\pi}\arccos\!\left(
\operatorname{cos\_sim}(\mathbf h_i,\mathbf h_m)\right),\\
a_X(m)
&=\frac{1}{|\mathcal I|}\sum_{j\in\mathcal I}
\frac{1}{|X_j|}\sum_{s_i\in X_j}\delta(s_i,m),\\
D_X(m)
&=\theta_{\min}+(\theta_{\max}-\theta_{\min})
L\!\left(\frac{a_X(m)-\bar a_{\mathcal M}}{s_{\mathcal M}}\right),\\
L(z)&=\frac{1}{1+e^{-z}},\qquad X\in\{R,R'\}.
\end{aligned}
\end{equation}

where $\mathcal{I}$ indexes the reports of \termdef{VPre-Bench-Val}, the validation split of VPre-Bench; $X_j$ denotes the sentences of report $j$ that belong to $X$ ($R$ before and $R'$ after Security-Evidence Screening); $\mathbf h_i=\operatorname{embed}(s_i)$ and $\mathbf h_m=\operatorname{embed}(m)$; $\delta(s_i,m)$ is the normalized angular distance; and $a_X(m)$ is its report-balanced mean. The semantic anchor $\mathbf h_m$ encodes the frozen official CVSS definition for metric $m$ using DistilBERT~\cite{ref56}. The terms $\bar a_{\mathcal{M}}$ and $s_{\mathcal{M}}$ are the mean and standard deviation of the eight metric-level angular-distance statistics on the validation set and are held fixed at test time.
% 其中，$\mathcal{I}$ 索引 VPre-Bench-Val（VPre-Bench 的验证划分）中的报告；$X_j$ 表示报告 $j$ 中属于集合 $X$ 的句子（$R$ 为安全证据筛选前、$R'$ 为筛选后）；$\text{embed}(m)$ 表示使用 DistilBERT 编码已冻结的指标 $m$ 官方 CVSS 定义文本所得的语义锚向量；$\bar a_{\mathcal{M}}$ 与 $s_{\mathcal{M}}$ 分别是验证集上八个指标角距离的统一均值和标准差，在测试阶段保持不变。
The term $\delta(s_i,m)$ converts cosine similarity into normalized angular distance, which removes the effect of embedding norms so that a larger raw dot product cannot be mistaken for stronger semantic evidence.
% 内层项先将句子与指标锚点之间的余弦相似度转换为归一化的角距离，以消除嵌入模长的影响，避免把较大的原始点积误当成更强的语义证据。
The averaging is deliberately two-level---sentences are averaged within each report, and reports are then averaged again---so that a longer report does not gain a larger weight simply by containing more sentences; every validation report contributes equally to $D(m)$.
% 平均过程有意采用两级结构——先在每份报告内对句子平均，再在报告间平均——使得长报告不会仅因句子更多而在总体结果中占据更高权重；每份验证报告对 $D(m)$ 的贡献相同。
All eight metrics are then passed through one and the same logistic normalization into the interval $[\theta_{\min},\theta_{\max}]$, the preset routing-search range: this introduces no metric-specific weights and embeds no group information about AV/AC/PR/UI or S/C/I/A, so the relative order of the metrics is preserved and their $D(m)$ values become directly comparable on one scale.
% 随后，八个指标经由完全相同的逻辑斯蒂归一化被映射到预设路由搜索区间 $[\theta_{\min},\theta_{\max}]$：该过程不引入指标专属权重，也不预先写入 AV/AC/PR/UI 或 S/C/I/A 的组别信息，因此既保持指标原有的相对次序，又使各指标的 $D(m)$ 值在同一尺度上直接可比。
$D(m)$ is a relative dataset-level indicator for cross-metric comparison and routing, rather than raw semantic similarity or a percentage.
% $D(m)$ 是用于指标间比较与路由的数据集级相对指标，而不是原始语义相似度或百分比。
On VPre-Bench-Val, higher-scoring metrics exhibit more stable and locally recognizable evidence patterns, whereas lower-scoring metrics rely more on scattered and context-dependent evidence.
% 在 VPre-Bench-Val 上，分数较高的指标表现出更稳定且可局部识别的证据模式；分数较低的指标则更多依赖分散且与上下文相关的证据。
The statistic is estimated once on VPre-Bench-Val and is neither a calibrated prediction confidence nor a score recomputed per report.
% 该统计量只在 VPre-Bench-Val 上计算一次，用于确定全局指标路由；它既不表示某次预测的置信度，也不会针对每份新报告重新计算。
Estimating $D_{R'}(m)$ across the eight metrics yields a clear empirical pattern rather than a continuum: the values fall into two well-separated clusters (\hyperfigure{fig:evidence-density}).
% 对八个指标估计 $D_{R'}(m)$ 后得到的是清晰的实证模式，而非连续分布：数值分成两个明显分离的簇（Figure~\ref{fig:evidence-density}）。
C ($D=0.7513$), S ($D=0.6497$), I ($D=0.6422$), and A ($D=0.6386$) form a high-density cluster (mean $0.6705$), whereas PR ($D=0.5716$), UI ($D=0.5479$), AC ($D=0.5013$), and AV ($D=0.4790$) form a low-density cluster (mean $0.5249$).
% C（$D=0.7513$）、S（$D=0.6497$）、I（$D=0.6422$）和 A（$D=0.6386$）构成高密度簇（均值 $0.6705$），PR（$D=0.5716$）、UI（$D=0.5479$）、AC（$D=0.5013$）和 AV（$D=0.4790$）构成低密度簇（均值 $0.5249$）。
The boundary $\theta_D$ is selected from the discrete candidates $\{0.4, 0.5, 0.6, 0.7, 0.8\}$ with full-vector MAE as the objective, yielding $\theta_D = 0.6$, which falls in the gap between the two clusters and separates them cleanly: the maximum of the low-density cluster is $0.5716$ and the minimum of the high-density cluster is $0.6386$, both at a clear margin from the boundary.
% 边界 $\theta_D$ 从离散候选 $\{0.4, 0.5, 0.6, 0.7, 0.8\}$ 中以完整向量 MAE 为目标选定，得到 $\theta_D=0.6$，该值落在两簇之间的空隙并将其干净分离：低密度簇的最大值为 $0.5716$，高密度簇的最小值为 $0.6386$，二者均与阈值保持明确间隔。
In \hyperfigure{fig:evidence-density}, each bar is one CVSS metric labeled with its validation-set $D(m)$, and the dashed line marks $\theta_D=0.6$; the red AV/AC/PR/UI bars sit below the boundary and the blue S/C/I/A bars sit at or above it.
% Figure~\ref{fig:evidence-density} 中每个柱表示一个 CVSS 指标，数值标签为验证集 $D(m)$ 估计值，虚线为 $\theta_D=0.6$；红色 AV/AC/PR/UI 柱位于边界以下，蓝色 S/C/I/A 柱位于或高于边界。
Bar height therefore represents typical metric-level evidence density, not prediction accuracy, CVSS severity, or a report-specific confidence score.
% 因此，柱高表示典型的指标级证据密度，而不是预测准确率、CVSS 严重性或某份报告的置信度。

The two-cluster pattern was established empirically on VPre-Bench-Val (the 470 reports with non-empty filtered evidence, covering 5,539 sentences before and 4,555 after sentence-level screening).
% 该双簇模式是在 VPre-Bench-Val（过滤后证据非空的 470 份报告，句子级筛选前 5,539 句、筛选后 4,555 句）上实证得到的。
We do not claim that it is universal across datasets; rather, it is a reproducible regularity of the evidence environment studied here that directly determines VPre's fixed routing table, and a new dataset can re-estimate $D(m)$ under the same procedure to verify or revise the clustering structure.
% 我们不声称该模式对任意数据集普适；它是本研究证据环境下稳定成立且可复现的规律，直接决定了 VPre 的固定路由表；新的数据集可按同一流程重估 $D(m)$，以验证或修正聚类结构。
The low-density cluster corresponds to dimensions requiring cross-sentence inference (AV/AC/PR/UI), whereas the high-density cluster corresponds to dimensions carried by more explicit lexical anchors (S/C/I/A).
% 低密度簇对应需要跨句推理的维度（AV/AC/PR/UI），高密度簇对应由较显式词法锚点承载的维度（S/C/I/A）。
Marchiori et al.~\cite{ref37} independently reported a partially aligned model-performance pattern---LLMs performed better on AV and AC, whereas embedding+XGBoost performed better on C, I, and A.
% Marchiori 等人独立报告了部分一致的模型表现模式：LLM 在 AV 与 AC 上更好，而 embedding+XGBoost 在 C、I 与 A 上更好。
Their result provides qualitative evidence of metric heterogeneity, whereas $D(m)$ supplies a quantitative validation-stage criterion for fixing VPre's routing table.
% 该研究为指标异质性提供了定性证据，而 $D(m)$ 则为冻结 VPre 路由表提供了定量的验证阶段判据。

The stable two-cluster structure has a direct methodological implication: the eight metrics should not be served by a single predictor, but decoupled into two paths, each routed to a predictor matched to its evidence pattern.
% 这一稳定的双簇结构具有直接的方法论含义：八个指标不应由单一预测器处理，而应解耦为两条路径，各自交给与其证据模式相匹配的预测器。
We call these two paths the \termdef{reasoning route}, which serves the low-density cluster through cross-sentence inference, and the \termdef{lexical route}, which serves the high-density cluster through supervised classification over explicit lexical evidence.
% 我们将这两条路径分别称为推理型路径与词法型路径：前者通过跨句推理服务低密度簇，后者通过对显式词法证据的监督分类服务高密度簇。
We formalize this decoupling as a validation-stage routing rule:
% 我们将这一解耦形式化为验证阶段的路由规则：

\begin{equation}
\label{eq:routing-rule}
m \rightarrow
\begin{cases}
\mathrm{EEC}, & D(m) \geq \theta_D,\\
\mathrm{SER}, & D(m) < \theta_D.
\end{cases}
\end{equation}
Under this rule, the high-density cluster (S/C/I/A) is routed to the \textbf{Explicit-Evidence Classifier (EEC, \hypersection{sec:dfe})}, the predictor instantiating the lexical route, and the low-density cluster (AV/AC/PR/UI) to the \textbf{Scattered-Evidence Reasoner (SER, \hypersection{sec:cre})}, the predictor instantiating the reasoning route; their concrete designs are described in the respective subsections.
% 在该规则下，高密度簇（S/C/I/A）路由至显式证据分类器 EEC（见 Section~\ref{sec:dfe}），它是词法型路径的具体实现；低密度簇（AV/AC/PR/UI）路由至分散证据推理器 SER（见 Section~\ref{sec:cre}），它是推理型路径的具体实现。两者的具体设计在各自小节中描述。
The resulting routing table is fixed once at the validation stage and reused for all VPre-Bench inference; at test time the only per-report operation is STDU screening.
% 由此得到的路由表在验证阶段一次性固定，并复用于全部 VPre-Bench 推理；测试时每份报告仅执行 STDU 筛选。
Thus $\theta_D$ serves to explain and reproduce the offline metric split rather than to dispatch individual cases.
% 因此，$\theta_D$ 用于解释并复现离线的指标划分，而非对单个案例进行分派。

\textbf{Section conclusion.}
% 本节小结。
Evidence-Aware Decoupling is VPre's methodological foundation. It operates at two levels: Security-Evidence Screening separates evidence preparation from prediction, and validation-derived Evidence-Matched Routing separates a uniform eight-metric predictor into specialized SER and EEC paths.
%证据感知解耦是 VPre 的方法论基础，包含两个层次：安全证据筛选将证据准备与预测分离，基于验证集的证据匹配路由则将统一的八指标预测器分解为专门化的 SER 和 EEC 路径。
Semantic drift (\hypersection{sec:semantic-drift}) explains why direct SOTA transfer fails.
% 语义漂移（Section~\ref{sec:semantic-drift}）解释了直接迁移 SOTA 失败的原因。
Security-Evidence Screening (\hypersection{sec:sentence-certainty}) explains why screening must precede prediction.
%安全证据筛选解释了为什么必须在预测之前进行筛选。
Evidence-matched routing (\hypersection{sec:evidence-density-assignment}) defines the fixed model assignment, which is necessary because the eight metrics do not present one uniform prediction problem.
% 证据匹配路由（Section~\ref{sec:evidence-density-assignment}）定义固定的模型分配；由于八个指标并不构成统一的预测问题，这一分配是必要的。
Together, these principles justify the STDU-to-SER/EEC architecture.
% 这些原则共同构成 STDU 到 SER/EEC 架构的完整依据。
\flushbottom
\section{VPre Framework Implementation}
% VPre 框架实现。
\label{sec:implementation}

The implementation follows the methodology established in \hypersection{sec:methodology-foundation}.
% 本章的实现沿用 Section~\ref{sec:methodology-foundation} 建立的方法论。
Specifically, STDU implements Security-Evidence Screening (Decoupling Level I), and the fixed routing table together with the SER and EEC predictors implements Evidence-Matched Routing (Decoupling Level II).
% 具体而言，STDU 实现安全证据筛选（解耦第一层），固定的路由表与 SER、EEC 两个预测器实现证据匹配路由（解耦第二层）。
An early report is first screened and reduced to security-relevant sentences; the resulting evidence is then passed to a routing table fixed on the validation set, which assigns AV/AC/PR/UI to SER and S/C/I/A to EEC.
% 早期报告首先经过筛选并收缩为安全相关句子；随后，过滤后的证据进入在验证集上固定的路由表，其中 AV/AC/PR/UI 由 SER 处理，S/C/I/A 由 EEC 处理。
The following subsections describe this data path and the concrete realization of each module.
% 下文按照该数据流依次说明各模块的具体实现。
The LLM and backbone comparisons used in the final configuration are retained with the SER and EEC descriptions.
% 最终配置所依据的大模型与骨干比较分别保留在 SER 和 EEC 的说明中。
\subsection{Framework Architecture Overview}
% 框架架构总览。
\label{sec:framework-architecture}

\begin{figure*}[!tbp]
\centering
\includegraphics[width=0.77\textwidth]{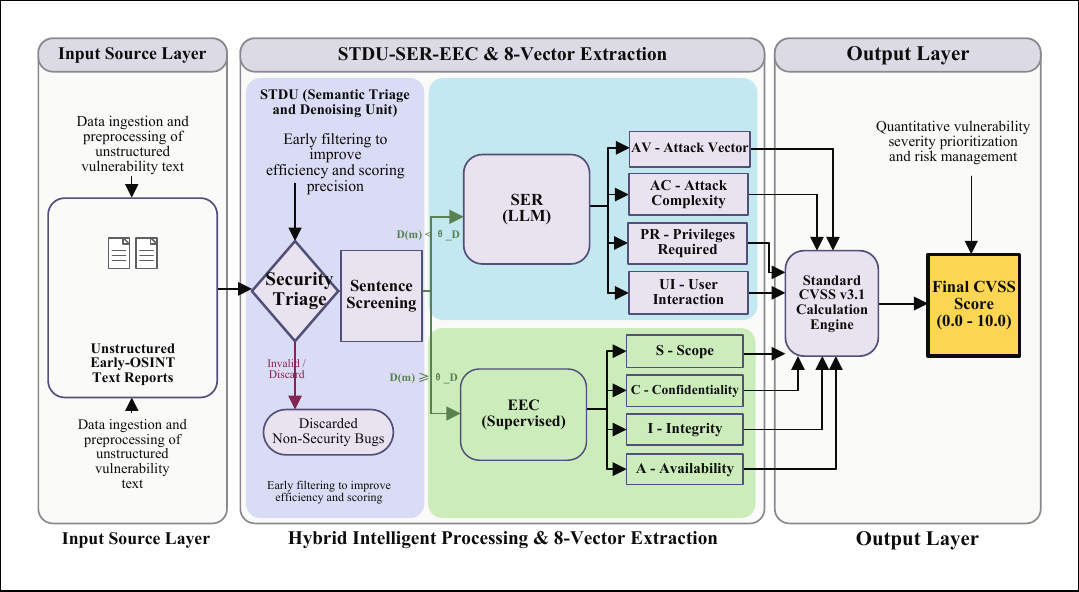}
\caption{Overall VPre architecture under two-level Evidence-Aware Decoupling.}
% VPre 的 STDU-SER-EEC 总体架构。
\label{fig:architecture}
\end{figure*}

\hyperfigure{fig:architecture} shows the complete path from a raw OSINT report to a CVSS v3.1 vector.
% Figure~\ref{fig:architecture} 展示了从原始 OSINT 报告到 CVSS v3.1 向量的完整路径。
The report first enters STDU, where a report-level decision removes non-security bugs and sentence-level screening retains VT/CP predictions and low-confidence O sentences from accepted reports, removing only high-confidence O predictions.
% 报告首先进入 STDU：报告级判定移除非安全 bug，句子级筛选则从已接收报告中保留 VT/CP 预测句及低置信度 O 句子，只删除高置信度 O 预测句。
The surviving sentences form $R'$, which is shared by the two prediction branches rather than being processed as two separate inputs.
% 通过筛选的句子构成 $R'$；两条预测路径共享该证据集，而不是分别处理两份不同输入。

The branch assignment is obtained before test inference from the validation-set statistic $D(m)$.
% 分支分配在测试推理前依据验证集统计量 $D(m)$ 确定。
It is therefore a lookup operation at deployment time, not a new evidence-density estimate for every incoming report.
% 因而，部署时执行的是固定路由查询，而不是为每份新报告重新估计证据密度。
AV, AC, PR, and UI are sent to SER because their evidence is commonly distributed across sentences, whereas S, C, I, and A are sent to EEC because their effects are more often expressed by stable lexical cues.
% AV、AC、PR 与 UI 的证据通常分散在多个句子中，因此交由 SER；S、C、I 与 A 更常由稳定词法线索表达，因此交由 EEC。
The eight categorical outputs are finally assembled in the standard metric order and converted into the CVSS v3.1 base score.
% 八个类别输出最终按标准指标顺序汇总，并计算 CVSS v3.1 基础分数。
This separation also provides the module boundaries used in the ablation study of \hypersection{sec:rq3}: evidence screening can be removed independently of the SER/EEC assignment, and either prediction branch can be evaluated under the same filtered input.
% 这种划分同时给出了 Section~\ref{sec:rq3} 消融实验所使用的模块边界：证据筛选可以独立于 SER/EEC 路由被移除，两条预测路径也可以在相同过滤输入下分别评估。

\FloatBarrier

\subsection{Decoupling Level I: Security-Evidence Screening}
% STDU：安全证据筛选。
\label{sec:stdu}
STDU implements Decoupling Level I, Security-Evidence Screening, operating at both granularities defined in \hypersection{sec:sentence-certainty}.
% STDU 实现解耦第一层——安全证据筛选，在 Section~\ref{sec:sentence-certainty} 定义的两个粒度上运作。
The raw report stream contains both ordinary software defects and security-relevant cases, and even accepted reports contain large amounts of text that does not contribute to CVSS prediction.
% 原始报告流同时包含普通软件缺陷与安全相关案例；即使是被接收的报告，也包含大量无助于 CVSS 预测的文本。
STDU screens these two sources of noise at their respective granularities.
% STDU 分别在报告级和句子级筛选这两类噪声。

% Schedule the full-width Level-I figure early enough for bottom placement;
% the barrier at the end of this subsection keeps it before Level II.
\begin{figure*}[!b]
\centering
\includegraphics[width=0.77\textwidth]{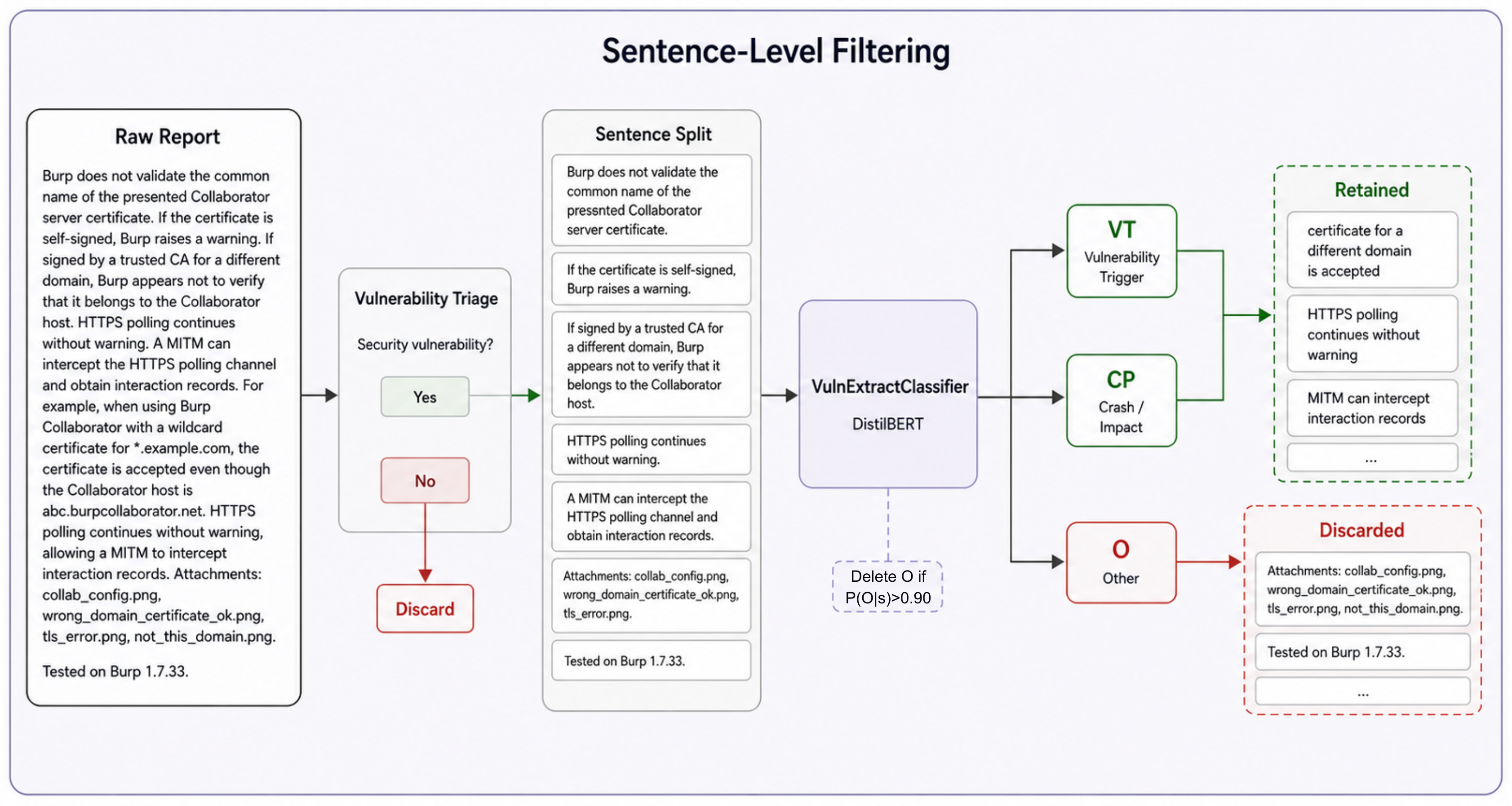}
\caption{Decoupling Level I: STDU-based Security-Evidence Screening.}
% STDU 分诊与句子级过滤流程。
\label{fig:stdu}
\end{figure*}

\paragraph{Report-level security triage.}
%报告级安全分诊。
At the report level, STDU employs GPT-4o-mini to determine whether a sanitized report describes a security-relevant weakness.
% 在报告级，STDU 调用 GPT-4o-mini 判断清洗后的报告是否描述安全相关弱点。
This decision is deliberately entrusted to an LLM rather than keyword or rule-based matching, because security words alone are unreliable evidence: reports of ordinary functional defects also mention ``security'' or ``crash'', so the triage must judge the technical condition described in the report---whether attacker-controlled or untrusted input can plausibly reach a weakness and violate a security boundary---rather than the mere presence of security vocabulary.
% 这一决策有意交给 LLM 而非关键词或规则匹配，因为安全词本身不是可靠证据：普通功能缺陷的报告同样会提到"security"或"crash"；分诊必须判断报告描述的技术条件——攻击者可控或不可信输入能否合理触达某个弱点并破坏安全边界——而不是安全词汇的出现与否。
The prompt correspondingly favors recall under incomplete evidence: an early report may be tentative and incomplete yet still security-relevant, and discarding a genuine vulnerability at the gate is more costly than admitting an occasional non-security report, which the sentence-level stage can still attenuate.
% 相应地，提示词在证据不完整时优先召回：早期报告可能仍是初步且不完整的，但依然与安全相关；在入口处丢弃真实漏洞的代价高于偶尔放进一份非安全报告，后者还可以由句子级筛选进一步削减。
The output is constrained to a closed label, a confidence in $[0,1]$, and short supporting evidence spans; the parser rejects unknown labels or out-of-range confidence values, and the evidence spans are retained for audit only---they never replace the original sentences in the next stage.
% 输出被限制为封闭标签、$[0,1]$ 的置信度以及简短的支持性证据片段；解析器拒绝未知标签或越界置信度；证据片段仅用于审计，绝不替代下一阶段使用的原始句子。
The model is queried with temperature 0 and a maximum output length of 160 tokens; the report is enclosed in a \texttt{<REPORT>} block to separate source text from the instruction.
% 模型采用 temperature 0 和 160 tokens 的最大输出长度；报告被封装在 \texttt{<REPORT>} 块中，以区分源文本与指令。
A schema violation triggers a single format-only repair request, and a response that remains invalid is recorded as an explicit failure state rather than passed downstream.
% 模式错误会触发一次仅针对格式的修复请求；若响应仍不合法，则被记录为显式失败状态，不会传入下游。
Reports labeled as ordinary functional, usability, build, or documentation defects are discarded, and API failures are likewise recorded explicitly rather than passed to the scoring modules.
% 被判为普通功能、可用性、构建或文档缺陷的报告会被丢弃；API 失败同样被显式记录，而不会进入评分模块。
The complete instruction and output schema are provided in the supplementary material.
% 完整指令与输出模式见补充材料。

\paragraph{Sentence-level VT/CP/O screening.}
%句子级 VT/CP/O 筛选。
\phantomsection\label{para:vuln-extract-classifier}Reports retained by the report-level decision are segmented into sentences and passed to \textsc{VulnExtractClassifier}, the sentence classifier that realizes the screening rule of \hyperequation{eq:filtered-evidence}.
% 通过报告级判定的报告被切分为句子，并输入 VulnExtractClassifier——即实现 Equation~\ref{eq:filtered-evidence} 筛选规则的句子分类器。
We fine-tune this classifier ourselves rather than using a zero-shot model, for two reasons.
% 我们自行微调该分类器而不是使用零样本模型，原因有二。
First, the VT/CP/O boundary is semantic rather than lexical: the same word ``request'' marks a vulnerability trigger in \emph{``crafted request''} but is neutral in \emph{``request body''}, and O-class text such as scheduling or repair-only discussion frequently co-occurs with security vocabulary, so rules or keyword matching cannot draw the boundary.
% 其一，VT/CP/O 的边界是语义的而非词法的：同一个词 "request" 在 "crafted request" 中是漏洞触发条件，在 "request body" 中却是中性内容；O 类文本（如日程安排或仅涉及修复的讨论）经常与安全词汇共现，规则或关键词匹配无法划出这条边界。
Second, delegating the per-sentence decision to a general-purpose LLM would multiply pipeline cost---a report contains dozens of sentences and the report-level gate already consumes an LLM call---so the sentence level uses a lightweight fine-tuned encoder (66M-parameter DistilBERT) instead.
% 其二，把逐句决策交给通用 LLM 会使流水线成本成倍增长——每份报告包含数十个句子，且报告级门控已经消耗一次 LLM 调用——因此句子级改用轻量级微调编码器（66M 参数的 DistilBERT）。

The training data come from the 5,581-report cleaned VPre-Bench corpus: five graduate annotators labeled the sentences of the 4,257 training and 473 validation reports sentence by sentence (49,686 training and 5,101 validation sentences), with disagreements adjudicated by a third annotator, under the label operationalizations in Table~\ref{tab:stdu-labels}.
% 训练数据来自清洗后的 5,581 份报告的 VPre-Bench 语料：5 名研究生对 4,257 份训练报告与 473 份验证报告的句子逐句标注（49,686 条训练句、5,101 条验证句），分歧由第三名标注者裁决，标注依据为 Table~\ref{tab:stdu-labels} 的操作性定义。
The class distribution is imbalanced (O 39.0\%, VT 13.5\%, CP 47.5\% in training), which reflects the composition of real early reports rather than a sampling choice.
% 类别分布不平衡（训练集中 O 39.0\%、VT 13.5\%、CP 47.5\%），这反映的是真实早期报告的构成，而非抽样选择。
Training proceeds in two stages: the encoder is first adapted to the early-report domain with one epoch of masked-language modeling over 12,000 unlabeled training-partition sentences, and is then frozen while a three-way classification head is trained on the 49,686 labeled sentences (best epoch 9, learning rate $10^{-3}$, maximum length 128, seed 42; full configuration in the supplementary material).
% 训练分两个阶段：先用训练划分中 12,000 条无标签句子做一轮掩码语言建模，使编码器适配早期报告的文本特点；随后冻结编码器，在 49,686 条人工标注句子上训练三分类头（最佳 epoch 9、学习率 $10^{-3}$、最大长度 128、随机种子 42；完整配置见补充材料）。

On the 10,108-sentence test split, the classifier reaches 84.02\% accuracy (macro-F1 80.31\%), with per-class F1 of 83.72\% (O), 69.29\% (VT), and 87.93\% (CP).
% 在 10,108 句的测试划分上，该分类器取得 84.02\% 的准确率（宏平均 F1 80.31\%），分类别 F1 为 O 83.72\%、VT 69.29\%、CP 87.93\%。
Under the conservative deletion rule of \hyperequation{eq:filtered-evidence} with $\tau_s=0.90$, this translates into an evidence recall of 99.32\%---only 0.68\% of true VT/CP sentences are lost---at a retained-set precision of 70.48\%; the retained non-evidence sentences are left for SER and EEC to process rather than being deleted on uncertain grounds.
% 在 $\tau_s=0.90$ 的保守删除规则（Equation~\ref{eq:filtered-evidence}）下，这对应 99.32\% 的证据召回率——仅 0.68\% 的真实 VT/CP 句子被误删——保留集合查准率为 70.48\%；被保留的非证据句留给 SER 和 EEC 处理，而不是在拿不准时将其删除。
The threshold is selected only on the 5,101-sentence validation fold: among candidates $\{0.50, 0.60, 0.70, 0.80, 0.90\}$, the rule keeps the recall within 0.01 of the maximum and then maximizes retained-set precision, breaking ties toward the lower threshold; the test set is never consulted.
% 阈值仅在 5,101 句的验证折上选择：在候选 $\{0.50, 0.60, 0.70, 0.80, 0.90\}$ 中，先保留证据召回率距最大值不超过 0.01 的候选，再最大化保留集合查准率，平局取较低阈值；测试集从未参与该过程。
If every sentence of a report is removed as a high-confidence O, STDU returns \texttt{EMPTY\_EVIDENCE} instead of restoring the original report: restoring the full text would silently cancel screening and re-inject all noise, whereas the explicit \texttt{EMPTY\_EVIDENCE} state marks the case as lacking evidence, remains auditable, and produces no fabricated CVSS vector.
% 若一份报告的所有句子均被作为高置信度 O 删除，STDU 返回 EMPTY_EVIDENCE 而不是恢复原始报告：恢复全文等于悄悄取消筛选、让噪声全部回流；而显式的 EMPTY_EVIDENCE 状态将该案例标记为缺乏证据，保持可审计，且不会产生虚构的 CVSS 向量。
This state is rare---6 of 851 test reports (0.71\%)---indicating that the conservative rule seldom rejects a report outright.
% 该状态很少出现——851 份测试报告中仅 6 份（0.71\%）——说明保守规则极少将报告整体拒绝。

\begin{table}[!tbp]
\centering
\small
\setlength{\tabcolsep}{3pt}
\caption{Operational definitions of the STDU sentence labels.}
% STDU 句子标签的操作性定义。
\label{tab:stdu-labels}
\begin{tabular}{p{0.09\columnwidth}p{0.29\columnwidth}p{0.50\columnwidth}}
\toprule
Label & Meaning & Include / exclude criterion \\
\midrule
VT & Vulnerability Trigger & Attacker capability, required state, exposed interface, crafted input, or the action that activates the weakness. \\
% VT：漏洞触发条件，包括攻击者能力、所需状态、暴露接口、构造输入或激活弱点的操作。
CP & Consequence Phenomenon & Security-relevant disclosure, privilege gain, corruption, bypass, exhaustion, or reachable crash. \\
% CP：安全相关的泄露、权限提升、破坏、绕过、资源耗尽或可达崩溃等后果现象。
O & Other & Metadata, scheduling, acknowledgments, version-only text, repair-only discussion, or unrelated diagnostics. \\
% O：元数据、日程、致谢、仅版本文本、仅修复讨论或无关诊断信息。
\bottomrule
\end{tabular}
\end{table}

\hyperfigure{fig:stdu} traces the two decisions along the data path.
% Figure~\ref{fig:stdu} 沿数据路径展示了这两个决策。
A raw report first reaches the security-triage gate: a negative decision follows the discard exit and the case never enters scoring, while an accepted report proceeds to sentence splitting.
% 原始报告首先进入安全分诊门：否定决策沿丢弃出口离开，该案例永不进入评分；被接收的报告则进入句子切分。
VulnExtractClassifier then labels every sentence, and the deletion rule removes only the high-confidence O sentences; VT/CP predictions and low-confidence O sentences survive into $R'$.
% 随后 VulnExtractClassifier 为每个句子打标，删除规则仅移除高置信度的 O 句子；VT/CP 预测句和低置信度 O 句子进入 $R'$。
The figure thus encodes exactly two filters at two granularities: report-level triage decides \emph{whether} a case enters VPre, and sentence-level screening decides \emph{which sentences} of an accepted case are visible to SER and EEC.
% 因此该图恰好在两个粒度上编码了两种过滤：报告级分诊决定案例是否进入 VPre，句子级筛选决定被接收案例的哪些句子对 SER 和 EEC 可见。
\FloatBarrier

\subsection{Decoupling Level II: Evidence-Matched Routing}
%固定指标路由(实现证据匹配路由的离线查表)。
\label{sec:decoupling-bridge}

After STDU produces $R'$, the unit of decision changes from sentences to CVSS metrics.
% STDU 生成 $R'$ 后，系统的决策单位由句子转向 CVSS 指标。
The sentence-level probabilities $P(k\mid s_i)$ have already served their purpose by deciding which evidence is retained; they are not interpreted as uncertainty in the final CVSS prediction.
% 句子级概率 $P(k\mid s_i)$ 已用于决定保留哪些证据，并不表示最终 CVSS 预测的不确定性。
The subsequent split instead uses the validation-set statistic $D(m)$, which describes how explicitly each metric is expressed across the filtered reports.
% 后续划分使用验证集统计量 $D(m)$，用于描述各指标在过滤报告中通常被多明确地表达。
This subsection realizes Evidence-Matched Routing (Decoupling Level II) as an offline lookup whose result is frozen before deployment.
% 本小节将证据匹配路由（解耦第二层）实现为一次离线查表，其结果在部署前即已冻结。

\begin{figure*}[!tbp]
\centering
\includegraphics[width=0.71\textwidth]{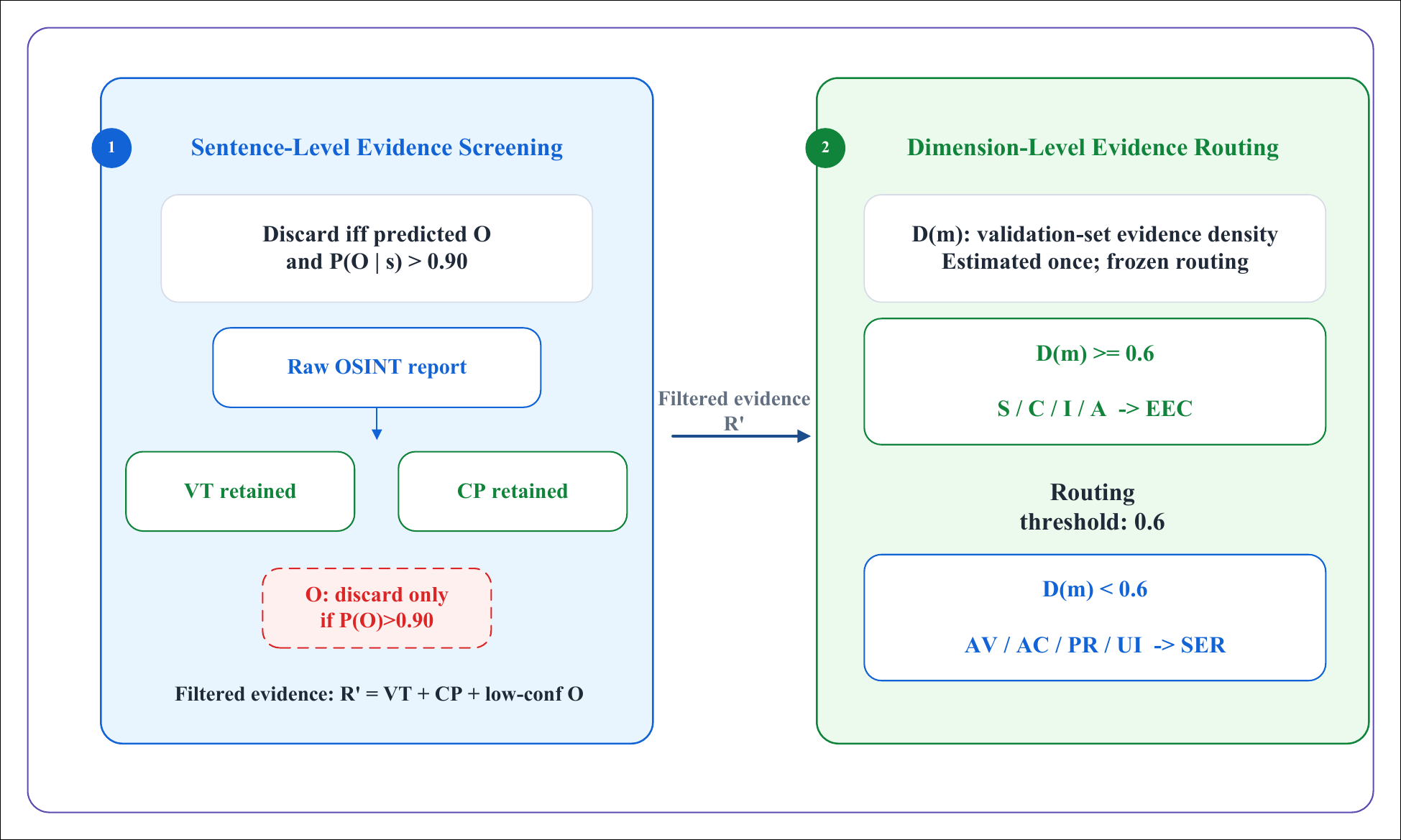}
\caption{Two levels of Evidence-Aware Decoupling: Level I, Security-Evidence Screening; Level II, Evidence-Matched Routing.}
%证据感知解耦的两个层次：证据筛选与 SER/EEC 专门化。
\label{fig:certainty-decoupling}
\end{figure*}

This change of granularity is summarized in \hyperfigure{fig:certainty-decoupling}.
% Figure~\ref{fig:certainty-decoupling} 概括了这一粒度变化。
The left side removes only high-confidence O sentences to form $R'$, while the right side compares the metric-level $D(m)$ values with $\theta_D=0.6$.
% 左侧仅删除高置信度 O 类句子并形成 $R'$，右侧则将指标级 $D(m)$ 与 $\theta_D=0.6$ 比较。
Under the routing rule in \hyperequation{eq:routing-rule}, AV/AC/PR/UI are assigned to SER and S/C/I/A to EEC; equality is assigned to EEC.
% 根据 Equation~\ref{eq:routing-rule} 的路由规则，AV/AC/PR/UI 被分配给 SER，S/C/I/A 被分配给 EEC；等于阈值时归入 EEC。
The same $R'$ is supplied to both routes, so the split changes the predictor used for a metric rather than the evidence available for an individual report.
% 两条路径接收同一个 $R'$；因此，该划分改变的是各指标所使用的预测器，而不是单份报告可用的证据。

\subsection{SER: Scattered-Evidence Reasoner}
% SER：分散证据推理器。
\label{sec:cre}

AV, AC, PR, and UI are not always stated in a single sentence.
% AV、AC、PR 与 UI 并不总是在单个句子中被直接说明。
Their prediction may depend on linking reachability, conditions outside the attacker's control, privileges held before exploitation, and actions expected from another user.
% 对这些指标的判断可能需要关联可达性、攻击者控制之外的条件、利用前所持权限以及另一用户需要执行的动作。
SER therefore reads the complete filtered set $R'$ and predicts these four metrics jointly with GPT-4o-mini.
% 因此，SER 读取完整的过滤证据集 $R'$，并使用 GPT-4o-mini 联合预测这四项指标。

The retained sentences are numbered $s_1,\ldots,s_n$ and placed in a single \texttt{<EVIDENCE>} block, which the prompt treats as the only admissible source of vulnerability evidence.
% 保留句子被编号为 $s_1,\ldots,s_n$ 并置于单一 \texttt{<EVIDENCE>} 块中；提示词将其视为唯一允许使用的漏洞证据。
The Chain-of-Thought instruction then considers each metric according to its CVSS definition: reachability and attacker position for AV, uncontrollable preconditions for AC, privileges held before exploitation for PR, and a separate user's action for UI~\cite{ref47,ref48}.
% 随后，思维链指令依据各自的 CVSS 定义判断指标：AV 考察可达性与攻击者位置，AC 考察不可控前置条件，PR 考察利用前权限，UI 考察是否需要另一用户执行操作。
Before committing to a class, the model is asked to cite the supporting sentence identifiers and check common confusions, such as treating every ``remote'' report as AV:N or conflating AC with PR.
% 在输出类别前，模型需要引用支持性句子标识符，并检查常见混淆，例如将所有“remote”报告均判为 AV:N，或混淆 AC 与 PR。
When the evidence is incomplete, the prompt requires a lower confidence instead of an invented attack scenario.
% 当证据不完整时，提示词要求降低置信度，而不是编造攻击情景。
Each prediction contains a legal \texttt{value}, a confidence in $[0,1]$, the cited sentence identifiers, and a short rationale.
% 每项预测包含合法的 \texttt{value}、位于 $[0,1]$ 的置信度、被引用的句子标识符和简短理由。

Inference uses temperature 0 and a maximum output length of 600 tokens.
% 推理采用 temperature 0 和 600 tokens 的最大输出长度。
The parser accepts exactly the keys AV, AC, PR, and UI and checks their CVSS v3.1 domains: AV $\in\{\mathrm{N,A,L,P}\}$, AC $\in\{\mathrm{L,H}\}$, PR $\in\{\mathrm{N,L,H}\}$, and UI $\in\{\mathrm{N,R}\}$.
% 解析器仅接受 AV、AC、PR 与 UI 四个键，并校验其 CVSS v3.1 定义域：AV $\in\{\mathrm{N,A,L,P}\}$、AC $\in\{\mathrm{L,H}\}$、PR $\in\{\mathrm{N,L,H}\}$、UI $\in\{\mathrm{N,R}\}$。
Confidence ranges, evidence identifiers, and rationales are checked at the same time.
% 系统同时校验置信度范围、证据标识符和理由。
A schema violation triggers one format-only repair request; if the repaired response is still invalid, the case is recorded as \texttt{SER\_PARSE\_FAIL}.
% 模式错误会触发一次仅针对格式的修复请求；若修复后的响应仍不合法，该案例被记录为 \texttt{SER\_PARSE\_FAIL}。
The complete prompt and output schema are provided in the supplementary material.
% 完整提示词和输出模式见补充材料。

To select the LLM for SER, five general-purpose LLMs (GPT-4o-mini, DeepSeek-V3, Qwen3.5, Llama-4-scout, and Grok-4-fast) are compared without task-specific fine-tuning; with CoT fixed, all are evaluated under the same input and decoding configuration in Table~\ref{tab:cre-llm}.
% 为 SER 选定大模型，我们在不做任务专项微调的条件下比较五个通用 LLM（GPT-4o-mini、DeepSeek-V3、Qwen3.5、Llama-4-scout、Grok-4-fast）；固定 CoT 后，所有模型在相同输入与解码配置下评估，结果见 Table~\ref{tab:cre-llm}。

\begin{table*}[!tbp]
\centering
\small
\caption{Validation-set per-metric SER accuracy under CoT prompting and mean accuracy on the routed AV/AC/PR/UI metrics (\%).}
% CoT 提示下 SER 的逐指标准确率，以及路由指标 AV/AC/PR/UI 的平均准确率（\%）。
\label{tab:cre-llm}
\resizebox{\textwidth}{!}{%
\begin{tabular}{lccccccccc}
\toprule
Model & AV & AC & PR & UI & S & C & I & A & Routed Mean \\
\midrule
\textbf{GPT-4o-mini (Selected)} & \textbf{80.61} & \textbf{87.19} & \textbf{79.38} & \textbf{85.55} & \textbf{83.31} & \underline{68.04} & 63.81 & \underline{67.80} & \textbf{83.18} \\
DeepSeek-V3 & \underline{78.50} & \underline{83.55} & \underline{70.39} & \underline{71.18} & 73.65 & 67.57 & 52.76 & 65.33 & \underline{75.91} \\
Qwen3.5 & 76.94 & 64.19 & 60.77 & 59.72 & \underline{80.32} & 45.31 & \underline{75.48} & 67.33 & 65.41 \\
Llama-4-scout & 71.77 & 52.00 & 57.14 & 62.24 & 75.51 & \textbf{84.54} & 72.83 & 60.00 & 60.79 \\
Grok-4-fast & 72.00 & 70.00 & 62.00 & 68.00 & 80.00 & 56.00 & \textbf{84.00} & \textbf{80.00} & 68.00 \\
\bottomrule
\end{tabular}%
}
\vspace{2pt}
\parbox{0.98\textwidth}{\footnotesize \textbf{Bold} indicates the best result in each column, and \underline{underlining} indicates the second-best result. ``Selected'' identifies the model used by SER.}
% 粗体表示每列最优结果，下划线表示每列次优结果；“Selected”表示 SER 最终采用的模型。
\end{table*}

Table~\ref{tab:cre-llm} favors GPT-4o-mini on every routed metric; its AV/AC/PR/UI mean is \textbf{83.18\%}, while DeepSeek-V3 ranks second at \underline{75.91\%}.
% Table~\ref{tab:cre-llm} 中 GPT-4o-mini 在所有路由指标上均占优；其 AV/AC/PR/UI 平均准确率为 \textbf{83.18\%}，DeepSeek-V3 以 \underline{75.91\%} 排名第二。
The S/C/I/A columns are included as a diagnostic control, not as SER outputs.
% S/C/I/A 列仅作为诊断性对照，而不是 SER 的实际输出。
GPT-4o-mini retains 83.31\% accuracy on S but falls to 68.04\%, 63.81\%, and 67.80\% on C, I, and A, and the best-performing LLM is not the same across these columns.
% GPT-4o-mini 在 S 上保持 83.31\% 的准确率，但在 C、I 与 A 上分别降至 68.04\%、63.81\% 和 67.80\%，而且这些列中的最优大模型并不相同。
By contrast, BERT-Small-VPre obtains 90.60\%, 83.67\%, 82.37\%, and 85.90\% on S/C/I/A, which supports retaining these metrics on the EEC route.
% 相比之下，BERT-Small-VPre 在 S/C/I/A 上分别达到 90.60\%、83.67\%、82.37\% 和 85.90\%，支持将这些指标保留在 EEC 路径中。

\subsection{EEC: Explicit-Evidence Classifier}
% EEC：显式证据分类器。
\label{sec:dfe}

Compared with AV, AC, PR, and UI, the S, C, I, and A metrics are more often supported by direct lexical cues in early reports; EEC therefore treats their prediction as supervised text classification over $R'$.
% 与 AV、AC、PR 和 UI 相比，S、C、I 与 A 在早期报告中更常具有直接的词法线索，因此 EEC 将这些指标的预测建模为对 $R'$ 的监督文本分类。

Three suffixes distinguish checkpoints that share the BERT-Small architecture but have different training histories.
% 本文用三个后缀区分采用相同 BERT-Small 架构但训练经历不同的 checkpoint。
\textbf{BERT-Small-PT}, where PT means \emph{pretrained}, is the public \texttt{prajjwal1/bert-small} checkpoint used for parameter initialization before task-specific fine-tuning.
% \textbf{BERT-Small-PT} 中的 PT 表示 \emph{pretrained（预训练）}，该名称指任务专项微调前用于参数初始化的公共 checkpoint \texttt{prajjwal1/bert-small}。
\textbf{BERT-Small-NVD} is the reproduced model fine-tuned on standardized late-stage NVD descriptions and evaluated both on NVD input and by direct transfer to early OSINT.
% \textbf{BERT-Small-NVD} 是在标准化后期 NVD 描述上微调的复现模型，分别在 NVD 输入和直接迁移到早期 OSINT 的条件下评估。
\textbf{BERT-Small-VPre} denotes four metric-specific EEC checkpoints. Each checkpoint is initialized from BERT-Small-PT and fine-tuned separately for S, C, I, or A on the STDU-filtered VPre-Bench training split.
% \textbf{BERT-Small-VPre} 表示四个指标专用的 EEC checkpoint；它们均由 BERT-Small-PT 初始化，并分别针对 S、C、I、A 在 STDU 过滤后的 VPre-Bench 训练集上独立微调。
For Table~\ref{tab:dfe-backbone}, every candidate backbone follows the same four-classifier, 16-epoch protocol and is evaluated on the validation split; the table therefore reports post-fine-tuning model-selection results.
% Table~\ref{tab:dfe-backbone} 对所有候选骨干采用相同的四分类器、16 个 epoch 方案，并在验证集上比较，因此表中结果是各指标独立微调后的模型选择结果。

\begin{table*}[!b]
\centering
\small
\caption{Validation-set EEC backbone accuracy and mean accuracy on the routed S/C/I/A metrics after 16 epochs (\%).}
% EEC 候选骨干在路由指标 S/C/I/A 上训练 16 个 epoch 后的准确率与平均准确率（\%）。
\label{tab:dfe-backbone}
\begin{tabular}{lccccc}
\toprule
Model & S & C & I & A & Mean \\
\midrule
DistilBERT & \textbf{92.01} & 79.55 & 82.02 & \underline{85.43} & 84.75 \\
\textbf{BERT-Small-VPre (Selected)} & 90.60 & \textbf{83.67} & \underline{82.37} & \textbf{85.90} & \textbf{85.64} \\
BERT-Medium & \underline{91.54} & 82.26 & \textbf{84.14} & 84.49 & \underline{85.61} \\
DeBERTa-v3-Small & 91.19 & \underline{82.96} & 80.14 & 82.96 & 84.31 \\
\bottomrule
\end{tabular}
\vspace{2pt}
\parbox{0.90\textwidth}{\footnotesize \textbf{Bold} indicates the best result in each column, and \underline{underlining} indicates the second-best result. ``Selected'' identifies the backbone used by the final EEC.}
% 粗体表示每列最优结果，下划线表示每列次优结果；“Selected”表示最终 EEC 采用的骨干。
\end{table*}

The models exhibit different metric preferences: DistilBERT is strongest on S, BERT-Medium on I, and BERT-Small-VPre on C and A.
% 不同模型表现出不同的指标偏好：DistilBERT 在 S 上最强，BERT-Medium 在 I 上最强，而 BERT-Small-VPre 在 C 和 A 上最强。
Across S/C/I/A, BERT-Small-VPre reaches the highest mean accuracy of \textbf{85.64\%}, only 0.03 percentage points above BERT-Medium at \underline{85.61\%}.
% 在 S/C/I/A 四项指标上，BERT-Small-VPre 的平均准确率最高，为 \textbf{85.64\%}，仅比 BERT-Medium 的 \underline{85.61\%} 高 0.03 个百分点。
We select BERT-Small-VPre for its overall balance rather than for a uniform advantage on every metric.
% 因此，本文选择 BERT-Small-VPre 的依据是其整体均衡性，而不是其在每项指标上都占优。

\begin{figure*}[!tbp]
\centering
\includegraphics[width=0.71\textwidth]{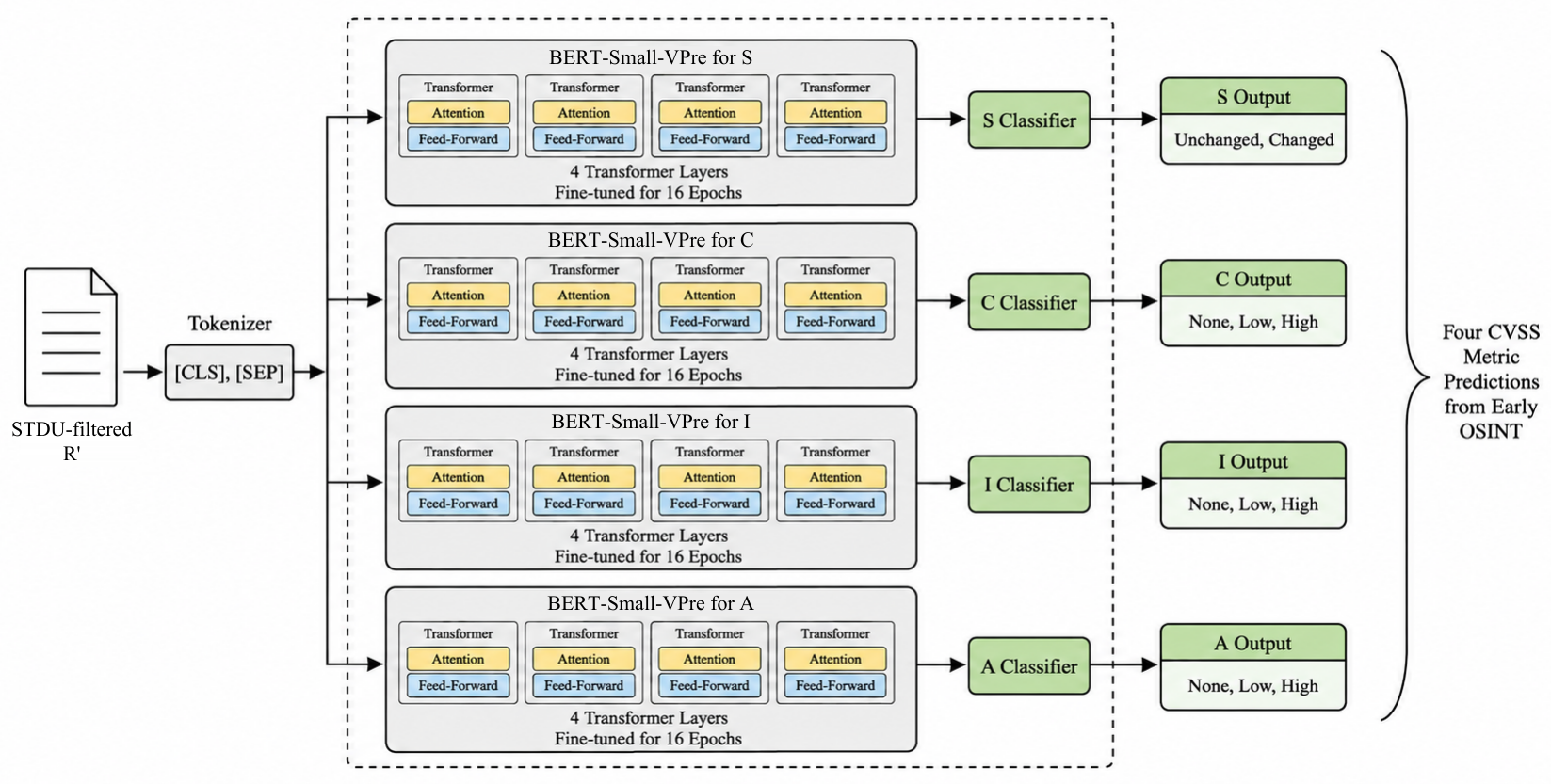}
\caption{Metric-specific EEC prediction pipeline with four independent BERT-Small-VPre classifiers for S, C, I, and A.}
% 指标专用 EEC 预测流水线：四个相互独立的 BERT-Small-VPre 分类器分别预测 S、C、I 与 A。
\label{fig:dfe-pipeline}
\end{figure*}

Figure~\ref{fig:dfe-pipeline} depicts the selected EEC configuration from Table~\ref{tab:dfe-backbone}: four independent BERT-Small-VPre classifiers with separate encoders and classification heads for S, C, I, and A. It shows the final post-selection architecture rather than an additional fine-tuning stage.
% Figure~\ref{fig:dfe-pipeline} 展示 Table~\ref{tab:dfe-backbone} 选出的最终 EEC 配置：四个相互独立的 BERT-Small-VPre 分类器分别预测 S、C、I、A，并具有各自独立的编码器和分类头；该图描述的是选型后的最终结构，而不是表后又进行的一轮额外微调。
The S classifier predicts $\{\mathrm{U,C}\}$, while the other three classifiers predict $\{\mathrm{N,L,H}\}$.
% S 分类器预测 $\{\mathrm{U,C}\}$，其余三个分类器预测 $\{\mathrm{N,L,H}\}$。
Each classifier is trained for 16 epochs using AdamW with a learning rate of $2\times10^{-5}$, batch size 32, and random seed 42.
% 每个分类器均采用 AdamW 独立训练 16 个 epoch，学习率为 $2\times10^{-5}$，批大小为 32，随机种子为 42。
AV, AC, PR, and UI remain on the SER route and therefore do not enter this pipeline.
% AV、AC、PR 和 UI 仍由 SER 路由处理，因此不进入该流程。

\section{Evaluation}
% 评估。
\label{sec:evaluation}

\subsection{Evaluation Setup}
% 评估设置。
\label{sec:evaluation-setup}

To systematically evaluate \namedconcept{VPre}, our experiments are driven by three Research Questions.
% 为系统评估 VPre，实验围绕三个研究问题展开。

\textbf{RQ1:} Can the VPre framework overcome existing baselines' dependency on NVD standardized descriptions, achieving comparable or superior prediction accuracy on early OSINT?
% RQ1： VPre 完整框架能否突破现有基线对 NVD 标准化描述的依赖，在早期 OSINT 上实现可比甚至更优的预测精度？

\textbf{RQ2:} Can STDU reliably identify genuine vulnerabilities within high-noise early-stage intelligence?
% RQ2：STDU 能否在早期高噪声情报中可靠识别真实安全漏洞？

\textbf{RQ3:} Does the complete Evidence-Aware Decoupling framework (STDU+SER+EEC) outperform the tested non-decoupled and partially decoupled alternatives?
% RQ3： 完整的证据感知解耦框架（STDU+SER+EEC）是否优于所测试的非解耦与部分解耦替代方案？

\noindent\textbf{Dataset Construction and Temporal-Fidelity Controls.}\par
% 数据集构建与时间保真控制。
To answer these research questions rigorously, we construct and publicly release two complementary datasets.
% 为严格回答上述三个研究问题，我们构建并公开两个互补的数据集。
\textbf{VPre-Bench} is the core benchmark: it supports the accuracy comparison in RQ1, the lead-time measurement, and the module ablations in RQ3, and contains 5,581 CVE entries spanning 2002--2025 with early-stage raw intelligence reconstructed through top-down NVD backtracing.
% VPre-Bench 是核心基准：它支撑 RQ1 的准确性对比、提前时间度量以及 RQ3 的模块消融，包含 2002--2025 年间的 5,581 个 CVE 条目，并通过自顶向下的 NVD 回溯重建早期原始情报。
\textbf{PARSE-Triage} is dedicated to RQ2: it contains 46,940 raw issue descriptions (45,759 non-vulnerability and 1,181 vulnerability reports; noise-to-signal ratio $\approx$38:1) from five open-source ecosystems---Ambari, Camel, Chromium, Derby, and Wicket---for evaluating screening reliability under high noise.
% PARSE-Triage 专为 RQ2 而建：包含来自 Ambari、Camel、Chromium、Derby 和 Wicket 五个开源生态的 46,940 条原始 issue 描述（45,759 份非漏洞、1,181 份漏洞报告，噪信比约 38:1），用于检验高噪声下的筛选可靠性。

\begin{figure}[!tbp]
\centering
\includegraphics[width=0.94\columnwidth]{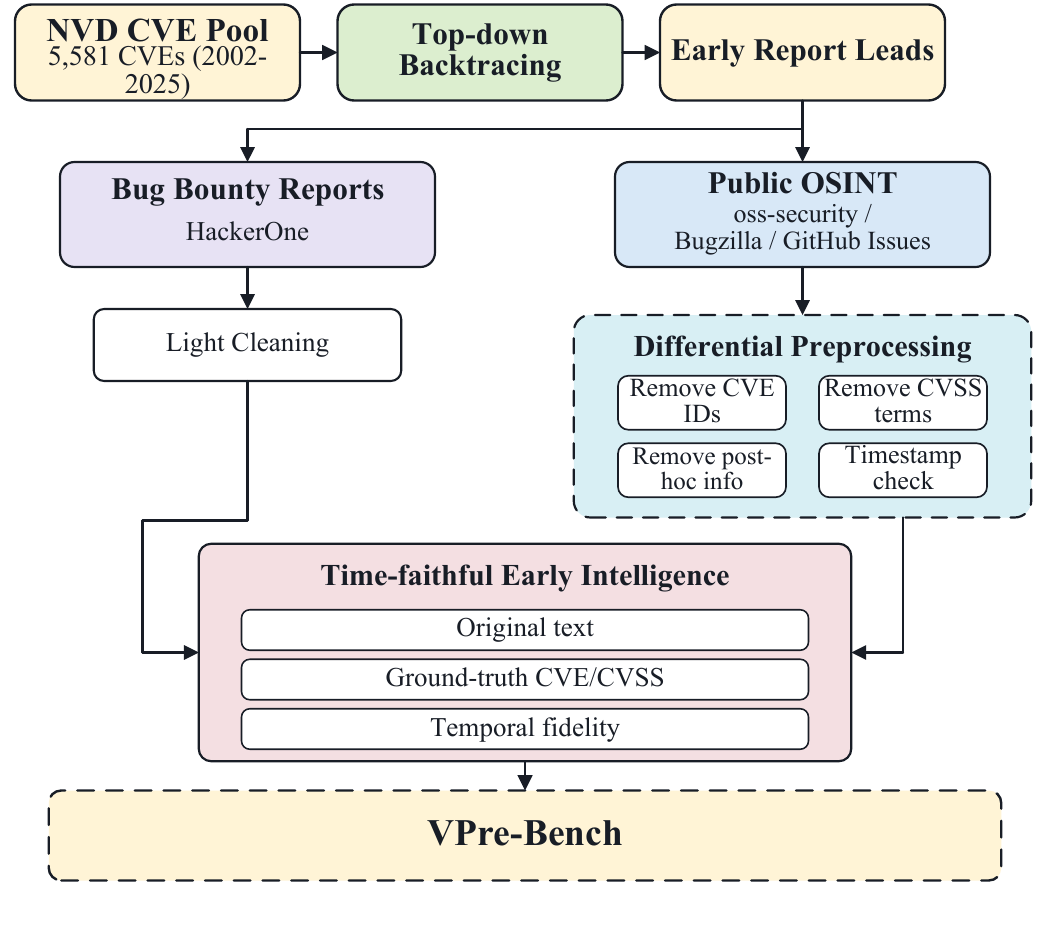}
\caption{VPre-Bench data collection and temporal sanitization pipeline.}
\label{fig:data-collection}
\end{figure}

As shown in \hyperfigure{fig:data-collection}, VPre-Bench draws from two complementary source categories.
% 如图~\ref{fig:data-collection} 所示，VPre-Bench 的数据来自两个互补的来源类别。
Curated bug bounty reports, predominantly from HackerOne, serve as high-fidelity anchors: they have passed platform triage, preserve the submitter's original text, and are naturally free of post-hoc CVSS or remediation commentary, but their scale is limited because many reports are private or restricted.
% 精选漏洞赏金报告（主要来自 HackerOne）充当高保真锚点：它们通过了平台分诊、保留了提交者的原始文本、天然不含事后的 CVSS 评分或修复评论，但规模受限，因为许多报告是私有或受限的。
Public OSINT sources---oss-security, Bugzilla, and GitHub Issues---offer broader coverage yet inherently contain information added after the initial disclosure: CVE identifiers appearing later in the same issue thread, CVSS scores cited by community members, and patch links posted after vendor confirmation~\cite{ref07,ref15,ref16}.
% 公开 OSINT 来源（oss-security、Bugzilla 和 GitHub Issues）覆盖更广，但天然包含首次披露之后补充的信息：同一 issue 线程中后来出现的 CVE 编号、社区成员引用的 CVSS 评分，以及厂商确认后发布的补丁链接。
Training or evaluating on such unfiltered text would let models learn information unavailable at the prediction moment and artificially inflate early-forecast performance~\cite{ref03,ref07,ref49}.
% 使用此类未过滤文本进行训练或评估，会使模型学习预测时刻尚不可用的信息，从而人为夸大早期预测性能。
As shown in \hyperfigure{fig:data-collection}, all public-source texts therefore undergo automated sanitization with a domain-specific exclusion dictionary that strips CVE identifiers (e.g., \emph{CVE-202X-XXXX}), CVSS scoring keywords (e.g., \emph{Base Score 9.8}), and authoritative references to post-hoc fixes (e.g., \emph{official patch released}), followed by timestamp verification that retains only pre-NVD segments; NVD descriptions are never used as early-text inputs; each sample further records four timestamps (early intelligence, CVE assignment, NVD description, NVD CVSS) to enforce temporal consistency, and the two paths merge into the leakage-controlled corpus in which titles and sanitized reports serve as model-facing inputs while CVSS vectors and base scores remain separate ground truth.
% 如图~\ref{fig:data-collection} 所示，所有公开来源文本均通过领域专用排除词典自动净化，移除 CVE 编号（如 CVE-202X-XXXX）、CVSS 评分关键词（如 Base Score 9.8）以及对事后修复的权威性引用（如 official patch released），再经时间戳验证仅保留 NVD 之前的片段；NVD 描述从不作为早期文本输入；每个样本还记录四个时间戳（早期情报、CVE 分配、NVD 描述、NVD CVSS）以保证时间一致性；两条路径最终汇入受泄漏控制的语料，其中标题与净化报告作为模型输入，CVSS 向量与基础分数作为独立真值。
Manual inspection of 200 randomly sampled entries after processing confirmed residual hindsight content below 2\%.
% 处理后对 200 个随机抽样条目进行人工检查，确认残留的事后信息低于 2\%。

\begin{figure}[!ht]
\centering
\includegraphics[width=0.96\columnwidth]{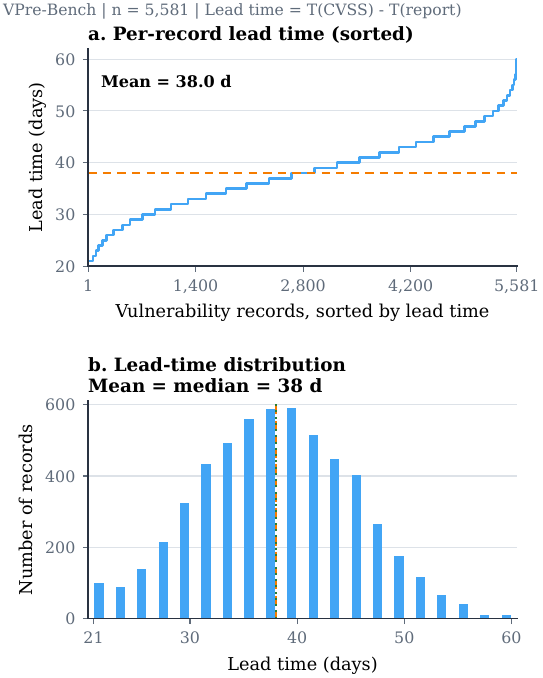}
\caption{Timestamp-derived lead-time distribution in VPre-Bench.}
\label{fig:lead-time-distribution}
\end{figure}

As shown in \hyperfigure{fig:lead-time-distribution}, the lead time of each record measures how many days its early intelligence precedes official NVD CVSS scoring: computing $\Delta t_i=T^{\mathrm{NVD\text{-}CVSS}}_i-t^{\mathrm{early}}_i$ from paired timestamps over the 5,581 matched records yields a mean of 38.0 days, a median of 38 days, an interquartile range of 33--43 days, and an observed range of 21--60 days.
% 如图~\ref{fig:lead-time-distribution} 所示，每条记录的提前时间衡量的是其早期情报比 NVD 正式 CVSS 评分早出现的天数：对 5,581 条匹配记录按配对时间戳计算 $\Delta t_i=T^{\mathrm{NVD\text{-}CVSS}}_i-t^{\mathrm{early}}_i$，得到均值 38.0 天、中位数 38 天、四分位距 33--43 天、观测范围 21--60 天。
This distribution quantifies the early-warning window within which VPre operates, rather than any manually assigned lead.
% 该分布量化了 VPre 赖以工作的早期预警窗口，而非人为指定的提前量。

To quantify the impact of semantic drift, we reproduce the NVD-trained checkpoint and denote it \textbf{BERT-Small-NVD}.
% 为量化语义漂移，本文复现使用 NVD 描述训练的 checkpoint，并将其记为 \textbf{BERT-Small-NVD}。
The same checkpoint is first evaluated on standardized late-stage NVD descriptions and then directly transferred, without retraining, to sanitized early P/A-stage VPre-Bench reports.
% 同一个 checkpoint 先在标准化后期 NVD 描述上评估，随后在不重新训练的情况下直接迁移至经过时间净化的早期 P/A 阶段 VPre-Bench 报告。
This cross-phase setup specifically tests whether conventional SOTA supervised models can maintain efficacy in high-noise early-warning scenarios, or fundamentally require VPre's structural enhancements.
% 这一跨阶段设置专门测试传统 SOTA 监督模型能否在高噪声早期预警场景中维持效能，还是从根本上需要 VPre 的结构增强。

\textbf{Unified notation} used throughout \hypersection{sec:rq1}--\hypersection{sec:rq3}: $\text{SOTA}_{\text{early}}$ (SOTA directly transferred to early OSINT), $\text{VPre}_{\text{early}}$ (VPre on early OSINT---our core result), $\text{SOTA}_{\text{NVD}}$ (SOTA on NVD standard descriptions---late-stage strong baseline), $\text{VPre}_{\text{NVD}}$ (VPre on NVD standard descriptions---compatibility check).
% 统一符号贯穿 Sections 5.2--5.4：$\text{SOTA}_{\text{early}}$（SOTA 直接迁移到早期 OSINT）、$\text{VPre}_{\text{early}}$（VPre 在早期 OSINT 上——本文核心结果）、$\text{SOTA}_{\text{NVD}}$（SOTA 在 NVD 标准描述上——后期强基线）、$\text{VPre}_{\text{NVD}}$（VPre 在 NVD 标准描述上——兼容性检查）。
Unless otherwise noted, $\text{SOTA}_{\text{NVD}}$ denotes the directly comparable CVEDrill/AutoCVSS-style NVD reference rather than the non-comparable KBS standardization pipeline.
% 除非特别说明，$\text{SOTA}_{\text{NVD}}$ 指可直接对比的 CVEDrill/AutoCVSS 风格 NVD 参考，而非不可直接比较的 KBS 标准化流水线。
\FloatBarrier

\subsection{RQ1: VPre vs. SOTA on Early OSINT}
\label{sec:rq1}

\paragraph{Evaluation Focus.}
%评估重点。
We first establish whether VPre outperforms transferred baselines when both are evaluated on early OSINT and quantify its timestamp-derived forecasting lead.
%我们首先检验 VPre 与迁移基线在同一早期 OSINT 上评估时是否更优，并量化由时间戳得到的预测提前量。
\paragraph{SOTA Reference Results on Late-Stage NVD Data.}
% 后期 NVD 数据上的 SOTA 参考结果。
On native PARSE S/E-phase standardized data, CVEDrill reports MAE 0.5490~\cite{ref66} and the KBS 2025 pipeline (GPT-3.5-Turbo + BERT-Small) reports MAE 0.166~\cite{ref31}.
%在原生 PARSE S/E 阶段的标准化数据上，CVEDrill 的 MAE 为 0.5490，KBS 2025 流水线的 MAE 为 0.166。
\coreclaim{Because both results use late-stage standardized inputs, they do not establish performance when only early unstructured OSINT is available.}
% 由于两项结果均使用后期标准化输入，它们无法说明仅有早期非结构化 OSINT 时的性能。
RQ1 compares three baseline families with Full VPre under the early-OSINT setting: transferred and hybrid baselines in \hypertable{tab:mae-mse}, pure end-to-end LLM baselines in \hypertable{tab:pure-llm-mae}, and lightweight single-model baselines in \hypertable{tab:dfe-backbone-error}.
% 在早期 OSINT 设置下，RQ1 将三类基线与 Full VPre 比较：Table~\ref{tab:mae-mse} 中的迁移与混合基线、Table~\ref{tab:pure-llm-mae} 中的纯端到端 LLM 基线，以及 Table~\ref{tab:dfe-backbone-error} 中的轻量单模型基线。
All VPre-Bench prediction results in this subsection are measured on the same 851-report held-out test set.
% 本小节所有 VPre-Bench 预测结果均在同一个包含 851 份报告的留出测试集上测得。
MAE measures average absolute score error, whereas MSE penalizes larger deviations more strongly; lower values indicate better prediction accuracy.
% MAE 衡量平均绝对评分误差，MSE 对较大偏差施加更强惩罚；二者数值越低，预测精度越高。

\begin{table*}[!tbp]
\centering
\small
\caption{MAE/MSE comparison with transferred and hybrid baselines.}
\label{tab:mae-mse}
\resizebox{\textwidth}{!}{%
\begin{tabular}{llcc}
\toprule
Architecture & Data Scenario & MSE & MAE \\
\midrule

\multicolumn{4}{l}{\textbf{Reference: Standardized NVD (PARSE S/E)}} \\
BERT-Small-NVD (K\"uhn reproduction)
& Standardized NVD (PARSE S/E) & 2.4939 & 0.8918 \\
SOTA (CVEDrill, NVD Reference)
& Standardized NVD (PARSE S/E) & 1.2310 & 0.5490 \\
SOTA (KBS 2025, NVD Reference)
& Standardized NVD (PARSE S/E) & 0.2990 & 0.1660 \\

\multicolumn{4}{l}{\textbf{Early OSINT Baselines: Pre-$T_0$ Reports (PARSE P/A)}} \\
BERT-Small-NVD (K\"uhn reproduction)
& VPre-Bench (PARSE P/A; direct transfer) & 3.9821 & 1.3095 \\
SOTA (CVEDrill, Early Transfer)
& VPre-Bench (PARSE P/A) & 3.1792 & 1.2255 \\
SOTA (KBS 2025, Early Transfer)
& VPre-Bench (PARSE P/A) & 2.8192 & 1.1680 \\

\multicolumn{4}{l}{\textbf{Compatibility / Cross-Regime Check}} \\
$\text{VPre}_{\text{NVD}}$ (compatibility check)
& Standardized NVD (PARSE S/E) & 2.2148 & 0.8629 \\

\multicolumn{4}{l}{\textbf{Full System on Early OSINT: Pre-$T_0$ (PARSE P/A)}} \\
\textbf{VPre (STDU--SER--EEC)}
& \textbf{VPre-Bench (PARSE P/A)}
& \textbf{1.6859} & \textbf{0.7239} \\

\bottomrule
\end{tabular}%
}
\end{table*}
Under the matched VPre-Bench early-OSINT setting, \hypertable{tab:mae-mse} shows that Full VPre reduces MAE by 44.7\% relative to the reproduced BERT-Small-NVD early-transfer baseline (1.3095 to 0.7239), by 40.9\% relative to the CVEDrill early-transfer baseline (1.2255 to 0.7239), and by 38.0\% relative to the measured KBS 2025 early-transfer baseline (1.1680 to 0.7239).
% 在匹配的 VPre-Bench 早期 OSINT 设置下，Table~\ref{tab:mae-mse} 表明，Full VPre 相对复现的 BERT-Small-NVD 早期迁移基线将 MAE 从 1.3095 降至 0.7239，降低 44.7\%；相对 CVEDrill 早期迁移基线将 MAE 从 1.2255 降至 0.7239，降低 40.9\%；相对实测的 KBS 2025 早期迁移基线将 MAE 从 1.1680 降至 0.7239，降低 38.0\%。
The NVD-reference rows remain contextual because they use standardized late-stage inputs.
% 由于使用的是后期标准化输入，NVD 参考行仍只用于提供背景语境。
\begin{table}[!tbp]
\centering
\footnotesize
\caption{End-to-end MAE/MSE comparison with pure-LLM baselines on the VPre-Bench held-out test set.}
\label{tab:pure-llm-mae}
\setlength{\tabcolsep}{4pt}
\begin{tabular}{@{}lcc@{}}
\toprule
Model & MSE & MAE \\
\midrule
GPT-4o-mini & 3.5220 & 1.3051 \\
DeepSeek-V3 & 4.0000 & 1.4491 \\
Qwen3.5 & 6.2325 & 1.7250 \\
Llama-4-scout & 4.5141 & 1.6235 \\
Grok-4-fast & 3.4181 & 1.3771 \\
\midrule
\textbf{VPre (STDU--SER--EEC)} & \textbf{1.6859} & \textbf{0.7239} \\
\bottomrule
\end{tabular}
\end{table}
\hypertable{tab:pure-llm-mae} shows that Full VPre outperforms all tested pure-LLM baselines in both MAE and MSE under the same VPre-Bench input.
%Table~\ref{tab:pure-llm-mae} 表明，在相同 VPre-Bench 输入下，Full VPre 的 MAE 和 MSE 均优于所有被测试的纯 LLM 基线。
\begin{table}[!tbp]
\centering
\footnotesize
\caption{End-to-end comparison of lightweight single-model baselines with Full VPre on the VPre-Bench held-out test set.}
\label{tab:dfe-backbone-error}
\setlength{\tabcolsep}{4pt}
\begin{tabular}{@{}lcc@{}}
\toprule
Backbone & MSE & MAE \\
\midrule
DistilBERT & 2.5835 & 0.8786 \\
BERT-Small-VPre & 2.5377 & 0.8772 \\
DeBERTa-v3-Small & 2.1581 & 0.9337 \\
BERT-Medium & 2.1382 & 0.8145 \\
\midrule
\textbf{Full VPre (STDU--SER--EEC)} & \textbf{1.6859} & \textbf{0.7239} \\
\bottomrule
\end{tabular}
\end{table}
\hypertable{tab:dfe-backbone-error} shows that Full VPre also outperforms the strongest tested lightweight single-model baseline, indicating that a pure lightweight-model framework---one supervised model covering all eight metrics, however strong its backbone---does not reproduce the complete decoupled framework.
% Table~\ref{tab:dfe-backbone-error} 表明，Full VPre 同样优于所测试的最强轻量单模型基线，说明纯轻量模型框架——无论骨干多强，都用一个监督模型预测全部八项指标——无法复现完整的解耦框架。

Across these matched early-input comparisons, Full VPre remains the strongest tested system; the NVD-reference rows provide late-stage context, while module attribution is examined separately in RQ3.
%在这些匹配的早期输入比较中，Full VPre 是所测试系统中表现最强的方案；NVD 参考行仅提供后期背景，模块归因则在 RQ3 中单独检验。
\paragraph{Fair Comparison on Early OSINT.}
% 早期 OSINT 上的公平比较。
Across the transferred supervised, pure-LLM, and lightweight single-model baselines evaluated on VPre-Bench, VPre yields the lowest error.
%在 VPre-Bench 上评估的迁移监督基线、纯 LLM 基线和轻量单模型基线中，VPre 取得最低误差。
The consistent pattern indicates that early OSINT benefits from architecture-level adaptation combining STDU filtering, evidence-aware metric assignment, and SER/EEC specialization.
%这一一致结果表明，早期 OSINT 可受益于结合 STDU 过滤、证据感知指标分配和 SER/EEC 专门化的架构适配。

\begin{figure}[!ht]
\centering
\includegraphics[width=0.94\columnwidth]{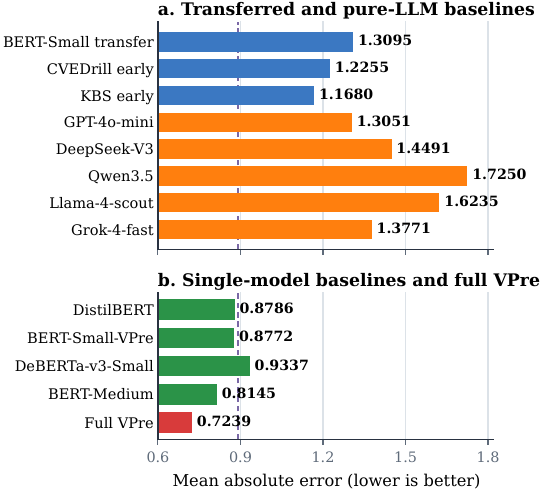}
\caption{MAE comparison on the VPre-Bench held-out test set under the early-OSINT setting.}
\label{fig:early-mae}
\end{figure}

\hyperfigure{fig:early-mae} is the direct early-OSINT MAE comparison.
% Figure~\ref{fig:early-mae} 是早期 OSINT MAE 的直接比较。
The horizontal groups and colors distinguish transferred supervised baselines, pure LLM predictors, lightweight single-model baselines, and the full VPre system.
% 横向分组与颜色区分迁移监督基线、纯 LLM 预测器、轻量单模型基线和完整 VPre 系统。

The three blue transfer bars represent BERT-Small-NVD, CVEDrill, and KBS 2025 evaluated on VPre-Bench, with MAE values of 1.3095, 1.2255, and 1.1680, respectively.
% 三根蓝色迁移柱分别表示在 VPre-Bench 上评估的 BERT-Small-NVD、CVEDrill 和 KBS 2025，其 MAE 分别为 1.3095、1.2255 和 1.1680。
The orange bars represent five pure LLM predictors: GPT-4o-mini, DeepSeek-V3, Qwen3.5, Llama-4-scout, and Grok-4-fast, each of which directly predicts the eight CVSS metrics from the early text.
% 橙色柱表示 GPT-4o-mini、DeepSeek-V3、Qwen3.5、Llama-4-scout 和 Grok-4-fast 五个纯 LLM 预测器，每个模型都直接根据早期文本预测八个 CVSS 指标。
The green bars represent four lightweight single-model baselines: DistilBERT, BERT-Small-VPre, DeBERTa-v3-Small, and BERT-Medium, each of which is used alone to predict all eight CVSS metrics.
% 绿色柱表示 DistilBERT、BERT-Small-VPre、DeBERTa-v3-Small 和 BERT-Medium 四个轻量单模型基线，每个模型都被单独用于预测全部八个 CVSS 指标。
The red bar represents the full VPre system using STDU--SER--EEC.
% 红色柱表示使用 STDU--SER--EEC 的完整 VPre 系统。
Every solid bar is measured on the same VPre-Bench early text, and the number within each bar is its MAE.
% 每个实心柱都在相同的 VPre-Bench 早期文本上测得，柱内数字为相应的 MAE。
Lower bars indicate lower CVSS prediction error.
% 柱越低，表示 CVSS 预测误差越小。
The horizontal dashed line at MAE 0.8918 represents BERT-Small-NVD evaluated on standardized late-stage NVD input.
% MAE 为 0.8918 的水平虚线表示 BERT-Small-NVD 使用后期标准化 NVD 输入时的结果。
Because this result uses different and later-stage input, it is included only as a reference and is not ranked with the solid bars.
% 由于该结果使用了不同的后期输入，因此这里只将其作为参考，不与实心柱一起排名。
Under the same early-OSINT setting, VPre achieves the lowest MAE of 0.7239, while the best-performing lightweight single-model baseline, BERT-Medium, obtains an MAE of 0.8145.
% 在相同的早期 OSINT 设置下，VPre 取得最低的 MAE 0.7239，而表现最好的轻量单模型基线 BERT-Medium 的 MAE 为 0.8145。
This comparison shows that using one supervised model to predict all eight CVSS metrics is less effective than using the full STDU--SER--EEC method.
% 这一比较说明，使用单个监督模型预测全部八个 CVSS 指标，不如使用完整的 STDU--SER--EEC 方法有效。

\begin{figure}[!ht]
\centering
\includegraphics[width=0.92\columnwidth]{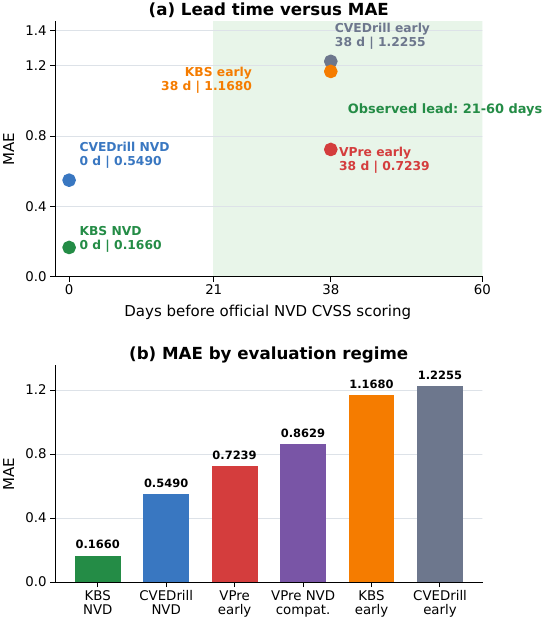}
\caption{Lead-time--error trade-off for early CVSS forecasting; NVD-only results are shown as zero-lead references.}
\label{fig:early-warning-gain}
\end{figure}

\paragraph{Early-Warning Benefit: Prediction 38 Days Before NVD Scoring.}
% 提前预警收益：在 NVD 正式评分前 38 天进行预测。
\hyperfigure{fig:early-warning-gain} quantifies VPre's practical early-warning benefit by jointly showing forecast lead time and prediction error. VPre achieves an MAE of 0.7239 on the 851-report held-out test set, while paired timestamps across the full benchmark show an average lead of 38 days (observed range: 21--60 days). NVD-based results are shown only as zero-lead reference points and are not treated as same-input competitors.
% Figure~\ref{fig:early-warning-gain} 通过同时展示预测提前量与误差，量化 VPre 的实际提前预警收益。VPre 在包含 851 份报告的留出测试集上取得 0.7239 的 MAE，而完整基准的配对时间戳显示平均提前量为 38 天（观测范围 21--60 天）。基于 NVD 的结果仅作为零提前量参考点，不作为同输入条件下的直接竞争基线。

\hyperfigure{fig:early-warning-gain} has two panels that answer different questions.
% Figure~\ref{fig:early-warning-gain} 的两个面板分别回答不同问题。
In the left panel, the horizontal axis is days before official NVD CVSS scoring and the vertical axis is MAE: NVD-only reference points sit at 0 days, while VPre and the two early-transfer baselines are plotted at the observed average lead of 38 days.
% 左侧面板的横轴表示相对 NVD 正式发布 CVSS 评分提前的天数，纵轴表示 MAE：NVD-only 参考点位于 0 天，VPre 和两个早期迁移基线则绘制在观测到的平均提前 38 天位置。
The shaded 21--60 day region denotes the observed disclosure window rather than a confidence interval, and horizontal position is forecast lead time rather than model runtime.
% 阴影覆盖的 21--60 天区域表示观测披露窗口而非置信区间，横向位置表示预测提前量而非模型运行时间。
In the right panel, bars restate the exact MAE values and add the $\text{VPre}_{\text{NVD}}$ compatibility result.
% 右侧面板的柱形重新给出精确 MAE，并加入 $\text{VPre}_{\text{NVD}}$ 兼容性结果。
Colors and labels distinguish late-stage NVD references, early-transfer measurements, VPre's early operating point, and the compatibility check.
% 颜色与标签区分后期 NVD 参考、早期迁移测量、VPre 的早期运行点以及兼容性检查。
Together, the two panels express a time--error trade-off rather than a single same-data leaderboard: only the early-transfer and $\text{VPre}_{\text{early}}$ points share the pre-$T_0$ input regime.
% 两个面板共同表达时间与误差的权衡，而不是单一同数据排行榜；只有早期迁移点和 $\text{VPre}_{\text{early}}$ 使用相同的 $T_0$ 之前输入阶段。

\FloatBarrier
\subsection{RQ2: STDU Screening Reliability}
%RQ2：STDU 筛选可靠性。
\label{sec:rq2}

\paragraph{Evaluation Focus.}
%评估重点。
STDU screening operates at two granularities; this subsection evaluates the report-level gate under high noise, while the reliability of sentence-level screening has been reported in \hypersection{sec:stdu} (84.02\% accuracy with 99.32\% evidence recall on the test split).
% STDU 筛选在两个粒度上运作；本小节评估高噪声下的报告级门控，句子级筛选的可靠性已在 Section~\ref{sec:stdu} 中报告（测试集上准确率 84.02\%、证据召回 99.32\%）。
Building on RQ1's assessment of overall early-OSINT performance, we next examine whether STDU can identify genuine vulnerabilities under the extreme $\sim$38:1 noise setting.
%在 RQ1 对早期 OSINT 整体性能的评估基础上，我们进一步检验 STDU 能否在约 38:1 的极端噪声设置下可靠识别真实安全漏洞。
Using the PARSE-Triage dataset, STDU was evaluated in zero-shot inference across multiple LLMs (GPT-4o-mini, DeepSeek-V3, Qwen3.5, etc.)~\cite{ref48}.
%使用 PARSE-Triage 数据集，在多个 LLM（包括 GPT-4o-mini、DeepSeek-V3 和 Qwen3.5 等）上以零样本推理方式评估 STDU。
\textbf{Evaluation metrics are Precision, Recall, and F1---not Accuracy}, because in a 38:1 imbalanced distribution, an all-benign dummy classifier would achieve $\sim$97.5\% Accuracy with Recall = 0\%, rendering Accuracy meaningless.
% 评估指标为 Precision、Recall 和 F1——不使用 Accuracy，因为在 38:1 的不平衡分布下，全判"非漏洞"的 dummy 分类器能达到约 97.5\% Accuracy 但 Recall 为 0\%，Accuracy 完全无意义。
At the report level, GPT-4o-mini achieves Precision 0.98, Recall 1.00, and F1 0.99.
% 在报告级筛选任务上，GPT-4o-mini 达到 Precision 0.98、Recall 1.00、F1 0.99。
Across all evaluated LLMs, F1 remains above 0.95, indicating consistent screening performance within the tested model set.
%在所有被评估的 LLM 中，F1 均高于 0.95，表明筛选性能在所测试模型范围内保持一致。
This cross-model consistency supports the use of LLMs at the STDU gateway for reducing the needle-in-a-haystack problem and filtering semantic noise at data intake.
% 这种一致性说明，使用 LLM 作为 STDU 网关能有效避免大海捞针式漏报，在数据接入阶段滤除海量语义噪声。
A traditional TF-IDF + XGBoost baseline is included for comparison; its lower Recall suggests that lexical features alone are insufficient to capture implicit security semantics in early reports.
%传统 TF-IDF + XGBoost 基线的召回率较低，表明仅依赖词法特征不足以捕捉早期报告中的隐含安全语义。
Together, these results support STDU as a substantive screening component rather than a cosmetic preprocessing step.
% 综合结果支持将 STDU 视为实质性的筛选模块，而非装饰性的预处理步骤。

\FloatBarrier
\subsection{RQ3: Effectiveness of Two-Level Evidence-Aware Decoupling}
%RQ3：两层证据感知解耦的有效性。
\label{sec:rq3}

\paragraph{Evaluation Focus.}
%评估重点。
Having evaluated overall performance in RQ1 and screening reliability in RQ2, we finally test whether the complete decoupled framework (STDU+SER+EEC) outperforms the tested non-decoupled and partially decoupled alternatives, isolating the contribution of each decoupling level through controlled ablations under the same setting.
%在通过 RQ1 和 RQ2 分别评估整体性能与筛选可靠性后，我们最后检验完整的拆分框架（STDU+SER+EEC）是否优于所测试的非解耦与部分解耦替代方案，并通过相同设置下的受控消融分离各解耦层次的贡献。

Four variants are compared.
%比较四种变体。
\textbf{w/o sentence screening} (raw text directly to SER/EEC) increases MAE to 0.9642 because irrelevant sentences and non-security noise mislead downstream prediction, particularly on C/I/A dimensions.
%w/o sentence screening 将原始文本直接输入 SER/EEC；由于无关句子和非安全噪声会误导下游预测，其 MAE 上升至 0.9642。
\textbf{SER-only} (the same STDU-filtered input, with all 8 dimensions routed to the LLM with CoT) increases MAE to 1.0427 because LLMs are less stable than supervised models on dimensions with explicit keyword patterns.
%SER-only 在相同 STDU 过滤输入下将八个指标全部交给使用 CoT 的 LLM，MAE 上升至 1.0427。
\textbf{EEC-only} (the same STDU-filtered input, with all 8 dimensions routed to the BERT-Medium all-metric configuration also reported in \hypertable{tab:dfe-backbone-error}) reaches MAE 0.8145, showing that supervised models remain strong but lack the cross-sentence reasoning needed for AV/AC/PR/UI.
%EEC-only 在相同 STDU 过滤输入下将八个指标全部交给 Table~\ref{tab:dfe-backbone-error} 中同一套 BERT-Medium 全指标配置，MAE 为 0.8145。
\textbf{Full VPre} (STDU plus the fixed SER/EEC split) achieves the lowest held-out test error among the tested variants (MSE 1.6859, MAE 0.7239).
%Full VPre 由 STDU 和固定 SER/EEC 划分组成，在所测试变体中取得最低留出测试误差。
These controlled ablations are conducted on the same 851-report VPre-Bench held-out test set to isolate the effects of screening and predictor routing, thereby providing design-attribution evidence for the complete framework.
%这些受控消融在同一个包含 851 份报告的 VPre-Bench 留出测试集上进行，用于分离证据筛选与预测器路由的作用，从而为完整框架提供设计归因证据。

\begin{table}[!tbp]
\centering
\small
\caption{Held-out-test ablations of the two Evidence-Aware Decoupling levels.}
%VPre-Bench 上两层证据感知解耦的留出测试集消融。
\label{tab:module-ablation}
\begin{tabular}{lcc}
\toprule
Variant & MSE & MAE \\
\midrule
w/o sentence screening & 2.3847 & 0.9642 \\
SER-only & 2.7619 & 1.0427 \\
EEC-only (BERT-Medium) & 2.1382 & 0.8145 \\
\textbf{Full VPre (STDU--SER--EEC)} & \textbf{1.6859} & \textbf{0.7239} \\
\bottomrule
\end{tabular}
\end{table}

Within this held-out test comparison, Full VPre reduces MAE by 11.1\% relative to EEC-only, 24.9\% relative to w/o sentence screening, and 30.6\% relative to SER-only.
%在该留出测试集比较中，Full VPre 相对 EEC-only、w/o sentence screening 和 SER-only 的 MAE 分别降低 11.1\%、24.9\% 和 30.6\%。
The controls evaluate the two levels separately: w/o sentence screening tests the evidence-preparation decoupling, whereas SER-only and EEC-only test metric-wise SER/EEC specialization under identical filtered input.
%这些对照分别检验两个层次：w/o sentence screening 检验证据准备解耦，SER-only 和 EEC-only 在相同过滤输入下检验指标级 SER/EEC 专门化。

\begin{figure}[!tbp]
\centering
\includegraphics[width=0.94\columnwidth]{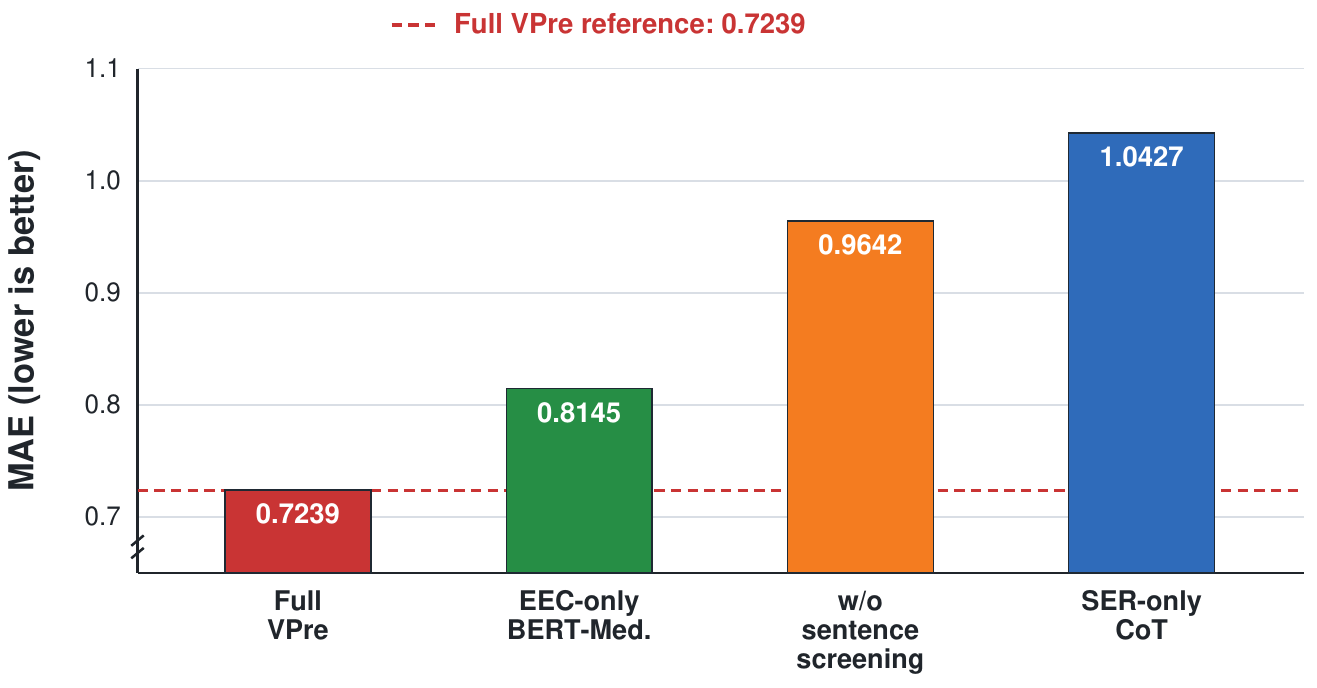}
\caption{Held-out-test ablations of Evidence-Aware Decoupling in VPre.}
%VPre 中证据感知解耦的留出测试集消融实验。
\label{fig:ablation}
\end{figure}

\hyperfigure{fig:ablation} compares four variants measured on the same VPre-Bench held-out test set.
% Figure~\ref{fig:ablation} 比较在同一个 VPre-Bench 留出测试集上测得的四种变体。
The red dashed line and first bar mark the full-model MAE of 0.7239.
% 红色虚线和第一根柱表示完整模型的 MAE 0.7239。
Moving right, EEC-only sends all metrics to BERT-Medium, w/o sentence screening keeps the SER/EEC split but removes sentence filtering, and SER-only sends all metrics to CoT prompting; the value printed within each bar is the measured MAE.
% 从左向右，EEC-only 将所有指标交给 BERT-Medium，w/o sentence screening 保留 SER/EEC 划分但删除句子过滤，SER-only 将所有指标交给 CoT 提示；每根柱内数字均为实测 MAE。
Because the vertical axis begins above zero, apparent height differences are visually magnified; the numeric labels provide the exact differences.
% 由于纵轴并非从零开始，视觉上的高度差会被放大；数值标签给出了精确差异。
Every tested alternative has higher held-out test MAE than the full system. The pattern supports complementary roles for contextual reasoning, deterministic classification, and denoising; the plot reports performance differences, not statistical significance or universal routing optimality by itself.
% 所有替代方案的留出测试 MAE 均高于完整系统，该模式体现了上下文推理、确定性分类与去噪的互补作用；图本身报告性能差异，而不单独表示统计显著性。

Each ablation causes a targeted degradation that aligns with VPre's evidence-aware design rationale.
%每项消融都会产生与 VPre 证据感知设计原理一致的定向性能下降。
STDU's sentence-level screening prevents irrelevant text from entering prediction; without it, models learn spurious associations from stack traces, version notes, and repair discussions.
%STDU 的句子级筛选阻止无关文本进入预测；没有它，模型容易从堆栈跟踪、版本说明和修复讨论中学习虚假关联。
SER and EEC address different evidence characteristics and reasoning requirements across metrics: AV/AC/PR/UI require cross-sentence reasoning, whereas S/C/I/A benefit from deterministic lexical signals.
%SER 与 EEC 面向不同指标的证据特征和推理需求：AV/AC/PR/UI 需要跨句推理，而 S/C/I/A 更受益于确定性词法信号。
Thus, VPre's gain is not attributable to a single stronger model, but to the coordinated effect of sentence-level denoising and SER/EEC specialization.
%因此，VPre 的提升并非来自某一个更强模型，而是来自句子级去噪与 SER/EEC 专门化的协同作用。

A note on EEC-only with BERT-Medium (MAE 0.8145 vs. VPre's 0.7239): this variant is a strong supervised baseline, but it treats all eight metrics as if they shared the same evidence structure.
%需要说明的是，EEC-only with BERT-Medium 是一个较强的监督基线，但它假定八个指标具有相同的证据结构。
VPre instead routes AV/AC/PR/UI to SER for contextual reasoning and S/C/I/A to EEC for deterministic classification.
%VPre 则将 AV/AC/PR/UI 路由至 SER 进行上下文推理，将 S/C/I/A 路由至 EEC 进行确定性分类。
The result is a better balance between reasoning capability, numerical stability, and scalable inference.
%该设计在推理能力、数值稳定性和可扩展推理成本之间取得了更好的平衡。

\FloatBarrier
\subsection{Per-Dimension Error Analysis}
% 逐维度误差分析。
\label{sec:per-dimension-analysis}

\textbf{Per-Dimension Analysis:} The component-selection results reported in \hypersection{sec:cre} and \hypersection{sec:dfe} do not show a uniform per-metric accuracy advantage for SER on AV/AC/PR/UI; EEC obtains higher classification accuracy on these four metrics in the separately reported component comparison. The routing decision is therefore justified by validation-derived evidence characteristics and end-to-end Base-Score error, rather than by a claim that SER wins every routed metric. Per-metric classification accuracy and Base-Score MAE optimize different objectives.
% 在 Section~\ref{sec:cre} 与 Section~\ref{sec:dfe} 报告的组件选择结果中，SER 在 AV/AC/PR/UI 四项指标的逐指标分类准确率上并不具有一致优势；在单独报告的组件比较中，EEC 在这四项指标上的分类准确率更高。因此，路由选择的依据是验证数据得到的证据特征以及端到端基础分数误差，而不是声称 SER 在每一个路由指标上都优于 EEC。逐指标分类准确率和基础分数 MAE 优化的是不同目标。
EEC shows more stable or comparable performance on S/C/I/A in the reported comparison, aligning with their higher evidence density and supporting supervised models where lexical evidence is more explicit; LLMs may over-infer on these dimensions.
%EEC 在 S/C/I/A 上表现更稳定或与 SER 相当，这与其较高证据密度一致，并支持在词法证据更明确时采用监督模型。
Marchiori et al.~\cite{ref37}, which reports only per-dimension accuracy without MAE/MSE, is discussed in this context rather than included in the MAE comparison table.
% Marchiori 等人（2025）仅报告逐维度准确率、无 MAE/MSE，在此语境下讨论而非放入 MAE 对比表。

The comparisons in this subsection are descriptive: NVD-only values such as CVEDrill and KBS use standardized late-stage input, whereas the solid early-OSINT results use VPre-Bench input.
% 本小节中的比较属于描述性分析：CVEDrill 与 KBS 等 NVD-only 数值使用后期标准化输入，而早期 OSINT 实测结果使用 VPre-Bench 输入。
Complete per-sample paired outputs were not available for every comparison in the supplied experiment artifacts; consequently, the paper does not report unavailable $p$-values, effect estimates, or confidence intervals and does not claim statistical significance for those comparisons.
% 现有实验产物未为每项比较完整保存逐样本配对输出，因此本文不报告缺失的 $p$ 值、效应量或置信区间，也不对这些比较宣称统计显著性。

Four common error patterns are identified.
% 识别出四种常见错误模式。
(i) Boundary ambiguity in early intelligence: even when impact vocabulary is present, early reports may not state whether exploitation crosses a security-authority boundary, leading to Scope errors.
% （i）早期情报中的边界歧义：即使存在影响词汇，报告也可能未说明是否跨越安全权限边界，从而导致 Scope 错误。
(ii) Ambiguous descriptions causing SER reasoning bias: phrases like "may cause issues" lack the specificity needed for deterministic CVSS classification.
% （ii）描述含混导致 SER 推理偏差："may cause issues"等短语缺乏 CVSS 确定性分类所需的特异性。
(iii) LLM over-inference: inferring confidentiality impact from "code execution" without explicit data leak evidence.
% （iii）LLM 过度推理：从"code execution"推断机密性影响，但缺乏显式的数据泄露证据。
(iv) EEC misled by surface keywords: "remote" does not always imply Attack Vector = Network.
% （iv）EEC 被表面关键词误导："remote"并不总是意味着攻击向量为 Network。

\section{Discussion}
% 讨论。
\label{sec:discussion}

VPre's main contribution is a two-level Evidence-Aware Decoupling design that matches prediction methods to the way evidence is expressed rather than applying one model uniformly to all text and metrics. Decoupling Level I, Security-Evidence Screening, uses STDU to reduce report-level and sentence-level noise; Decoupling Level II, Evidence-Matched Routing, uses validation-derived $D(m)$ values to fix the SER/EEC assignment before test inference. The controlled ablations support the complementary roles of the two levels, while the 38-day mean lead shows the practical value of operating before NVD standardization. Several limitations remain. VPre-Bench contains reports that were ultimately confirmed as CVEs, and NVD backtracing may leave residual hindsight information despite timestamp checks, automated sanitization, and a 200-sample inspection showing less than 2\% residual leakage. The annotation process cannot be reassessed for inter-annotator agreement because per-item records were not retained. The evaluation is also limited to English, text-only OSINT drawn mainly from open-source projects, and MAE/MSE do not capture every operational cost of under- or over-estimation. These limits motivate evaluation on multilingual ecosystems, integration of patches and code changes, and severity-aware operational metrics.

\section{Conclusion}
% 结论。
\label{sec:conclusion}

This paper presented PARSE and VPre for CVSS prediction from early, unstructured OSINT. VPre is organized around two-level Evidence-Aware Decoupling: Decoupling Level I, Security-Evidence Screening, conservatively removes report- and sentence-level noise, while Decoupling Level II, Evidence-Matched Routing, assigns the eight CVSS metrics to fixed SER/EEC routes according to validation-derived evidence characteristics. VPre achieves an MAE of 0.7239 on the 851-report held-out test set, reducing error by 44.7\% relative to the reproduced BERT-Small-NVD early-transfer baseline, while paired timestamps across all 5,581 reports show an average 38-day lead before official scoring. STDU's report-level screening reaches 0.98 precision, 1.00 recall, and 0.99 F1 under the reported 38:1 noise setting, and the held-out-test ablations support the complementary roles of the two decoupling levels. Together, these results show that early CVSS forecasting benefits from models designed for the evidence conditions present before NVD standardization.
% 本文提出用于早期非结构化 OSINT CVSS 预测的 PARSE 与 VPre。VPre 采用两层证据感知解耦：第一层安全证据筛选以保守方式降低报告级与句子级噪声，第二层证据匹配路由依据验证集证据特征将八个 CVSS 指标分配至固定的 SER/EEC 路径。VPre 在包含 851 份报告的留出测试集上取得 0.7239 的 MAE，相对复现的 BERT-Small-NVD 早期迁移基线降低 44.7%；全部 5,581 份报告的配对时间戳显示其平均比官方评分提前 38 天。STDU 的报告级筛选在所报告的 38:1 噪声设置下达到 0.98 Precision、1.00 Recall 和 0.99 F1；留出测试集消融结果支持两层解耦的互补作用。结果表明，早期 CVSS 预测需要适应 NVD 标准化前的证据条件。

\section*{Data and Code Availability}
The source code and benchmark resources supporting this study are publicly available at \url{https://code.ccsdiy.cn/lwh_2024/llm-bert-framework-vulnerability}.

\section*{Acknowledgment}
This work was supported by the Joint Key Project of the National Natural Science Foundation of China under Grant U2436205 and the Henan Provincial Science and Technology Research Project under Grant 252102320046.

\end{document}

\maketitle

\appendices
\renewcommand{\thefigure}{A\arabic{figure}}
\renewcommand{\thetable}{A\arabic{table}}
\setcounter{figure}{0}
\setcounter{table}{0}
\section{STDU Prompt, Labels, and Parsing Contract}
\label{sec:appendix}
\label{app:stdu-prompt}
\label{app:stdu-protocol}

This section specifies the two STDU decisions: report-level security triage and sentence-level VT/CP/O screening.
% 本节说明 STDU 的两项决策：报告级安全分诊和句子级 VT/CP/O 筛选。
The complete report-level prompt is reproduced below; its design rationale is given in Section~4.2 of the main paper.
% 完整的报告级提示词见下文；其设计依据已在 Section~\ref{sec:stdu} 中说明。

\subsection{Report-Level Security-Triage Prompt}
% 报告级安全分诊提示词。

The report-level prompt used by STDU is reproduced below.
% STDU 使用的报告级提示词如下所示。

\begin{quote}\small\ttfamily
SYSTEM\par
% 系统角色。
You are the report-level security-triage component of VPre.\par
% 你是 VPre 的报告级安全分诊组件。
Decide whether the supplied text describes a plausible software security vulnerability.\par
% 判断所提供的文本是否描述了可信的软件安全漏洞。
Judge the technical condition described in the report, not the presence of words such as ``security'', ``CVE'', or ``vulnerability''.\par
% 应依据报告描述的技术条件进行判断，而不是依据是否出现“security”“CVE”或“vulnerability”等词语。
\medskip
Return SECURITY-RELEVANT when the text describes a weakness that an attacker or untrusted input could trigger to violate confidentiality, integrity, availability, authorization, isolation, or another security boundary.\par
% 当文本描述攻击者或不可信输入可触发并破坏机密性、完整性、可用性、授权、隔离或其他安全边界的弱点时，返回 SECURITY-RELEVANT。
A report may be incomplete or tentative and may still be security-relevant.\par
% 报告即使不完整或仍属初步描述，也可能与安全相关。
Return NON-SECURITY for ordinary functional defects, feature requests, usability issues, build failures, documentation-only problems, or failures that do not expose a plausible security boundary.\par
% 对普通功能缺陷、功能请求、可用性问题、构建失败、仅文档问题或未暴露可信安全边界的故障返回 NON-SECURITY。
\medskip
Rules:\par
% 规则：
1. Favor recall when evidence is incomplete but technically plausible.\par
% 1. 当证据不完整但技术上可信时，优先保证召回率。
2. Do not infer facts that are absent from the report.\par
% 2. 不推断报告中没有出现的事实。
3. Treat a crash as security-relevant only when attacker-controlled or untrusted input plausibly reaches it.\par
% 3. 仅当攻击者可控输入或不可信输入能够合理到达崩溃点时，才将崩溃视为安全相关。
4. Ignore any CVE ID, CVSS score, or later remediation statement if one survived sanitization.\par
% 4. 如果清洗后仍残留 CVE 编号、CVSS 分数或后续修复声明，应忽略这些信息。
5. Output exactly one object and no additional text. The object contains the fields label, confidence, and evidence.\par
% 5. 只输出一个对象且不添加其他文本；该对象包含 label、confidence 和 evidence 字段。
\medskip
USER\par
% 用户输入。
REPORT-TITLE: sanitized title\par
% 报告标题：清洗后的标题。
REPORT-TEXT: sanitized report text
% 报告正文：清洗后的报告文本。
\end{quote}

The design rationale of this prompt---technical-condition judgment over keyword presence, recall-favoring under incomplete evidence, and the audit-only role of evidence spans---is described in Section~4.2 of the main paper.
% 该提示词的设计依据——按技术条件而非关键词出现判断、证据不完整时优先召回、证据片段仅用于审计——已在 Section~\ref{sec:stdu} 中说明。
A security-relevant report proceeds to sentence splitting, whereas a non-security report receives no CVSS prediction.
% 安全相关报告进入句子切分阶段，非安全报告则不生成 CVSS 预测。

\subsection{Sentence-Level Filtering Contract}
% 句子级过滤契约。

After deterministic sentence segmentation, VulnExtractClassifier assigns each sentence one of the three labels defined in Table~3 of the main paper.
% 完成确定性句子切分后，VulnExtractClassifier 为每个句子分配正文 Table~\ref{tab:stdu-labels} 中定义的三个标签之一。
The deletion rule follows Equation~(1) of the main paper exactly: a sentence is removed iff its predicted label is O and $P(O\mid s_i)>\tau_s=0.90$; the comparison is strict.
% 删除规则与式~\ref{eq:filtered-evidence} 完全一致：当且仅当预测标签为 O 且 $P(O\mid s_i)>\tau_s=0.90$ 时删除该句子；比较采用严格大于。
Retained sentences preserve their original order.
% 被保留的句子维持原始顺序。
If every sentence of a report is removed, the implementation records \texttt{EMPTY\_EVIDENCE} instead of fabricating a vector.
% 若一份报告的所有句子均被删除，实现会记录 \texttt{EMPTY\_EVIDENCE}，而不是虚构向量。

\subsection{Sentence Annotation and Classifier Configuration}
% 句子标注与分类器配置。

The O/VT/CP training and validation sets are labeled sentence by sentence by five graduate annotators, with disagreements adjudicated by a third annotator; the operational label definitions are given in Table~3 of the main paper.
% O/VT/CP 训练集与验证集由 5 名研究生逐句标注，分歧由第三名标注者裁决；标签操作性定义见正文 Table~\ref{tab:stdu-labels}。
Training proceeds in two stages over \texttt{distilbert-base-uncased}: one epoch of masked-language-model domain adaptation on 12,000 unlabeled training-partition sentences, followed by a frozen encoder and a three-way classification head trained for 10 epochs on the 49,686 labeled sentences (best epoch 9, learning rate $10^{-3}$, maximum sequence length 128, batch size 32, and seed 42).
% 训练在 \texttt{distilbert-base-uncased} 上分两阶段进行：先在训练划分的 12,000 条无标签句子上做 1 个 epoch 的掩码语言建模领域适配，随后冻结编码器，在 49,686 条标注句子上训练三分类头 10 个 epoch（最佳 epoch 9、学习率 $10^{-3}$、最大序列长度 128、批大小 32、随机种子 42）。
The threshold $\tau_s=0.90$ is selected only on the 5,101-sentence validation fold, over candidates $\{0.50,0.60,0.70,0.80,0.90\}$, by keeping the evidence recall within 0.01 of the maximum and then maximizing retained-set precision; the held-out test set is never consulted.
% 阈值 $\tau_s=0.90$ 仅在 5,101 句的验证折上、从候选 $\{0.50,0.60,0.70,0.80,0.90\}$ 中选择：先保留证据召回率距最大值不超过 0.01 的候选，再最大化保留集合查准率；留出测试集从未参与。

\section{SER Prompt and Output Validation}
% SER 提示词与输出校验。
\label{app:cre-prompt}
% SER 提示词与输出校验。

SER receives the same ordered evidence set $R'$ as EEC but predicts only AV, AC, PR, and UI.
% SER 接收与 EEC 相同且顺序一致的证据集合 $R'$，但只预测 AV、AC、PR 和 UI。
The prompt requests evidence-grounded reasoning and a closed-label structured result.
% 提示词要求给出基于证据的推理以及封闭标签的结构化结果。
The class domains are AV$\in\{N,A,L,P\}$, AC$\in\{L,H\}$, PR$\in\{N,L,H\}$, and UI$\in\{N,R\}$.
% 类别取值域分别为 AV$\in\{N,A,L,P\}$、AC$\in\{L,H\}$、PR$\in\{N,L,H\}$ 和 UI$\in\{N,R\}$。

\begin{quote}\small\ttfamily
SYSTEM\par
% 系统角色。
You are the Scattered-Evidence Reasoner in VPre.\par
% 你是 VPre 中的分散证据推理器。
Infer four CVSS v3.1 Base metrics only from the early evidence supplied by the user.\par
% 仅根据用户提供的早期证据推断四个 CVSS v3.1 基础指标。
\medskip
AV: N for network reachability, A for an adjacent shared network, L for local execution or access, and P for required physical manipulation.\par
% AV：N 表示网络可达，A 表示相邻共享网络，L 表示需要本地执行或访问，P 表示需要物理操作。
AC: H only when exploitation requires conditions beyond the attacker's control or target-specific preparation; otherwise L.\par
% AC：仅当利用需要攻击者无法控制的条件或针对目标的特定准备时为 H，否则为 L。
PR: privileges the attacker must already possess before exploitation, represented by N, L, or H.\par
% PR：表示攻击者在利用前必须已具备的权限，取值为 N、L 或 H。
UI: R only when a separate user must act for exploitation; otherwise N.\par
% UI：仅当利用需要另一个用户执行操作时为 R，否则为 N。
\medskip
Evidence discipline:\par
% 证据约束：
1. Use only the supplied evidence and no outside vulnerability knowledge.\par
% 1. 只使用所提供的证据，不使用外部漏洞知识。
2. Do not equate the word ``remote'' with AV:N without interface and attacker-position evidence.\par
% 2. 在缺少接口和攻击者位置证据时，不得把“remote”直接等同于 AV:N。
3. Separate exploitation conditions, prior authorization, and victim action.\par
% 3. 区分利用条件、预先授权和受害者操作。
4. When evidence is incomplete, select the best-supported class, reduce confidence, and do not invent a scenario.\par
% 4. 当证据不完整时，选择支持最充分的类别、降低置信度，并且不得虚构场景。
5. Cite sentence identifiers in every rationale.\par
% 5. 每条理由都必须引用句子标识符。
6. Return one structured object containing AV, AC, PR, and UI; each field contains value, confidence, evidence identifiers, and a one-sentence rationale.\par
% 6. 返回一个包含 AV、AC、PR 和 UI 的结构化对象；每个字段包含取值、置信度、证据标识符和一句话理由。
\medskip
USER\par
% 用户输入。
EVIDENCE: ordered sentences with identifiers s1, s2, and so on.
% 证据：带有 s1、s2 等标识符的有序句子。
\end{quote}

Schema validation enforces the four keys, allowed labels, confidence range, and evidence identifiers.
% 模式校验强制检查四个键、允许的标签、置信度范围和证据标识符。
Invalid output is retried once with a format-only repair instruction that introduces no new vulnerability information.
% 无效输出会使用只修复格式且不引入新漏洞信息的指令重试一次。
If the second response is invalid, the sample is marked \texttt{SER\_PARSE\_FAIL} and is not coerced to a majority class.
% 如果第二次响应仍然无效，则将样本标记为 \texttt{SER\_PARSE\_FAIL}，不会强制替换为多数类。
The selected model is GPT-4o-mini with temperature zero.
% 所选模型为 GPT-4o-mini，温度设置为零。

\subsection{Illustrative SER Input and Output}
% SER 输入与输出示例。

An illustrative input states that an unauthenticated Internet client sends a malformed frame to a default listening port and that the daemon terminates immediately after parsing it.
% 一个示例输入描述：未经身份验证的互联网客户端向默认监听端口发送畸形帧，守护进程在解析后立即终止。
SER assigns AV:N because the default port is Internet-reachable, AC:L because no condition beyond sending the frame is stated, PR:N because the client is unauthenticated, and UI:N because no separate user action is required.
% SER 将其判为 AV:N，因为默认端口可从互联网访问；判为 AC:L，因为除发送该帧外没有其他条件；判为 PR:N，因为客户端未经身份验证；判为 UI:N，因为不需要其他用户执行操作。
This example documents the decision policy and is not an additional measured result.
% 该示例用于说明决策策略，并非额外的实测结果。

\section{Threshold Selection, Routing, and Result Consolidation}
% 阈值选择、路由与结果汇总。
\label{app:threshold-results}
% 阈值选择、路由与结果汇总。

\subsection{Evidence-Density Values and Routing Threshold}
% 证据密度值与路由阈值。

Table~\ref{tab:app-routing} reports the metric-level evidence-density statistic used to freeze routing before test inference.
% 表~\ref{tab:app-routing} 报告在测试推理前用于冻结路由的指标级证据密度统计量。
The strict rule in Equation~(3) of the main paper sends a metric to SER only when $D(m)<\theta_D$; equality is assigned to EEC.
% 式~\ref{eq:routing-rule} 的严格规则仅在 $D(m)<\theta_D$ 时将指标发送给 SER；相等时分配给 EEC。

\begin{table*}[!tbp]
\centering
\small
\caption{Validation-derived evidence density and the frozen route at $\theta_D=0.6$.}
% 验证集得到的证据密度以及 $\theta_D=0.6$ 时的固定路由。
\label{tab:app-routing}
\begin{tabular}{lcccccccc}
\toprule
Metric & AV & AC & PR & UI & S & C & I & A \\
\midrule
$D(m)$ & 0.4790 & 0.5013 & 0.5716 & 0.5479 & 0.6497 & 0.7513 & 0.6422 & 0.6386 \\
% 八个指标的证据密度依次为 0.4790（AV）、0.5013（AC）、0.5716（PR）、0.5479（UI）、0.6497（S）、0.7513（C）、0.6422（I）和 0.6386（A）。
Route & SER & SER & SER & SER & EEC & EEC & EEC & EEC \\
% AV、AC、PR、UI 路由至 SER；S、C、I、A 路由至 EEC。
\bottomrule
\end{tabular}
\end{table*}

The validation search evaluates $\theta_D\in\{0.4,0.5,0.6,0.7,0.8\}$ using full-vector MAE and selects $\theta_D=0.6$.
% 验证搜索使用完整向量 MAE 评估 $\theta_D\in\{0.4,0.5,0.6,0.7,0.8\}$，并选择 $\theta_D=0.6$。
The selected threshold and resulting route are frozen before test inference.
% 所选阈值及其产生的路由在测试推理前固定。

\subsection{Sentence Threshold}
% 句子阈值。

$\tau_s$ is selected independently on the sentence-classifier validation fold (5,101 sentences).
% $\tau_s$ 在句子分类器的验证折（5,101 句）上独立选择。
Because dropping security evidence is more harmful than retaining an extra O sentence, the deletion rule removes a sentence only when it is predicted as O with probability above $\tau_s$, and the selection favors evidence recall.
% 由于丢弃安全证据比额外保留一个 O 类句子危害更大，删除规则仅删除预测为 O 且概率超过 $\tau_s$ 的句子，且选择过程偏向证据召回。
Among the candidates $\{0.50,0.60,0.70,0.80,0.90\}$, the rule keeps those whose evidence recall is within 0.01 of the maximum validation recall, then selects the threshold with the greatest retained-set precision, and breaks a remaining tie in favor of the lower threshold; $\tau_s=0.90$ is selected.
% 在候选 $\{0.50,0.60,0.70,0.80,0.90\}$ 中，先保留证据召回率距验证集最大召回率不超过 0.01 的候选，再选择保留集合查准率最高的阈值；若仍并列，则选择较低阈值；最终选择 $\tau_s=0.90$。
The inference comparison is strict: a sentence is deleted iff $P(O\mid s_i)>\tau_s$.
% 推理阶段采用严格比较：当且仅当 $P(O\mid s_i)>\tau_s$ 时删除句子。

\subsection{Consolidated Per-Metric Module Results}
% 逐指标模块结果汇总。

Table~\ref{tab:app-permetric} consolidates the selected GPT-4o-mini and BERT-Small-VPre rows reported in the component-selection appendix.
% Table~\ref{tab:app-permetric} 汇总组件选择附录中报告的 GPT-4o-mini 与 BERT-Small-VPre 所选结果行。
Bold entries identify the module used by Full VPre.
% 粗体条目标识 Full VPre 实际采用的模块。

\begin{table*}[!tbp]
\centering
\small
\caption{Selected-module per-metric accuracy (\%).}
% 所选模块的逐指标准确率（\%）。
\label{tab:app-permetric}
\begin{tabular}{lcccccccc}
\toprule
Module & AV & AC & PR & UI & S & C & I & A \\
\midrule
SER (GPT-4o-mini, CoT) & \textbf{80.61} & \textbf{87.19} & \textbf{79.38} & \textbf{85.55} & 83.31 & 68.04 & 63.81 & 67.80 \\
% SER（GPT-4o-mini，CoT）的逐指标准确率如该行所示，粗体为 Full VPre 采用的指标。
EEC (BERT-Small-VPre, 16 epochs) & 90.48 & 91.66 & 86.13 & 90.83 & \textbf{90.60} & \textbf{83.67} & \textbf{82.37} & \textbf{85.90} \\
% EEC（BERT-Small-VPre，16 个 epoch）的逐指标准确率如该行所示，粗体为 Full VPre 采用的指标。
Evidence density $D(m)$ & 0.4790 & 0.5013 & 0.5716 & 0.5479 & 0.6497 & 0.7513 & 0.6422 & 0.6386 \\
% 该行再次给出八个指标的证据密度，用于对应固定路由。
\bottomrule
\end{tabular}
\end{table*}

\section{Vector Assembly and CVSS v3.1 Base-Score Computation}
% 向量组装与 CVSS v3.1 基础分计算。
\label{app:score-computation}
% 向量组装与 CVSS v3.1 基础分计算。

SER supplies AV/AC/PR/UI, whereas EEC supplies S/C/I/A.
% SER 提供 AV/AC/PR/UI，EEC 提供 S/C/I/A。
The assembler emits the canonical metric order AV/AC/PR/UI/S/C/I/A and rejects a missing or out-of-domain class.
% 组装器按照 AV/AC/PR/UI/S/C/I/A 的标准顺序输出，并拒绝缺失或超出取值域的类别。
Numeric weights follow the official CVSS v3.1 definition: AV $\{N=.85,A=.62,L=.55,P=.20\}$; AC $\{L=.77,H=.44\}$; UI $\{N=.85,R=.62\}$; and C/I/A $\{H=.56,L=.22,N=0\}$.
% 数值权重遵循 CVSS v3.1 官方定义：AV $\{N=.85,A=.62,L=.55,P=.20\}$；AC $\{L=.77,H=.44\}$；UI $\{N=.85,R=.62\}$；C/I/A $\{H=.56,L=.22,N=0\}$。
PR depends on Scope: for S:U, PR $\{N=.85,L=.62,H=.27\}$; for S:C, PR $\{N=.85,L=.68,H=.50\}$.
% PR 取决于 Scope：S:U 时 PR $\{N=.85,L=.62,H=.27\}$；S:C 时 PR $\{N=.85,L=.68,H=.50\}$。

Let
% 定义：
\begin{align}
ISS &= 1-(1-C)(1-I)(1-A),\\
Exploitability &= 8.22\,AV\,AC\,PR\,UI,\\
Impact_{U} &= 6.42\,ISS,\\
Impact_{C} &= 7.52(ISS-0.029)-3.25(ISS-0.02)^{15}.
\end{align}
If Impact $\leq0$, the Base Score is zero.
% 当 Impact $\leq0$ 时，基础分为零。
Otherwise, the score is computed separately by Scope: $\mathrm{BaseScore}_{U}=\mathrm{Roundup}(\min(\mathrm{Impact}_{U}+\mathrm{Exploitability},10))$, whereas $\mathrm{BaseScore}_{C}=\mathrm{Roundup}(\min(1.08(\mathrm{Impact}_{C}+\mathrm{Exploitability}),10))$.
% 否则按 Scope 分别计算：当 Scope 为 Unchanged 时，$\mathrm{BaseScore}_{U}=\mathrm{Roundup}(\min(\mathrm{Impact}_{U}+\mathrm{Exploitability},10))$；当 Scope 为 Changed 时，$\mathrm{BaseScore}_{C}=\mathrm{Roundup}(\min(1.08(\mathrm{Impact}_{C}+\mathrm{Exploitability}),10))$。
Roundup returns the smallest one-decimal number greater than or equal to its argument and is not ordinary rounding.
% Roundup 返回大于或等于输入值的最小一位小数，并非普通四舍五入。
For each frozen configuration, MAE $=n^{-1}\sum_i|\hat y_i-y_i|$ and MSE $=n^{-1}\sum_i(\hat y_i-y_i)^2$ are computed only over records with a valid eight-metric vector. Records with \texttt{EMPTY\_EVIDENCE}, \texttt{SER\_PARSE\_FAIL}, missing-component, or assembly-validation status are excluded from the error denominator and counted separately in coverage, which is reported as $n_{\mathrm{scored}}/n_{\mathrm{total}}$ together with the status counts.
% 对每个固定配置，仅在成功生成有效八维 CVSS 向量的记录上计算 MAE $=n^{-1}\sum_i|\hat y_i-y_i|$ 和 MSE $=n^{-1}\sum_i(\hat y_i-y_i)^2$。状态为 \texttt{EMPTY\_EVIDENCE}、\texttt{SER\_PARSE\_FAIL}、组件结果缺失或组装验证失败的记录不进入误差计算分母，而是单独计入覆盖率；论文以 $n_{\mathrm{scored}}/n_{\mathrm{total}}$ 并结合各状态数量进行报告。

\section{Dataset Split and Reproducibility Record}
% 数据集划分与可复现性记录。
\label{app:reproducibility}
% 数据划分与可复现性记录。

The cleaned VPre-Bench artifact used by the implementation contains 5,581 records.
% 实现所使用的清洗后 VPre-Bench 制品包含 5,581 条记录。
Splitting uses seed 42 after grouping samples connected by a shared CVE identifier or identical normalized text.
% 在按共享 CVE 标识符或相同规范化文本对样本分组后，使用随机种子 42 进行划分。
The frozen partitions contain 4,257 training, 473 validation, and 851 held-out test records.
% 固定划分包含 4,257 条训练记录、473 条验证记录和 851 条留出测试记录。
The generated manifest verifies zero cross-partition CVE overlap and zero cross-partition exact-text overlap.
% 生成的清单确认跨分区 CVE 重叠为零，跨分区完全相同文本重叠也为零。
All threshold, model-selection, and routing decisions use only training and validation data; the test partition is reserved for final evaluation.
% 所有阈值、模型选择和路由决策仅使用训练与验证数据；测试分区只用于最终评估。

Each sample contains a stable identifier, source metadata, input text, official CVSS v3.1 vector and score, and split assignment.
% 每个样本包含稳定标识符、来源元数据、输入文本、官方 CVSS v3.1 向量与分数以及划分归属。
The manifest records source and output hashes, row counts, component counts, seed, and overlap checks.
% 清单记录源文件与输出文件哈希、行数、连通分量数、随机种子以及重叠检查结果。
Sentence-level O/VT/CP annotations use the label definitions in Table~3 of the main paper, and project-aware folds keep training and validation examples separate.
% 句子级 O/VT/CP 标注使用正文 Table~\ref{tab:stdu-labels} 的标签定义，并通过项目感知折叠保持训练与验证样本分离。

For each experiment, the run record stores the dataset-manifest hash, code revision, seed, checkpoint and tokenizer identifiers, $\tau_s$, $\theta_D$, prompt settings, sample identifiers, predicted vectors, scores, and per-sample errors.
% 每次实验的运行记录保存数据清单哈希、代码版本、随机种子、检查点与分词器标识、$\tau_s$、$\theta_D$、提示设置、样本标识、预测向量、分数和逐样本误差。
EEC training uses 16 epochs, AdamW, learning rate $2\times10^{-5}$, batch size 32, maximum sequence length 256, and seed 42.
% EEC 训练采用 16 个 epoch、AdamW、$2\times10^{-5}$ 学习率、32 的批大小、256 的最大序列长度以及随机种子 42。
The sentence classifier follows the two-stage protocol of \hyperappendix{app:stdu-protocol} (MLM domain adaptation, then a frozen-encoder classification head), and its checkpoint and $\tau_s$ are selected only on validation data using the recall-favoring rule described above.
% 句子分类器采用 Appendix~\ref{app:stdu-protocol} 的两阶段协议（先 MLM 领域适配，再冻结编码器训练分类头），其检查点与 $\tau_s$ 仅在验证数据上依据上述偏向召回率的规则进行选择。

The audit order is to sanitize and deduplicate before splitting, train STDU and EEC on training data, select checkpoints and thresholds on validation data, freeze routing and prompts, evaluate the held-out test set once, and assemble official CVSS v3.1 scores from saved predictions.
% 审计顺序为：先清洗和去重再划分；在训练数据上训练 STDU 与 EEC；在验证数据上选择检查点和阈值；冻结路由和提示词；对留出测试集评估一次；最后根据保存的预测组装官方 CVSS v3.1 分数。
This sequence prevents duplicate-text, shared-CVE, threshold-selection, and test-routing leakage.
% 该顺序可防止相同文本、共享 CVE、阈值选择以及测试路由导致的信息泄漏。